\documentclass{elsart}
\usepackage{graphicx} 
\journal{Nuclear Physics A}

\begin{document}

\begin{frontmatter}

\title{The PHOBOS Perspective on Discoveries at RHIC}

\author[ANL]{B.B.Back}
\author[BNL]{M.D.Baker}
\author[MIT]{M.Ballintijn}
\author[BNL]{D.S.Barton}
\author[BNL]{B.Becker}
\author[UIC]{R.R.Betts}
\author[UMD]{A.A.Bickley}
\author[UMD]{R.Bindel}
\author[INP]{A.Budzanowski}
\author[MIT]{W.Busza}
\ead{PHOBOS Spokesperson:busza@mit.edu}
\author[BNL]{A.Carroll}
\author[BNL]{Z.Chai}
\author[MIT]{M.P.Decowski}
\author[UIC]{E.Garc\'{\i}a}
\author[INP]{T.Gburek}
\author[ANL,BNL]{N.K.George}
\author[MIT]{K.Gulbrandsen}
\author[BNL]{S.Gushue}
\author[UIC]{C.Halliwell}
\author[UoR]{J.Hamblen}
\author[UoR]{A.S.Harrington}
\author[BNL]{M.Hauer}
\author[BNL]{G.A.Heintzelman}
\author[MIT]{C.Henderson}
\author[UIC]{D.J.Hofman}
\author[UIC]{R.S.Hollis}
\author[INP]{R.Ho\l y\'{n}ski}
\author[BNL,UIC]{B.Holzman}
\author[UIC]{A.Iordanova}
\author[UoR]{E.Johnson}
\author[MIT]{J.L.Kane}
\author[MIT,UIC]{J.Katzy}
\author[UoR]{N.Khan}
\author[UIC]{W.Kucewicz}
\author[MIT]{P.Kulinich}
\author[NCU]{C.M.Kuo}
\author[MIT]{J.W.Lee}
\author[NCU]{W.T.Lin}
\author[UoR]{S.Manly}
\author[UIC]{D.McLeod}
\author[UMD]{A.C.Mignerey}
\author[BNL,UIC]{R.Nouicer}
\author[INP]{A.Olszewski}
\author[BNL]{R.Pak}
\author[UoR]{I.C.Park}
\author[MIT]{H.Pernegger}
\author[MIT]{C.Reed}
\author[BNL]{L.P.Remsberg}
\author[UIC]{M.Reuter}
\author[MIT]{C.Roland}
\author[MIT]{G.Roland}
\author[MIT]{L.Rosenberg}
\author[UIC]{J.Sagerer}
\author[MIT]{P.Sarin}
\author[INP]{P.Sawicki}
\author[BNL]{H.Seals}
\author[BNL]{I.Sedykh}
\author[UoR]{W.Skulski}
\author[UIC]{C.E.Smith}
\author[BNL]{M.A.Stankiewicz}
\author[BNL]{P.Steinberg}
\author[MIT]{G.S.F.Stephans}
\author[BNL]{A.Sukhanov}
\author[NCU]{J.-L.Tang}
\author[UMD]{M.B.Tonjes}
\author[INP]{A.Trzupek}
\author[MIT]{C.M.Vale}
\author[MIT]{G.J.van~Nieuwenhuizen}
\author[MIT]{S.S.Vaurynovich}
\author[MIT]{R.Verdier}
\author[MIT]{G.I.Veres}
\author[MIT]{E.Wenger}
\author[UoR]{F.L.H.Wolfs}
\author[INP]{B.Wosiek}
\author[INP]{K.Wo\'{z}niak}
\author[ANL]{A.H.Wuosmaa}
\author[MIT]{B.Wys\l ouch}
\author[MIT]{J.Zhang}
\collab{PHOBOS collaboration \\ www.phobos.bnl.gov}

\address[ANL]{Argonne National Laboratory, Argonne, IL 60439-4843, USA}
\address[BNL]{Brookhaven National Laboratory, Upton, NY 11973-5000, USA}
\address[INP]{Institute of Nuclear Physics PAN, Krak\'{o}w, Poland}
\address[MIT]{Massachusetts Institute of Technology, Cambridge, MA 02139-4307, USA}
\address[NCU]{National Central University, Chung-Li, Taiwan}
\address[UIC]{University of Illinois at Chicago, Chicago, IL 60607-7059, USA}
\address[UMD]{University of Maryland, College Park, MD 20742, USA}
\address[UoR]{University of Rochester, Rochester, NY 14627, USA}

\date{\today}

\vspace{0.30in}
\begin{abstract}

This paper describes the conclusions that can be drawn
from the data taken thus far with the PHOBOS detector at RHIC. In the most central
Au+Au collisions at the highest beam energy, evidence is found for the
formation of a very high energy density system whose description in terms of 
simple hadronic degrees of freedom is inappropriate.  Furthermore, the 
constituents of this novel system are found to undergo a significant level of interaction.  The 
properties of particle production at RHIC energies are shown to follow a 
number of simple scaling behaviors, some of which continue trends found at 
lower energies or in simpler systems.  As a function of centrality, the 
total number of charged particles scales with the number of participating 
nucleons.   
When comparing Au+Au at different 
centralities, the dependence of the yield on the number of participants 
at higher $p_{_T}$ ($\sim$4~GeV/c) is very similar to that at low transverse momentum.  The measured 
values of charged particle pseudorapidity density and elliptic flow were found 
to be independent of energy over a broad range of pseudorapidities when 
effectively viewed in the rest frame of one of the
colliding nuclei, a property we describe as ``extended longitudinal 
scaling''.  Finally, the centrality and energy dependences of several 
observables were found to factorize to a surprising degree.

\end{abstract}



\end{frontmatter}

\section{Introduction}
\label{Sec1}

Currently, there exists a good understanding of the basic building blocks of
normal matter, and of the fundamental forces or interactions between them.  The bulk
of hadronic matter is comprised of partons (quarks and gluons) bound into  
neutrons,
protons, and subsequently nuclei by the strong force mediated by the field quanta, the gluons.
The fundamental interactions between these partons are
described by the theory of quantum chromodynamics (QCD) \cite{Kas93} and are
reasonably well understood.  However, because of the strength and non-Abelian nature 
of the interactions, finding solutions to the QCD equations remains
notoriously difficult.  As a result, the current understanding of the phase
structure of strongly interacting matter (what phases exist, what are the
properties of the matter in each phase, and what is the nature of the
transitions between phases) is only partly based on theoretical QCD calculations.  Instead, it is
driven, to a large extent, by experiment. Among many examples of the
significance of the properties of QCD ``matter'' is the fact that more than
98\% of the mass of all normal hadronic matter in the universe arises from the
interactions (i.e. the gluons and the sea quarks), not from the (current) mass of the valence quarks in the hadrons~\cite{PDG04}.  This
mass is generated predominantly by the lower energy interactions which are most
difficult to study quantitatively.  Areas of impact outside nuclear physics
include the evolution of the early universe, as well as the overall properties and interior structure of compact stars and 
stellar remnants. Both theory and experiment suggest the existence of a very
rich ``condensed matter'' governed by QCD.

At very short distances ($\ll$hadronic sizes) the QCD coupling constant
between partons is weak and decreases as the distance between the partons
decreases, a phenomenon known as ``asymptotic freedom'' \cite{Gro73,Pol73,Nob04}.
An expected consequence of asymptotic freedom is that a system created by
heating the vacuum to high temperatures should have the properties of an almost ideal
relativistic gas in which color is deconfined (first pointed out by \cite{Col75} using the term 
``quark soup'', see also \cite{Shu80,Sat81,Gro81}).  The high 
temperature of this medium entails an extremely high concentration of partons, 
whose thermodynamics follows the Stefan-Boltzmann law.   Such a system has 
traditionally been
designated the Quark-Gluon Plasma (QGP), a term proposed in \cite{Shu80}.  To specifically recognize its ideal,
weakly interacting nature, we use the term wQGP.  The current consensus is that
the whole universe was in the
wQGP state at an early stage
following the big bang.

At another extreme, it is known that the only stable configuration of
strongly interacting matter at low temperatures and densities is the multitude
of varieties of color neutral objects, namely the hadrons, as well as
conglomerates of hadrons such as atomic nuclei.  In addition, the QCD
Lagrangian (and the wQGP solution of that Lagrangian) is understood to have a
higher symmetry than the observed hadron states.  The solutions of QCD at
temperatures and densities which correspond to normal matter, i.e.\ the world
of hadrons and nuclei, spontaneously break this
so-called ``chiral symmetry'' (see, for example, \cite{Pag75,Mar78,Pis84}).  
The questions of what forms and
phases of QCD matter exist between the two extremes and what symmetries,
properties, and interactions characterize these phases, are currently the
subject of very active theoretical and experimental research (see, for example, \cite{Raj00b}).

On both the experimental and the theoretical fronts, there are very few tools
available for the study of QCD matter as a function of density and
temperature. To date, the most fruitful approach to the theoretical study of
high temperature QCD has been the use of numerical calculations based on the
techniques of lattice gauge theory.  These calculations suggest that at
low baryon densities there is a phase difference in QCD matter below and above
a critical temperature $T_c\sim$150--200~MeV or energy density 
$\sim$1~GeV/fm$^3$ (see, for example, \cite{Kar02a}, 
which quotes a $T_c$ of 175~MeV and an energy density of 700~MeV/fm$^3\pm$50\%).  
At another extreme, theoretical progress has
been made in recent years in the understanding of cold, ultra-dense, QCD matter
which must be in some color superconducting state
\cite{Sch03,Alf01,Raj00a}.  For example, there are indications that a dense,
cold system of equal numbers of u, d and s quarks can form a ``color-flavor locked''
superconducting phase.  This regime is currently out of range of experimentation using
accelerators, but such phenomena might be manifested in the dense cores of neutron
stars and, therefore, might be open to study through astronomical observation.
The possible connection of QCD and neutron stars has a long history (see, for example, \cite{Bay76,Cha77}). 

The most useful experimental approach in the area of high temperature QCD
matter is the detailed analysis of heavy ion collisions.  
In fact, the suggestion of the use of heavy ion collisions to create high density states of matter
predates the full development of QCD \cite{Cha73}.
The value of
$\sim$1~GeV/fm$^3$ is not much higher than the energy density inside nucleons 
($\sim$500~MeV/fm$^3$) and nuclei ($\sim$150~MeV/fm$^3$), and it is also
comparable to estimates of the initial energy density created in hadronic
collisions at high energy accelerators.  In heavy ion collisions at
relativistic velocities, there is both compression of the baryonic matter in
the nuclei and also the release of a large amount of energy within a small volume from
the almost simultaneous collisions of many nucleons.  One or the other, or
both, of these consequences of the interactions have the potential to produce 
new forms or phases of QCD matter.  This is one of the prime reasons
why in the past few decades much effort has been spent studying
collisions of heavy ions at higher and higher energies.
Extensive information can be found in the proceedings of the Quark Matter series
of conferences \cite{QM04} and in recent reviews \cite{Jac04,Hwa03,Kol03,Ris04}.
The conditions created may be similar to those of the early universe at about
10~$\mu$sec after the big bang.
Another important aspect of such studies is the extraction of valuable
information about the mechanisms of particle production in small and large
systems at high energies.

The most recent experimental facility for the study of heavy ion collisions is
the Relativistic Heavy Ion Collider (RHIC) at Brookhaven National
Laboratory.  Since the inception of the physics program in July, 2000, four
experiments at RHIC, namely BRAHMS, PHENIX, PHOBOS, and STAR, have studied
collisions of p+p, d+Au, and Au+Au at center-of-mass collision energies per
incident nucleon pair, $\sqrt{s_{_{NN}}}$, from 19.6 to 200~GeV.  
Note that, for technical reasons discussed in Appendix~\ref{SecAppB-1}, 
$\sqrt{s_{_{NN}}}$ for d+Au was actually larger by about 0.35\% but, for 
simplicity, this tiny difference is omitted in the text and figure labels of
this document. 
Data from all
four detectors are being studied to get a better understanding of the physics
of heavy ion collisions, and, in particular, to search for evidence of the
creation of new forms of QCD matter \cite{Wpa04}. To the best of our
knowledge, where there is overlap, there are no major differences in the
data and extracted results obtained by the four experiments at RHIC.  The
level of agreement is a testament to the quality of the detectors and the
analyses performed by the collaborations and is a great strength of the whole
RHIC research program.  This paper summarizes the most important results
obtained to date by the PHOBOS collaboration and the conclusions that can be
drawn from PHOBOS results, augmented where necessary by data from other
experiments.

One of the most important discoveries at RHIC is the evidence that, in central
Au+Au collisions at ultra-relativistic energies, an extremely high energy 
density system
is created, whose description in terms of
simple hadronic degrees of freedom is inappropriate.  
Furthermore, the constituents of this system experience a significant level of interaction with each other inside the medium.
These conclusions are
based on very general and, to a large extent, model independent arguments.

It is not claimed that the observed phenomena are unique to RHIC energies. Nor is it
claimed that there is direct evidence in the data analyzed so far for color
deconfinement or chiral symmetry restoration.  It should be noted that
interpretations of the data which invoke a high density of gluons or other
non-hadronic components are certainly consistent with, and could be construed
to provide at least circumstantial evidence for, deconfinement.  Also, the
definition of the concept of deconfinement is not so clear when the particles in
the medium interact significantly.  No convincing evidence has been found for the creation at
RHIC of the wQGP, in contrast to the expectations of a large part of the heavy ion
community in the era before the start of the RHIC physics program.  This
expectation may have partly resulted from a misinterpretation of the lattice results.
The calculations reveal that the pressure and energy density reach 70--80\% of
the Stefan-Boltzmann value (i.e.\ the value for a non-interacting gas) for
temperatures above the critical temperature (see, as one recent example,
\cite{Kan03}).  This observation was typically assumed to imply the presence
of a weakly interacting system although questions were occasionally raised (for one early example, see \cite{Plu84}).  More recently, this conclusion has been
seriously challenged (see, for example, \cite{Shu04a,Shu04c}).  
As an aside, some string theory models which have been shown to be related to QCD 
can be solved exactly in the strong-coupling limit and yield a result comparable
to $\sim$75\% of the Stefan-Boltzmann value \cite{Hor91,Gub98}.
This recent reversal of opinion was to a large degree driven by the experimental
results from RHIC.  Recent lattice QCD studies have shown that the quarks do retain
a degree of correlation above the critical temperature (see, for example,
\cite{Kar03,Shu04b}).  However, at extremely high energy density (for example, the
very early universe), the theoretical expectation remains that the system will
become weakly interacting \cite{Kaj03}.

Another equally interesting result from RHIC arose from the studies of the
mechanism of particle production in nuclear collisions.  Specifically, it has
been discovered that much of the data in this new regime can be expressed in terms of simple
scaling behaviors. 
Some of these behaviors had been noted in data at lower energies or for
simpler systems.  These observations suggest either the existence of strong
global constraints or some kind of universality in the mechanism of the
production of hadrons in high energy collisions, possibly connected to ideas of
parton saturation.  The data strongly suggest that the initial geometry and
very early evolution of the system establish conditions which determine the
final values of many observables.  The most concise formulation of this
discovery is the statement that the overall properties of the data appear to be
much simpler than any of the models invoked to explain them.  A full
exploration and detailed analysis of all aspects of the data will be required
for a complete understanding of the properties of QCD physics in the
interesting regime probed by heavy ion collisions at relativistic velocities.

Section~\ref{Sec2} of this paper describes the derived properties of the state
formed shortly after the collisions at RHIC, Sect.~\ref{Sec3} describes the
evidence that the constituents of this state interact significantly, and Sect.~\ref{Sec4}
discusses the broad range of scaling behaviors that have been discovered.

As a useful reference, the PHOBOS detector and its properties are briefly
described in Appendix~\ref{SecAppA}.  
Variables used in the description of the data, in particular those relating to
event characterization, are defined in Appendix ~\ref{SecAppB}.  The precise determination of the
collision impact parameter or centrality is critical to heavy ion physics in
general and the PHOBOS program in particular.  Appendix~\ref{SecAppC}
describes how centrality and the biases associated with triggering and various
elements of the data analysis are derived from measurements and simulations
for the various colliding systems and beam energies.

\section{Properties of the initial state produced at RHIC} 
\label{Sec2}

The primary goal of the RHIC accelerator was the study of QCD matter under
extreme conditions.  In particular, it was expected that the center-of-mass
energies more than an order of magnitude higher than achieved at the SPS accelerator at CERN would
lead to the creation of a system with significantly higher energy density.  An
additional consequence of the higher beam energy compared to the SPS was the
displacement of the projectile baryons a factor of two farther apart in
rapidity.  This was expected to lead to a lower baryon chemical potential in
the high energy density region at midrapidity.  Although progress has been
made recently in lattice calculations which include the effects of
a non-zero baryon chemical potential (see, for
example, \cite{Kan03,Kar04,Eli04,Eji04,Fod04,For04,Dav04} and references therein), the most extensively studied system remains one with a
value close to zero (see, for example, \cite{Lae03,Kar02b} and references therein).  Therefore, creation of
a system with a lower baryon chemical potential might offer the potential for
more reliable comparisons of experimental data to the fundamental QCD
predictions.  This section describes the conclusions that can be drawn from
PHOBOS data concerning these two critical properties of the state formed in
collisions of heavy ions at RHIC.

\subsection{Energy density} 
\label{Sec2-A}

In very high energy heavy ion interactions, the maximum energy density occurs
just as the two highly Lorentz contracted nuclei collide.  Clearly this system
is very far from being equilibrated and, as a result, the value of the energy
density, although well defined, may not be very interesting.  In any reference
frame, the potentially more interesting quantity is the energy density carried
by particles which are closer to equilibrium conditions, i.e.\ those particles
which have, on average, comparable longitudinal and transverse momenta.  These conditions
are roughly equivalent to restricting the particles to a range of
pseudorapidity $|\eta|\le$1.  Unfortunately, there are no direct
measures of energy density and, therefore, it must be inferred from the
properties of the detected particles.  PHOBOS data have been used to
investigate what range of initial energy densities are consistent with the
observations.  Studies of pseudorapidity and transverse momentum distributions,
as well as elliptic flow, have been combined to constrain assumptions about the
energy in the system and the time evolution of the volume from which the
particles emanate.

\begin{figure}[ht]
\begin{center}
 
\includegraphics[width=0.95\textwidth]{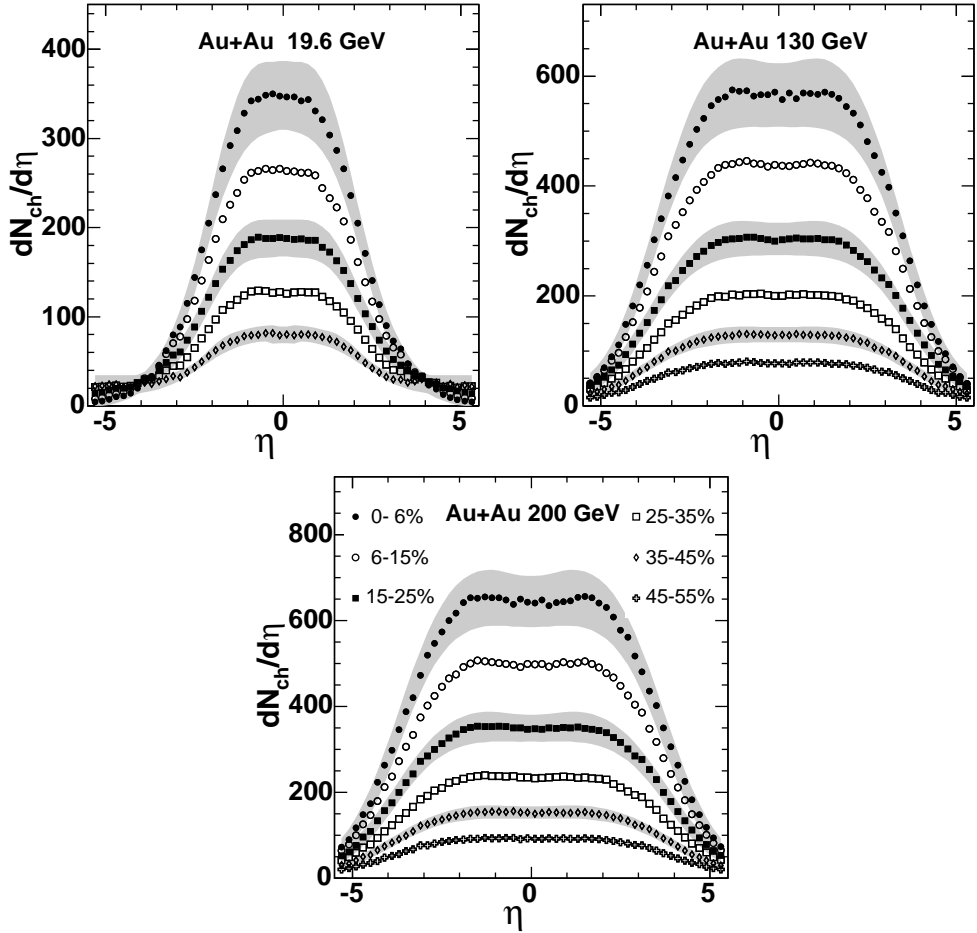}

\caption{ \label{WP1_dNdeta_eta_AuAu_19_130_200} 
Pseudorapidity density of charged particles emitted in Au+Au collisions at
three different values of the nucleon-nucleon center-of-mass energy \cite{Bac03c}.  Data are
shown for a range of centralities, labeled by the fraction of the total
inelastic cross section in each bin, with smaller numbers being more central.
Grey bands shown for selected centrality bins indicate the typical systematic uncertainties (90\% C.L.).  Statistical errors are smaller than the symbols.}

\end{center}
\end{figure}

\begin{figure}[ht]
\begin{center}

\includegraphics[width=0.95\textwidth]{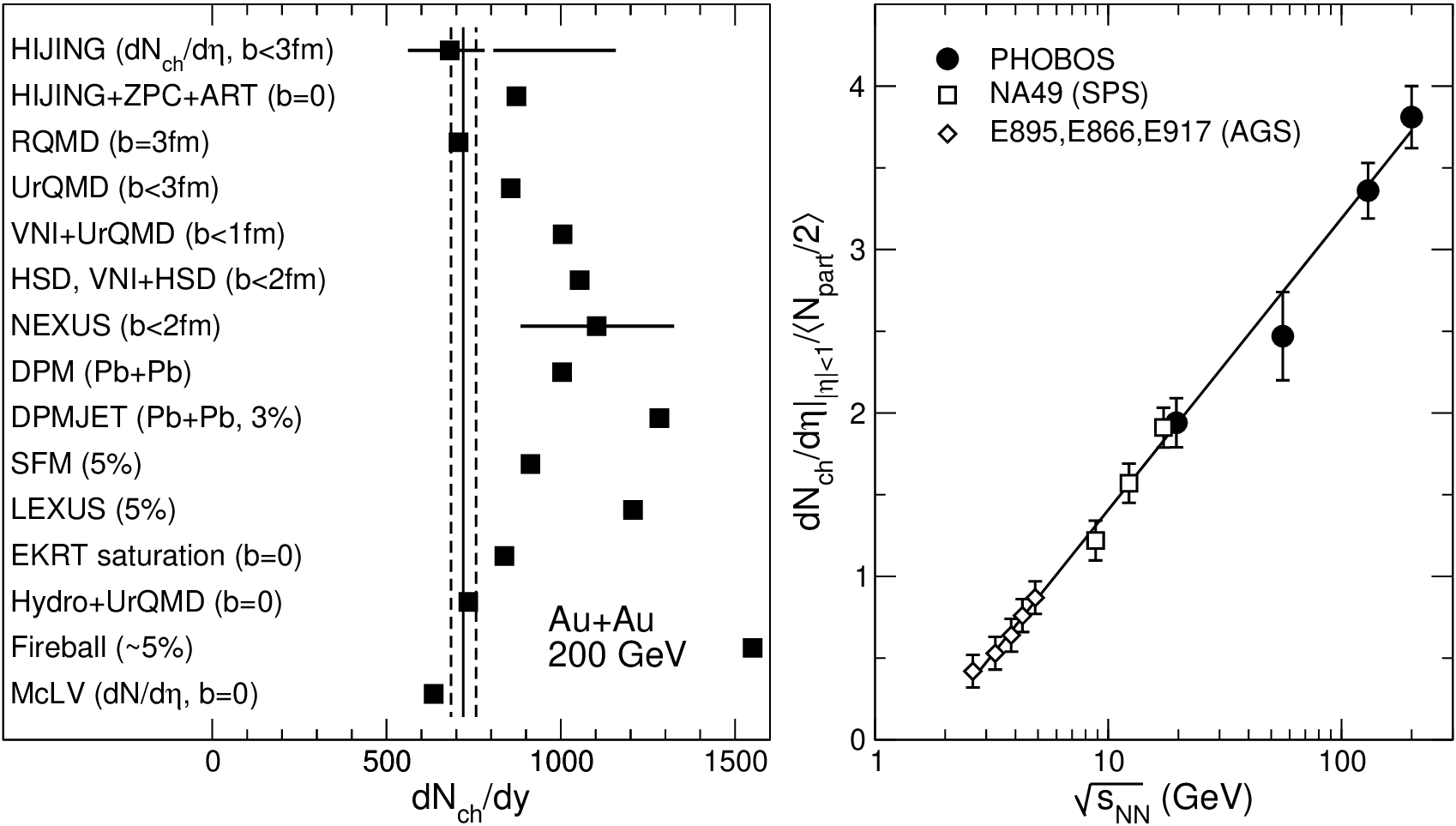}

\caption{ \label{WP2_dNdetaMidPred_sqrts} 
(Left panel) Results of PHOBOS measurements of the charged particle density near
midrapidity in central Au+Au at $\sqrt{s_{_{NN}}}$=200~GeV 
\cite{Bac03c,Bac02b,Bac02c,Bac04e} (shown by the
vertical line with the dashed lines denoting the systematic uncertainty)
compared to theoretical predictions. This panel is adapted from \cite{Esk02}.  
From top to bottom, the references for the models are 
\cite{Wan99,Zha99,Sor99,Ble99,Bas99a,Cas99,Dre99,Cap99,Ran99,Arm00,Jeo00,Esk00,Bas99b,Sta99,Kra01}.
See text for discussion.
(Right panel) Normalized
pseudorapidity density of charged particles emitted within $|\eta|\le$1 in
central Au+Au (AGS \cite{Ahl98a,Bac02e,Ahl00,Kla03,Dun99} and PHOBOS at RHIC \cite{Bac03c,Bac00a,Bac02a,Bac02b,Bac02c,Bac04e}) and Pb+Pb (SPS \cite{Afa02,Ant04}) collisions as a function of
nucleon-nucleon center-of-mass energy.  
See text for discussion.}

\end{center}
\end{figure}

Figure~\ref{WP1_dNdeta_eta_AuAu_19_130_200} shows distributions of charged particle pseudorapidity
densities, \linebreak
$dN_{ch}/d\eta$, for Au+Au collisions at $\sqrt{s_{_{NN}}}$=19.6,
130, and 200~GeV for various centralities \cite{Bac03c}.  The produced particle
densities are at their maximum near midrapidity and increase with both
collision energy and centrality.  The right panel of Fig.~\ref{WP2_dNdetaMidPred_sqrts} is a
compilation of the evolution of the midrapidity charged particle density,
$dN_{ch}/d\eta\rfloor_{|\eta|\le1}$, per participating nucleon pair, $N_{part}/2$,
as a function of collision energy from PHOBOS \cite{Bac03c,Bac00a,Bac02a,Bac02b,Bac02c,Bac04e}
and lower energy heavy ion reactions at the SPS \cite{Afa02,Ant04} and AGS \cite{Ahl98a,Bac02e,Ahl00,Kla03,Dun99}.  
The PHOBOS data are for the 6\% most central Au+Au interactions.
For most of the SPS and AGS data, the $dN_{ch}/d\eta$ values were obtained using
sums of $dN/dy$ results for a variety of identified particles.
The data follow a simple logarithmic extrapolation from lower
energies as shown by the line drawn to guide the eye.
The PHOBOS apparatus allows several independent techniques to be used to measure 
centrality and the number of particles emitted near midrapidity, all of which provide
results that differ by no more than a small fraction of their separate systematic 
errors.  The values of  $dN_{ch}/d\eta\rfloor_{|\eta|\le1}$ per participating nucleon pair, 
1.94$\pm$0.15, 2.47$\pm$0.27, 3.36$\pm$0.17 and 3.81$\pm$0.19
for the 6\% most central Au+Au collisions at 19.6, 56, 130, and 200~GeV, 
respectively, represent weighted averages of the published results.  It is notable
that multiplicity measurements were initially obtained by PHOBOS and later
confirmed by the other experiments at every new beam energy and species
provided during the first three RHIC runs, from the first Au+Au
collisions \cite{Bac00a} through the d+Au collisions \cite{Bac04j}.

It is interesting to note that the measured midrapidity charged particle
density at RHIC is lower than the prediction of most models (see the left panel of
Fig.~\ref{WP2_dNdetaMidPred_sqrts}, as well as \cite{Esk02,Bas99a}. From top to 
bottom, the references for the models are 
\cite{Wan99,Zha99,Sor99,Ble99,Bas99a,Cas99,Dre99,Cap99,Ran99,Arm00,Jeo00,Esk00,Bas99b,Sta99,Kra01}).
The authors of \cite{Esk02} quoted a factor of 1.1 for converting $dN/d\eta$ to 
$dN/dy$ for comparison of data and theory.  For consistency, the PHOBOS 
$dN_{ch}/d\eta$ has been multiplied by the same factor to obtain the value shown
in the figure.

Among the models which predicted a value close to that seen in the data were two which
invoked the concept of saturation in either the initial state \cite{Kra01} or the produced partons \cite{Esk00}. 
Related concepts were used in more recent formulations which describe the formation
of a Color Glass Condensate (CGC).  This newer CGC model successfully
related the pseudorapidity and energy dependences of charged particle
production to the gluon structure function measured in e+p collisions
\cite{Kha01a}. It should be noted that this model also made predictions for the 
properties of particle production at high $p_{_T}$ in d+Au collisions \cite{Kha03,Kha04} 
which agreed qualitatively with the pattern of hadron suppression in the d+Au 
data at middle to forward rapidities \cite{Bac03d,Bac04f,Ars04}, but which cannot
explain the excess of particle production at high $p_{_T}$ for backward rapidities \cite{Liu04,Acc04}.
The search for other evidence for possible parton saturation
effects remains a topic of interest at RHIC but a more detailed discussion is
beyond the scope of this paper.

Before attempting to make detailed estimates of the energy density, it is
important to stress that the midrapidity particle density at the top RHIC
energy is about a factor of two higher than the maximum value seen at the SPS
\cite{Bac02b} and there is evidence that the transverse energy per particle has not
decreased \cite{Adc01,Ada04c}.  Thus, with little or no model dependence, 
it can be inferred
that the energy density has increased by at least a factor of two from
$\sqrt{s_{_{NN}}}$=17 to 200~GeV. 

\begin{figure}[ht] 
\begin{center}

\includegraphics[width=0.65\textwidth]{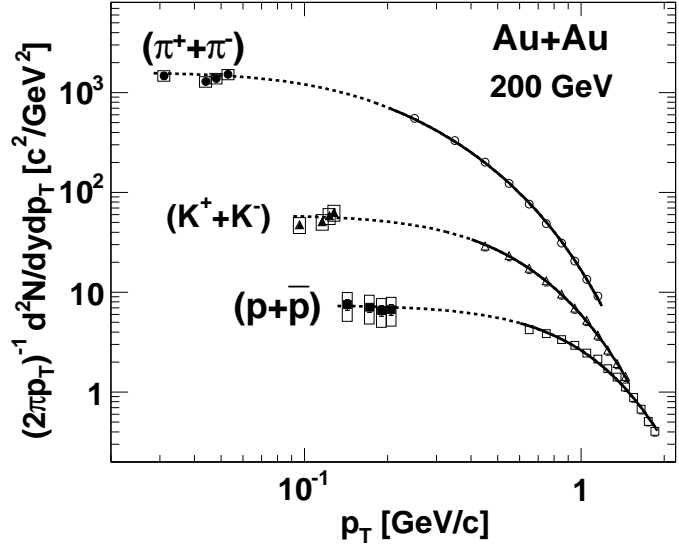}

\caption{ \label{WP3_LowPt_pT_lines}
Transverse momentum distributions of identified charged particles
emitted near midrapidity in central Au+Au collisions at
$\sqrt{s_{_{NN}}}$=200~GeV.  Invariant yield data shown are from PHENIX at higher momenta \cite{Adl04a} and
PHOBOS at lower momenta \cite{Bac04c}.  Boxes around the PHOBOS data indicate systematic
uncertainties.  Fits to PHENIX measurements are shown by solid curves $(\propto
1/[e^{(m_{_T}/T_i)}+\epsilon]$, where $\epsilon=-$1 and +1 for mesons and baryons, respectively,
$m_{_T}$ is the transverse mass, and $T_i$ is the fit parameter for each species). Note that the
extrapolations (dashed curves) of the fit to the data at higher momenta are
consistent with the low momentum yields.}

\end{center}
\end{figure}

In addition to the measured particle multiplicities, estimating the energy
density more precisely requires knowledge of the average energy per particle,
as well as the volume from which they originate.  PHOBOS data for the
transverse momentum distribution of charged particles \cite{Bac04a} can be used
to find a mean transverse momentum but these data only extend down to a few
hundred MeV/c.  Alternatively, Fig.~\ref{WP3_LowPt_pT_lines} compares
identified particle yields at very low transverse momentum measured by PHOBOS
\cite{Bac04c} to PHENIX data \cite{Adl04a} for higher momenta.  Both data sets
are for particles emitted near midrapidity in central Au+Au collisions at
$\sqrt{s_{_{NN}}}$=200~GeV.  The PHOBOS data clearly demonstrate that the fits
shown hold over the full range of transverse momentum and that extrapolation
should give a correct value for the average.  
The low momentum identified particle data shown in Fig.~\ref{WP3_LowPt_pT_lines} 
are in non-overlapping regions of $p_{_T}$ for the three different species.
Thus, without additional assumptions it is not possible to merge them into a
low $p_{_T}$ charged particle value for comparison to PHOBOS spectra for charged 
particles at higher $p_{_T}$.

Accounting for the yields of the
various particles, an average transverse momentum for all charged particles of
$\langle p_{_T} \rangle\sim$500~MeV/c can be derived.  The value found from the
PHOBOS unidentified charged particle distributions is the same to within 5\%.  Averaging over
the pions, kaons, and nucleons, and assuming the yields for the unobserved
neutral particles, an average transverse mass, $m_{_T}$, of $\sim$570~MeV/c$^2$
can be extracted.  Under the assumption of a spherically symmetric distribution
in momentum space, which would have equal average transverse and longitudinal
momenta, the average energy per particle is equal to the transverse mass
($m_{_T}$) at midrapidity (i.e.  $\langle E^2 \rangle = \langle m_{_0}^2 +
p_{_T}^2 + p_{_\parallel}^2 \rangle \approx \langle m_{_0}^2 + p_{_T}^2 \rangle
\rfloor_{\eta=0}$).  Alternatively, assuming that transverse momentum is
independent of pseudorapidity, the contribution due to the longitudinal
momentum can be found by averaging $p_{_\parallel} = p_{_T}\cot(\theta)$.  Over
the range $0<\eta<1$, this results in $\langle p_{_\parallel}^2\rangle$ which
is approximately 30--40\% of $\langle p_{_T}^2\rangle$ and would, therefore, raise
the average energy by about 10--15\%.  Since there are significant theoretical
uncertainties in this and other elements of the calculation, and we are
interested in a lower limit, a rounded estimate of 600~MeV per particle will be
used.

The total energy in the system created near midrapidity in central Au+Au collisions at
$\sqrt{s_{_{NN}}}$=200~GeV can be found from 
\[ E_{tot}=2E_{part}dN_{ch}/d\eta\rfloor_{|\eta|\le1}f_{neut}f_{4\pi},\] where
$E_{part}$ is the average energy per particle,
$dN_{ch}/d\eta\rfloor_{|\eta|\le1}$=655$\pm$35(syst) is the midrapidity charged particle density for the 6\% most central collisions, 
f$_{neut}$ is a factor of 1.6
to roughly account for undetected neutral particles, and the factor of 2 integrates over
$-$1$\le\eta\le$$+$1.  One further issue to consider is
that there are particles
with similar total momentum in the center-of-mass system but which are not
traveling predominantly in the transverse direction.  The correction for these
additional particles, f$_{4\pi}$, is trivially estimated from the fraction of
solid angle outside
$\theta=40^\circ$--$140^\circ$ (i.e.\ outside $|\eta|\leq 1$) and equals
about 1.3.  It should be stressed that this methodology does not suggest that
the entire distribution of particles is isotropic; in fact, the data shown in
Fig.~\ref{WP1_dNdeta_eta_AuAu_19_130_200} clearly contradict any such idea.
Instead, the goal is to obtain the energy density for the component of the
distribution which is consistent with isotropic emission from a source at
midrapidity.  
Combining all of these terms, the total energy contained in all
particles emitted near midrapidity, with transverse and longitudinal momenta consistent
with emission from an equilibrated source, is about 1600~GeV.  This is roughly 4\%
of the total energy of 39.4~TeV in the colliding system.

Converting this to a density in the rest frame of the system consisting of these particles requires knowledge of the volume within which this
energy is contained at the earliest time of approximate equilibration.  For
central collisions, a transverse area equal to that of the Au nuclei
($\approx$150~fm$^2$) can be assumed, but which value to use for the longitudinal
extent is not as clear.  One extreme is to take the very first instant when the
two Lorentz contracted nuclei overlap (longitudinal size $\approx$0.1~fm),
which yields an upper limit on the energy density in excess of 100~GeV/fm$^3$.
There is, however, no reason to assume that at such an early instant the system is in any
way close to equilibrium.  A second commonly-used assumption is that proposed
by Bjorken \cite{Bjo83}, namely a transverse size equal to the colliding nuclei
and a longitudinal size of 2~fm (corresponding to a time of the order of $\tau\sim$1~fm/c since the collision) which implies
an energy density of about 5~GeV/fm$^3$.
\footnote{The frequently-used Bjorken approximation for the energy density with 
the same information from the data used here would yield a value of about 
4~GeV/fm$^3$.}
Finally, the elliptic flow results
discussed below suggest that an upper limit of the time for the system to reach
approximate equilibrium is of the order of 1--2~fm/c.  Using the upper range of
this estimate and further conservatively assuming that the system expands during this time in both the longitudinal and transverse directions 
(with expansion velocities $\beta_{_\parallel}\approx$1 and $\beta_{_\perp}\approx$0.6), 
one obtains a lower limit of the energy density
produced when the system reaches approximate equilibrium at RHIC of
$\geq$3~GeV/fm$^3$.  Even this very conservative estimate is about six
times the energy density inside nucleons and about twenty times the energy
density of nuclei.  Therefore, this is a system whose description in terms of
simple hadronic degrees of freedom is inappropriate.

\subsection{Baryon chemical potential} 
\label{Sec2-B}

\begin{figure}[ht]
\begin{center}

\includegraphics[width=0.85\textwidth]{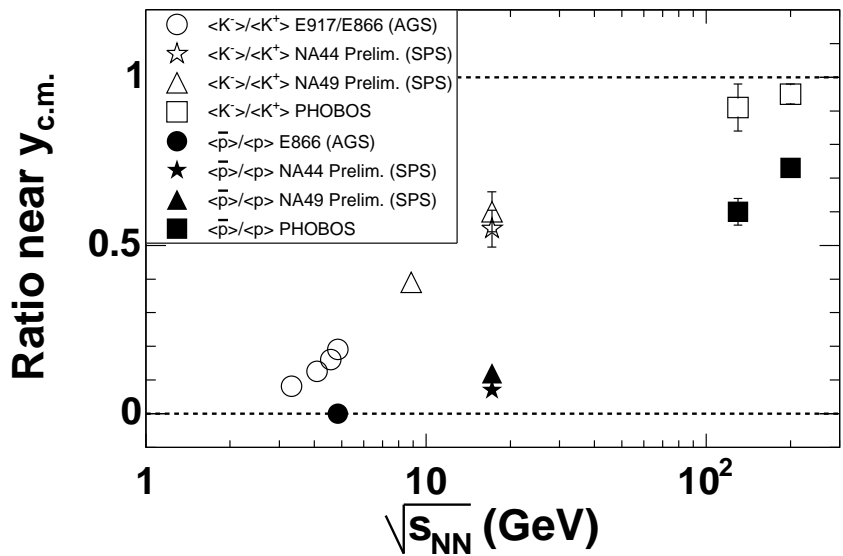}

\caption{ \label{WP4_pbarp_kmkp_sqrts}
Ratios of identified antiparticles over particles measured near midrapidity
in central collisions of Au+Au (AGS \cite{Ahl00,Ahl99,Ahl98b} and PHOBOS at RHIC \cite{Bac01a,Bac03a}) and Pb+Pb (SPS \cite{Bea96,Bac99}) as a function of
nucleon-nucleon center-of-mass energy.  Error bars are statistical only.}

\end{center}
\end{figure}

Turning to the baryon chemical potential, $\mu_{_B}$, early results regarding
this property of the high energy density medium produced at RHIC came from the
measurement of the ratios of charged antiparticles to particles near
midrapidity for central collisions.  
In the simplest Boltzmann approximation, the ratio of antiprotons to protons is proportional to 
$e^{-2\mu_B/T}$, where $T$ is the temperature at the time of chemical freezeout.  
Using particle yields to deduce properties of the system is a concept that long
predates QCD and heavy ion collisions \cite{Fer50,Hag65,Hag67}.
Figure~\ref{WP4_pbarp_kmkp_sqrts} compares the antiparticle to particle
ratios for both protons and kaons measured at RHIC by PHOBOS \cite{Bac01a,Bac03a} to the corresponding numbers found at lower
energies \cite{Ahl00,Bea96,Bac99,Ahl99,Ahl98b}.  Clearly, the systems formed at RHIC are much closer to
having equal numbers of particles and antiparticles than was true at lower
energies.  The measured value of
0.73$\pm$0.02(stat)$\pm$0.03(syst) for the antiproton to proton ratio near
midrapidity for central Au+Au collisions at $\sqrt{s_{_{NN}}}$=200~GeV
\cite{Bac03a} indicates that these collisions are approaching a very low value
of $\mu_{_B}$.  
Within the framework of thermal models, these ratios can be used to extract the baryon
chemical potential \cite{Bec01}.  Assuming a hadronization temperature of
165~MeV, a value of $\mu_{_B}$=27~MeV was found for central Au+Au at
$\sqrt{s_{_{NN}}}$=200~GeV.  This baryon chemical potential is an order of
magnitude lower than was obtained for Pb+Pb data at
$\sqrt{s_{_{NN}}}$=17.2~GeV from the SPS \cite{Bec96,Bra99}.  Although the
system created near midrapidity at RHIC cannot be described as completely free of net baryons, it is
clearly approaching the environment treated in most lattice calculations.
 
\subsubsection{Comparison of particle ratios in Au+Au and d+Au}
\label{Sec2-B-1}

In addition to the higher center-of-mass energies, a critical element of the
design of RHIC was the ability to collide asymmetric systems.  This capability
was first exploited with the collision of deuterons with gold nuclei at
$\sqrt{s_{_{NN}}}$=200~GeV.  It is hoped that analysis of such simpler
systems will serve as critical ``control'' experiments to aid in the
understanding of the more complicated nucleus-nucleus data.  As a first
example, this section presents a study of the antiparticle to
particle ratios.

\begin{figure}[ht]
\begin{center}

\includegraphics[width=0.75\textwidth]{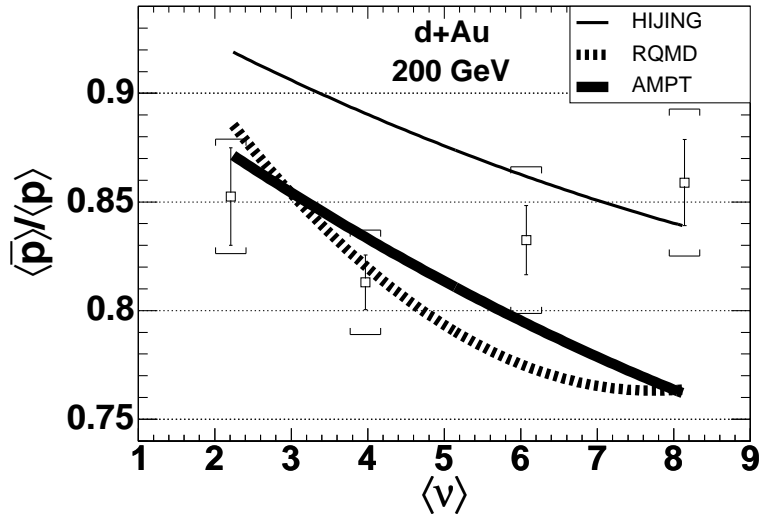}

\caption{ \label{WP5_pbarp_nu_dAu_models} 
The ratio of antiprotons to protons emitted in a rapidity region spanning
approximately $0.0<y<0.8$ (where positive rapidity is in the direction of the
deuteron projectile) for d+Au collisions at $\sqrt{s_{_{NN}}}$=200~GeV \cite{Bac04b}.  Data
are shown for 4 centrality ranges.  The parameter $\langle\nu\rangle$ is the average number
of collisions suffered by each participant in the deuteron ($N_{coll}/N^d_{part}$).
Statistical and point-to-point systematic uncertainties are shown as
bars and brackets, respectively.  The results of several models \cite{Gyu94,Sor95,Lin01,Zha00} are
shown for comparison.}

\end{center}
\end{figure}

As described above, particle ratios can be used to extract information about
the properties of the system, in particular the chemical potentials.  The
measured values of these parameters are established at the point of chemical
freeze-out when inelastic interactions between the produced particles cease.
However, the properties of the early evolution of the system can clearly
influence final conditions.  Of particular interest in this regard is the ratio
of antiprotons to protons measured at midrapidity.  This ratio can be
interpreted as reflecting the interplay of two mechanisms, namely
the transport of baryons from the two projectile nuclei to midrapidity and the
production of antibaryon-baryon pairs in the interaction.  By studying ratios
as a function of centrality in d+Au, the effect of multiple collisions of the
nucleons in the deuteron can be explored.  The surprising result is shown in
Fig.~\ref{WP5_pbarp_nu_dAu_models} \cite{Bac04b}.

The simple expectation, supported by various model calculations 
(HIJING \cite{Gyu94}, RQMD \cite{Sor95}, and AMPT \cite{Lin01,Zha00}) was that
the proportion of antiprotons near midrapidity would fall slowly with
collision centrality as the deuteron participants suffered more collisions and,
consequently, were effectively transported closer to the center-of-mass rapidity.  In
contrast, the data show a ratio which is consistent with being the same at all
centralities.  At present, no simple
explanation or interpretation of the observed particle ratios is known.

The d+Au data at RHIC serve an important function as a
control experiment since an extended volume of
high density matter is presumably not formed in these collisions.  Understanding the basic
mechanisms of baryon transport and baryon pair production will clearly be
critical to a full description of heavy ion interactions.  

\subsection{Nature of the transition to the high density regime}
\label{Sec2-C}

The transition to the high density state at RHIC has not been observed to
create abrupt changes in any observable studied to date, including, among
others, charged particle multiplicity, elliptic flow, HBT, as well as derived
quantities such as energy density and freeze-out parameters.  This lack of a
dramatic change in character may make it more difficult to delineate the exact
boundaries of the onset of significant influence from non-hadronic degrees of
freedom.  However, this observation may be consistent
with the expectations concerning the nature of the phase transition from the
most recent lattice QCD calculations \cite{Kan03,Kar04,Lae03,Ber04}, which predict a rapid
crossover in the region of the phase diagram believed to be relevant for the
systems created near midrapidity at RHIC. It should be noted that the lack of 
dramatic shifts in observables does not necessarily rule out the presence of
a phase transition with different characteristics (see, for example, the
discussion in \cite{Kol03}).

It should be noted that indications of possible non-monotonic behavior in the
energy evolution of some quantities were reported in the range
$\sqrt{s_{_{NN}}}$=5--10~GeV at the CERN SPS (see, for example, \cite{Bec04}
and references therein).  The extracted properties of the environment created
near midrapidity in these lower energy collisions are significantly different
from those found near midrapidity at RHIC, with energy densities at least a
factor of 3--4 smaller and baryon chemical potentials an order of magnitude or
more larger.  A discussion of these results at lower energy falls outside the scope
of this paper but future work in this area might prove important to the
full exploration of the QCD phase diagram.

\section{Strength of interactions in the high energy density medium}
\label{Sec3}

In early discussions of the high density systems formed in RHIC
collisions, the expectation was that a deconfined state of quarks and gluons
would be weakly interacting.  This interpretation arose at least partly from
the na\"{\i}ve assumption that any matter that attained a large fraction of
the Stefan-Boltzmann limit for the pressure would act like a gas \cite{Shu04a}.  One of the
most dramatic early discoveries at RHIC is the clear indication that the nature
of the systems formed is very far from weakly interacting.  Evidence for this
conclusion is found in the magnitude of elliptic flow and in the centrality
dependence of particle production at high transverse momentum.  The former
provides information on the manner in which particle production depends on the
shape of the incident system and the latter explores how the spectrum of the
produced particles is impacted by the medium.  Additional evidence is provided
by the yields of particles at
very low transverse momentum, a measurement unique to PHOBOS.

\begin{figure}[ht] 
\begin{center}

\includegraphics[width=0.65\textwidth]{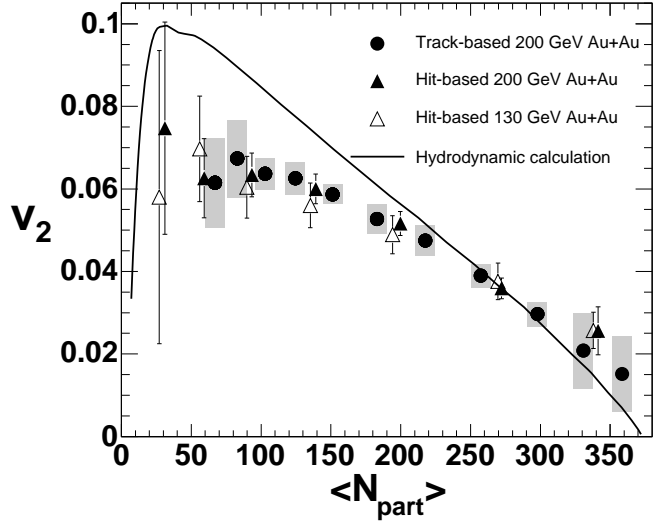}

\caption{ \label{WP6_v2_Npart_AuAu_130_200}
Elliptic flow of charged particles near midrapidity ($|\eta| <1$) as a
function of centrality in Au+Au collisions at $\sqrt{s_{_{NN}}}$=200~GeV
using two different methods \cite{Bac04h} (closed circles and triangles, see text for
details) and at $\sqrt{s_{_{NN}}}$=130~GeV (open triangles) \cite{Bac02d}.  Grey boxes
show the systematic errors (90\% C.L.) for the 200~GeV data.  The curve shows the prediction
from a relativistic hydrodynamics calculation \cite{Huo04}.}

\end{center}
\end{figure}

\begin{figure}[ht]
\begin{center}

\includegraphics[width=0.65\textwidth]{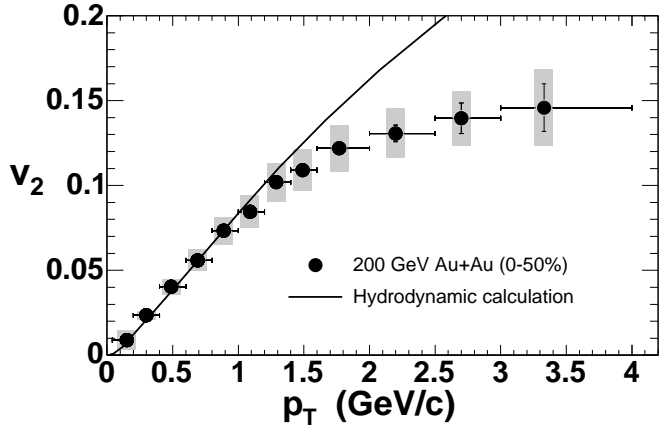}

\caption{ \label{WP7_v2_pT_AuAu_200}
Elliptic flow of charged particles emitted near midrapidity ($0<\eta<1.5$) in
the 50\% most central Au+Au collisions at $\sqrt{s_{_{NN}}}$=200~GeV as a
function of transverse momentum \cite{Bac04h}.  Grey boxes show the systematic uncertainties
of the data (90\% C.L.).  The curve is the prediction of a relativistic hydrodynamics calculation \cite{Huo04}.}

\end{center}
\end{figure}

Figure~\ref{WP6_v2_Npart_AuAu_130_200} shows PHOBOS measurements of the magnitude of elliptic flow,
$v_2$, near midrapidity ($|\eta|\le1$) in Au+Au collisions at
$\sqrt{s_{_{NN}}}$=130 \cite{Bac02d} and 200~GeV \cite{Bac04h} as a function of centrality, denoted by
$\langle $N$_{part}\rangle$.  Two different methods of determining the flow
signal, one based on counting hits in the multiplicity detector and one based
on counting tracks in the spectrometer \cite{Bac04h}, were used at the higher
beam energy.  Similar results were first shown for RHIC data by the STAR
collaboration \cite{Ack01}.  Figure~\ref{WP7_v2_pT_AuAu_200} shows data from the track-based
method in the rapidity interval $0<\eta<1.5$ for the 50\% most central Au+Au
collisions at $\sqrt{s_{_{NN}}}$=200~GeV as a function of transverse momentum,
$p_{_T}$ \cite{Bac04h}.  Data in both figures are compared to the predictions
of a hydrodynamical calculation \cite{Huo04}.  These results show that elliptic
flow is unexpectedly large at RHIC energies.  Over a wide range of centrality
and transverse momentum, the value near midrapidity is as large as that calculated under the
assumption that a boost-invariant relativistic hydrodynamic fluid was formed.

When two nuclei collide with non-zero impact parameter, the lenticular (or
almond-shaped) overlap region has an azimuthal spatial asymmetry (see right
panel of Fig.~\ref{WP35_collide_diagram}).  However, if the particles do not interact
after their initial production (presumably with azimuthally uniform momenta),
the asymmetrical shape of the source region will have no impact on the
azimuthal distribution of detected particles.  Therefore, observation of
azimuthal asymmetry in the outgoing particles is direct evidence of interactions between the
produced particles.  In addition, the interactions must have occurred at relatively early times, since
expansion of the source, even if uniform, will gradually erase the magnitude
of the spatial asymmetry.

Qualitatively, it is clear that an asymmetric system of interacting particles
will have azimuthally varying pressure gradients which can alter the observed particle
directions.  Hydrodynamical models can be used to calculate a quantitative
relationship between a specific initial source shape and the distribution of
emitted particles (see, for example, \cite{Huo04}).  Due to the ideal nature of the
fluid assumed in these models (not to be confused with the non-interacting ideal {\em gas}), the resulting final asymmetry is generally
assumed to be an upper limit for a specific starting condition.  From the
strength of the observed elliptic flow and from the known dimensions of the
overlap region in Au+Au collisions, it can be conservatively estimated that the
pressure build-up in the initially formed medium must have occurred in a time
less than about 2~fm/c (with a best-fit value from flow and other data of 0.6~fm/c) \cite{Kol03}.  Thus, the presence of a large flow signal carries
several important implications, the first of which, a limit on the timescale
for equilibration, has been used previously in the discussion of energy
density.  In addition, one can conclude that at these early times the initially
produced particles must already be interacting significantly,  corresponding more closely to the conditions in a fluid rather than
 a gas.

Additional indirect evidence that the constituents of the system produced in
heavy ion collisions at RHIC are interacting significantly is provided by the observed yield of particles with very small
transverse momentum ($\leq$100~MeV/c) \cite{Bac04c}, shown previously in
Fig.~\ref{WP3_LowPt_pT_lines}.  Recall that the production of particles with $p_{_T}$ as
low as 30~MeV/c was consistent with extrapolations from a fit to the
distribution in the range of a few hundred MeV/c to a few~GeV/c.  If, in RHIC
collisions, a medium of weakly interacting particles was initially produced,
one could expect an enhancement in the production of particles with wavelengths
up to the overall size of the collision volume (i.e.\ coherent pion production)
\cite{Bus93}.  In essence, the observation that there is no such excess is another
manifestation of the high pressure gradient and significant level of interaction present in the medium,
which gives rise to the large magnitude of the elliptic flow signal seen at
RHIC.  These properties would also produce large radial flow so that any
particles initially produced with low velocity would subsequently be
accelerated by the interactions.

The study of the yield of particles with large transverse momentum can be used
to more directly explore the level of interactions present in the medium
produced in $\sqrt{s_{_{NN}}}$=200~GeV Au+Au collisions at RHIC.  Presuming
that high momentum transfer processes are induced via relatively short-range
interactions, one may expect QCD factorization theorems, proven for simpler processes, to continue to hold and, therefore,
a particular hard process can be induced by any binary collision in the overall
nucleus-nucleus interaction \cite{Col85,Col88}.  This is the motivation for the
nuclear modification factor, $R_{AA}$, defined in Appendix~\ref{SecAppB-3} and
first studied at RHIC by PHENIX \cite{Adc02,Adc03}, which measures how effective each
particular binary collision is for inducing a hard scattering process.  Strong deviations
from unity indicate violations of factorization, which may be caused by initial
or final state effects.  In their pioneering work, the PHENIX collaboration
showed that in central collisions of Au+Au at $\sqrt{s_{_{NN}}}$=130~GeV there
was significant suppression of the yield of high transverse momentum particles
compared to the p+p data scaled by the number of binary collisions, $N_{coll}$.  

\begin{figure}[ht] 
\begin{center}

\includegraphics[width=0.95\textwidth]{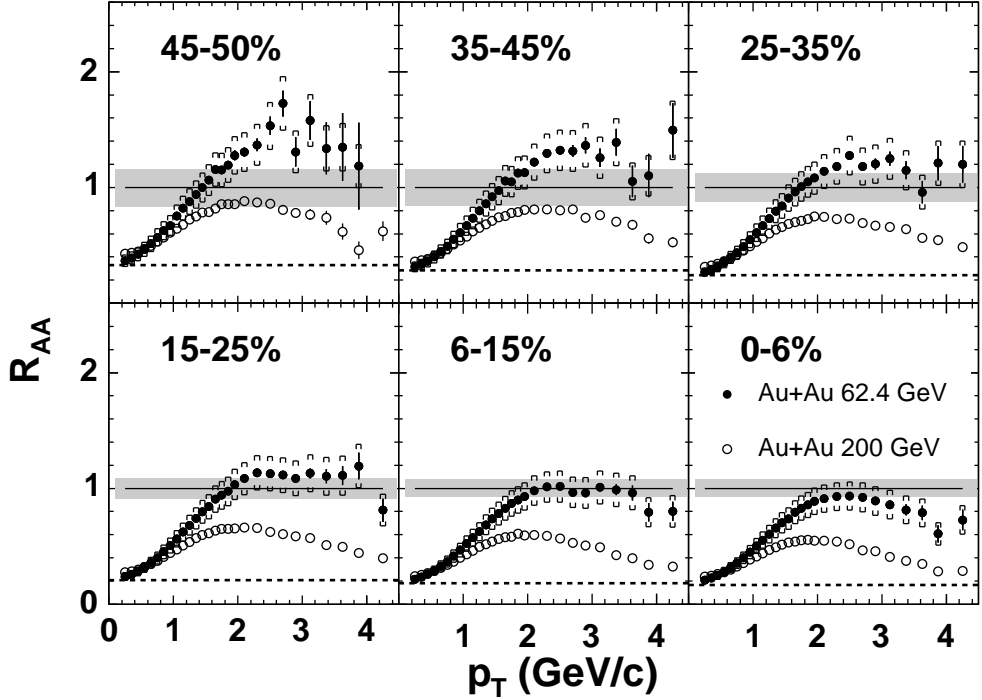}

\caption{ \label{WP8_RAA_pT_AuAu_64_200}
Nuclear modification factor, $R_{AA}$, as a function of transverse momentum for
Au+Au collisions at $\sqrt{s_{_{NN}}}$=62.4 (closed symbols) and 200~GeV (open symbols), for six centrality
ranges \cite{Bac04a,Bac04d}.  Centrality is expressed as a fraction of the total inelastic cross
section with smaller numbers being more central.  Bars and brackets show
statistical and systematic uncertainties, respectively.  
The grey bands show the systematic error in the overall scale due to $N_{coll}$.  
The solid (dashed)
line shows the expectation for scaling with $N_{coll}$ ($N_{part}$/2) times p+p data
(See discussion in Appendix~\ref{SecAppB-3}).
}

\end{center}
\end{figure}

The PHOBOS collaboration has confirmed that a similar effect is present in
Au+Au collisions at 200~GeV \cite{Bac04a}, and has also performed the first similar
studies at $\sqrt{s_{_{NN}}}$=62.4~GeV \cite{Bac04d}, see Fig.~\ref{WP8_RAA_pT_AuAu_64_200}.
More importantly, as discussed later and shown in Fig.~\ref{WP31_RPCNpart_Npart_AA_6pT}, the
yields for Au+Au interactions at $\sqrt{s_{_{NN}}}$=200~GeV, which span a range of more than a factor of five in the
number of participants, were found to scale with the number of participants,
when compared with central Au+Au collisions, to within $\leq$25\% at all transverse
momenta.  The fact that data up to $p_{_T}$ of 4~GeV/c show much the same
scaling as at low momentum clearly demonstrates that any scaling of the yield due to hard
processes with the number of binary collisions is completely obliterated.  Note
the significant difference in the magnitudes and overall shapes of $R_{AA}$ at the two energies 
shown in Fig.~\ref{WP8_RAA_pT_AuAu_64_200}, as well
as the fact that the difference is similar at all centralities.  Additional
discussion of this interesting observation, as well as other scaling properties
of the data, can be found in Sect.~\ref{Sec4}.

It is important to note that, except where specifically mentioned, the reference
p+p data in this and all other cases of comparison to RHIC data is for
inelastic collisions.  This choice is made for consistency rather than being
strongly motivated by physics considerations.  In most cases, the difference
between the yield in non-single diffractive (NSD) and inelastic measurements is
about 10\% or less.

\begin{figure}[ht] 
\begin{center}

\includegraphics[width=0.70\textwidth]{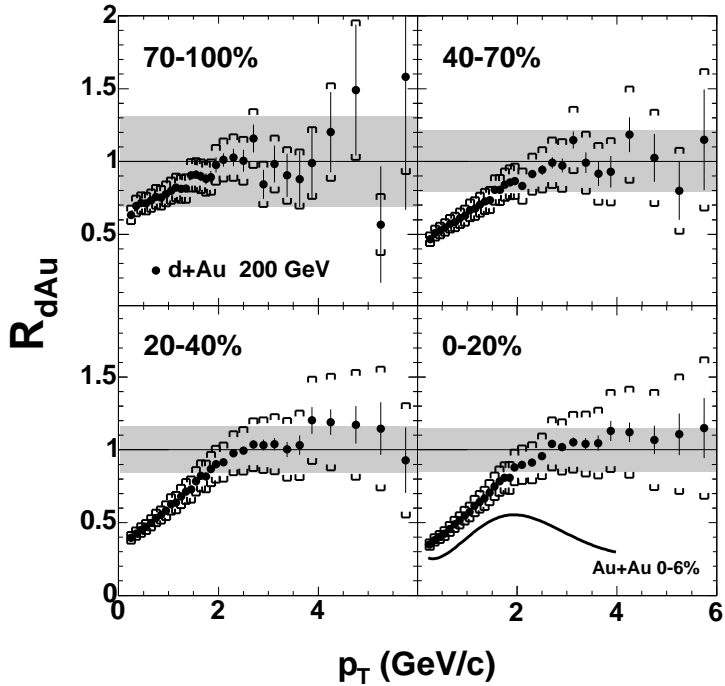}

\caption{ \label{WP9_RdAu_pT_4cent_AuAuLine}
Nuclear modification factor, $R_{dAu}$, as a function of transverse momentum
for d+Au collisions at $\sqrt{s_{_{NN}}}$=200~GeV, for four centrality ranges \cite{Bac03d}.
Centrality is expressed as a fraction of the total inelastic cross section with
smaller numbers being more central.  Bars  and brackets show statistical and
systematic uncertainties, respectively.  The shaded area shows the uncertainty
(90\% C.L.) in $R_{dAu}$ due to the systematic uncertainty in $N_{coll}$ and
the scale uncertainty in the proton-proton data.  In the bottom right panel,
the nuclear modification factor, $R_{AA}$, for the 6\% most central Au+Au 
collisions at the same energy \cite{Bac04a} is shown as a dark curve for comparison.}

\end{center}
\end{figure}

As mentioned above, the observed suppression of hard processes could result
from some modification in the initial state (see, for example, \cite{Kha03}), as
well as from interactions in the dense medium formed after the collision.  To
investigate this possibility, similar data were taken for d+Au collisions at
the same energy.  Figure~\ref{WP9_RdAu_pT_4cent_AuAuLine} shows the nuclear modification factor,
$R_{dAu}$, measured by PHOBOS in d+Au at $\sqrt{s_{_{NN}}}$=200~GeV, in four
different impact-parameter ranges \cite{Bac03d} and the similar modification factor,
$R_{AA}$, in central Au+Au collisions at the same energy \cite{Bac04a}.  Note
the dramatic difference between the results for central d+Au and Au+Au
collisions at higher transverse momentum shown in the lower right panel of the
figure.  For 2~GeV/c~$\leq p_{_T}\leq$~6~GeV/c the yield of  charged particles in d+Au is
consistent with binary collision scaling of p+p data, whereas in Au+Au
collisions the yield is clearly suppressed.

The observation that the data points at higher $p_{_T}$ in
Fig.~\ref{WP9_RdAu_pT_4cent_AuAuLine} are similar at all centralities and all lie near unity
may be evidence for binary collision scaling at higher $p_{_T}$ in d+Au.
However, this interpretation is
unclear since the characteristics of the data may be a consequence of the interplay of an enhancement
(similar to the so-called ``Cronin effect'' \cite{Acc04,Cro75,Ant79,Acc02,Vit03}), and some suppression, due to
either energy loss in the final state or parton saturation effects in the
initial state.  Furthermore, several effects complicate the assumed connection 
between binary collision scaling and the magnitude and centrality independence 
of $R_{dAu}$.
First, it should be noted that the number of participants
and the number of collisions do not deviate as much with centrality in d+Au as
in Au+Au.  Using the number of participant pairs as the scaling variable (i.e.\
using $R_{dAu}^{N_{part}}$ defined in Appendix~\ref{SecAppB-3}) would raise the
values at all transverse momenta by an average factor of about 1.65.  However,
the factor would differ only by 29\%, 14\%, and 6\% for centrality bins of 70--100\%,
40--70\%, and 20--40\%, respectively, compared to the 0--20\% data.  These shifts
are comparable to, or smaller than, the systematic uncertainties in the overall
scale of the modification factors.  Thus, the observation of similar values of $R_{dAu}$ at
all centralities does not necessarily imply scaling with the number of
collisions.

\begin{figure}[ht]
\begin{center}

\includegraphics[width=0.75\textwidth]{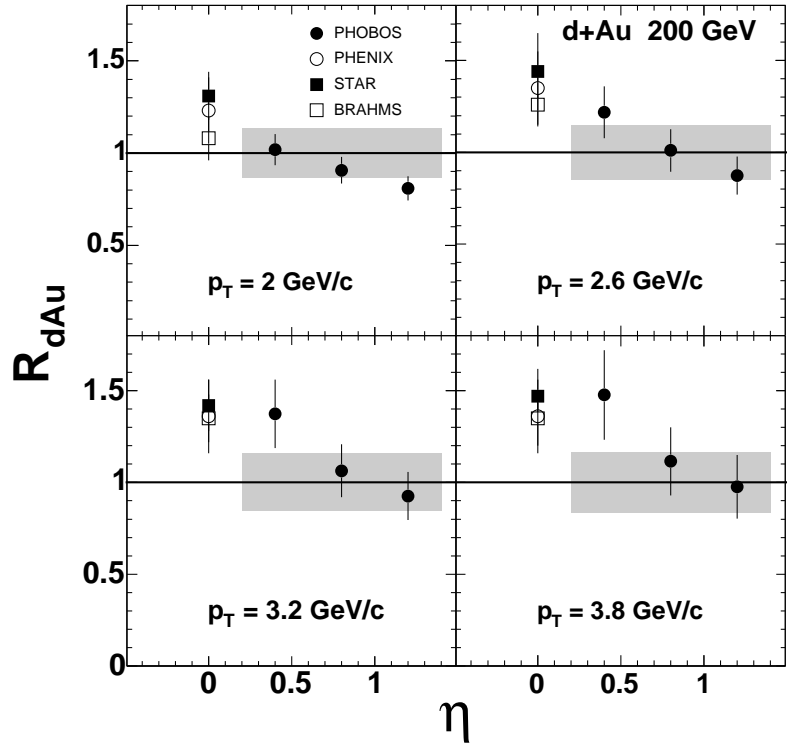}

\caption{ \label{WP10_RdAu_eta_4pT} 
Nuclear modification factor, $R_{dAu}$, for four different values of $p_{_T}$
as a function of pseudorapidity in d+Au collisions at
$\sqrt{s_{_{NN}}}$=200~GeV.  PHOBOS results away from midrapidity
\cite{Bac04f} are compared to data near $\eta$=0 from BRAHMS \cite{Ars03},
PHENIX \cite{Adl03b}, and STAR \cite{Ada03}.  For the PHOBOS points, the error
bars are the point-to-point systematic errors (90\% C.L.).  The systematic errors in the
overall scale of the PHOBOS $R_{dAu}$ are shown as grey bands.}

\end{center}
\end{figure}

To further complicate the interpretation, the value of the nuclear modification
factor was found to depend on the pseudorapidity of the emitted
particles.  This was first inferred from the comparison of the 
PHOBOS results \cite{Bac03d} to those of the other experiments \cite{Ars03,Adl03b,Ada03}.  It can also be seen from 
PHOBOS results directly as shown in Fig.~\ref{WP10_RdAu_eta_4pT} \cite{Bac04f}.  Data from 
BRAHMS suggest that this
trend may continue to higher positive pseudorapidity \cite{Ars04} while
preliminary PHENIX data suggests that $R_{dAu}$ may even continue rising for
negative pseudorapidity (i.e.\ towards the Au projectile rapidity)
\cite{Liu04}.  The trend seen in the PHOBOS and BRAHMS data has been
interpreted as support for the CGC model, but this conclusion is far from
clear and the PHENIX data at negative pseudorapidity  remain even more poorly understood
\cite{Acc04}. For this reason, the observation of the particular value of
$R_{dAu}$=1 at higher $p_{_T}$ is a consequence of the PHOBOS acceptance and
again does not necessarily imply $N_{coll}$ scaling. 

Therefore, the important feature is not the possible scaling of the particle
yields in d+Au with $N_{coll}$ times p+p yields, but instead the very significant
difference between the transverse momentum dependence of the d+Au and Au+Au
nuclear modification factors.  The larger system appears to lead to a strong
suppression while the smaller system does not.  Very similar results were
reported simultaneously by all four RHIC experiments \cite{Bac03d,Ars03,Adl03b,Ada03}.  
Part of the difference in the behavior of the two colliding
systems may be attributed to initial state effects.  However, it is difficult
to avoid the conclusion that the majority of the difference in $R_{AA}$
compared to $R_{dAu}$ results from the impact of the high energy density matter
on the yield of particles with $p_{_T}$ in this measured range.  Clearly, the
constituents of the medium produced in the
central Au+Au collisions experience a significant level of interaction.  
Since, as discussed above, the system at this early stage
cannot be primarily hadronic in nature, one can conclude that the high energy
density matter created at RHIC interacts very significantly with high $p_{_T}$
partons (or with whatever constituents comprise the dominant degrees of freedom
at this early stage).  It certainly does not appear to be a weakly interacting
parton or hadron gas.

In related measurements, the STAR experiment has studied back-to-back
correlations of high $p_{_T}$ particles.  Measuring the yield of particles as a
function of the azimuthal angle relative to a very high $p_{_T}$ trigger
particle, a suppression was found in particles emitted on the opposite side
\cite{Adl03a}.  This suppression was found to depend on the azimuthal angle of
the trigger particles with respect to the reaction plane \cite{Ada04b}.  One
strength of the correlation analysis is that it is essentially self-normalizing
in the sense that the result does not depend on any assumptions about the
scaling of the primary production of particles.  One can interpret this as
additional support for the conclusions that are being drawn from the single particle data.

Further evidence that the system may be both non-hadronic in nature and also
characterized by a significant level of interaction comes from flow data for identified particles.  PHENIX
\cite{Adl03c} and STAR \cite{Ada04a} have measured the elliptic flow and its
dependence on transverse momentum for a variety of mesons and baryons.  These
data appear to be consistent with an interpretation that the flow of produced
particles results from the recombination of quarks which are themselves flowing
\cite{Mol03}.  The impact of this flow of quarks is that the $v_2$ parameter
divided by the number of valence quarks scales as a function of the transverse
momentum, also divided by the number of valence quarks.  It should be noted
that this recombination model only holds for elliptic flow at higher values of
$p_{_T}\geq 1-2$~GeV/c.  If this interpretation is correct, it lends support to
the presumption that the system has a component of constituent quarks which
experience significant interactions early in the evolution of the collision.

In conclusion, the data from RHIC collisions provide strong evidence for the
creation of a very high energy density, low baryon chemical potential, medium which
cannot simply be described in terms of hadrons and whose constituents
experience significant interactions with each other.

\section{Simple scaling behaviors of particle production}
\label{Sec4}

The wide range of systems and energies provided by the RHIC accelerator,
combined with the unique capabilities of the PHOBOS detector, has allowed a
study of the properties of particle production over a very broad range of
pseudorapidity and transverse momentum for a wide variety of initial conditions.  
This work continues a long history of investigations to understand particle 
production under a variety of conditions.
In the process of this study, a
surprising result was discovered.  It emerged that an enormous span of data for charged particles emitted in d+Au and Au+Au collisions at RHIC energies 
could, to a large extent, be described using only a few simple unifying
features.  Some of these scaling behaviors had been observed previously,
either at lower energies or for less complicated systems than heavy ion
collisions.  Although a direct theoretical connection between these observed 
trends in the data and the nature of the systems created is not presently
apparent, it is clear that the unifying features must reflect important
aspects of the dynamics of the evolution starting from the earliest stages 
of the collision. In addition, these observations shed light on broader 
aspects of particle production under a variety of conditions.
This section describes the extent to which these scaling
behaviors and other unifying features apply to charged particle production at RHIC
energies.

In order to achieve the broadest possible coverage in pseudorapidity and
transverse momentum, most of these measurements rely on detection techniques
which do not differentiate between the production of different species of
particles.  Therefore, it is generally not known at this time to what extent
the production of any specific particle exhibits the scaling behaviors
described in this section.  However, the degree to which one particular species
deviates from any of the observed dependencies must be compensated by the sum
of all the other species, a correspondence between particle types that is
interesting in itself.  The occurrence of such balancing could contain
important information about the global influences on the processes taking place
during particle production.

In a wide variety of systems (hadron+A up to A+A), the total number of emitted
charged particles is observed to have a very simple dependence on energy and
centrality.  In all cases, the total multiplicity appears to scale linearly
with the number of participant pairs, $N_{part}$/2.  
It should be noted that throughout this document the generic term 
``participant pairs'' refers simply to the total number of participants divided
by 2, i.e.\ a quantity that is unity in p+p, and does not imply a matched pair
from the two colliding species.
The total multiplicity of
charged particles emitted in hadron+A (including p+A) and d+A is equal to
$N_{part}$/2 times the multiplicity observed in p+p.  In contrast, for heavier
nucleus-nucleus interactions, the constant of proportionality is the
multiplicity produced in e$^+$+e$^-$ annihilations, which is approximately
equal to that measured in p+p at twice the center-of-mass energy.  This is
suggestive of a universal energy dependence of charged particle multiplicities
in strong interactions.  Centrality, as reflected by the number of participants (both the
total number and, for asymmetric systems, the number in each of the nuclei)
appears to have a strong influence on the shape of the pseudorapidity
distributions.  In addition, the yield of high transverse momentum particles 
($p_{_T}\geq$4~GeV/c) shows a dependence on the number of participants that is 
surprisingly similar to that for low momentum particles when
comparing Au+Au at different centralities.

Over a broad range of emission angles, the distributions of pseudorapidity
density and the elliptic flow signal, when measured as a function of the
variable $\eta^{\prime}=\eta -y_{beam}$ (i.e.\ when 
shifted by $y_{beam}$ and thereby effectively viewed in the approximate rest
frame of one of the colliding particles), appear to be identical both in shape
and magnitude at all beam energies over a large range of $\eta^{\prime}$.  The details of the shape of the
distributions depend on the impact parameter, but again in an
energy-independent way.  In addition to this extended longitudinal scaling, no
evidence is seen for a boost invariant central plateau in the pseudorapidity distributions of either particle
multiplicity or elliptic flow.

Another aspect of the centrality dependence is the observation that many differences 
between data for Au+Au and p+p, for example in the multiplicity per participant or in the 
shape of the transverse momentum distributions, persist essentially unchanged over a 
centrality range corresponding to a number of participants that spans a factor of 5 or more.
Finally, many properties of particle production exhibit separate dependences on
the energy and centrality of the collisions which factorize to a surprising
degree.  In other words, the centrality dependence of data such as
pseudorapidity density and transverse momentum spectra was found to be
identical even at center-of-mass energies separated by up to an order of
magnitude.  

\subsection{Energy dependence of total multiplicity}
\label{Sec4-A}

\begin{figure}[t]
\begin{center}

\includegraphics[width=0.95\textwidth]{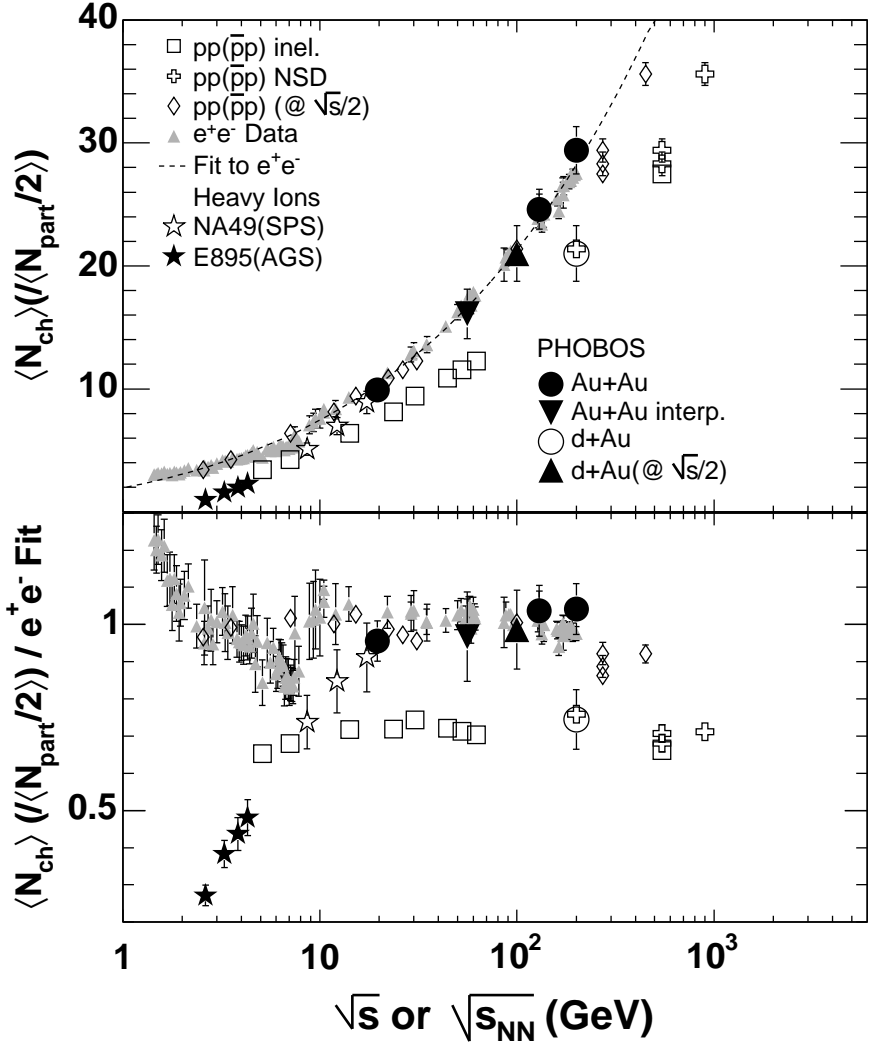}

\caption{ \label{WP11_Ntot_sqrts_AA_pp_ee}
(Top panel) Normalized total multiplicities of charged particles emitted in e$^+$+e$^-$,
p($\bar{p}$)+p (compiled from references in \cite{Eid04}), d+Au \cite{Bac04j}, Au+Au (AGS \cite{Kla01} and PHOBOS at RHIC \cite{Bac03e,Bac03c}) and Pb+Pb (SPS \cite{Afa02})
collisions at a variety of nucleon-nucleon center-of-mass energies.
Nucleus-nucleus data are all for central collisions and the multiplicities have
been divided by the number of participating nucleon pairs.  (Bottom
panel) The values for all systems are shown divided by a fit to the e$^+$+e$^-$
data.  
}

\end{center}
\end{figure}

The most basic observable in the study of multiplicity is the total number of
produced particles.  Collisions at RHIC extend the center-of-mass energy
range available in heavy ion interactions by more than an order of magnitude.
Section~\ref{Sec2-A} described the energy dependence of the midrapidity
particle density.  In this section, the total integrated particle yield is
discussed.  As is clearly shown in Fig.~\ref{WP1_dNdeta_eta_AuAu_19_130_200}, the PHOBOS
multiplicity detector extends over a uniquely broad range of pseudorapidity and, therefore, the
extrapolation to account for missing regions of solid angle is small even at
the highest RHIC energy.  The total multiplicity of charged particles per
participant pair in A+A collisions over a wide range of energies \cite{Bac03c,Afa02,Bac03e,Kla01} is shown in
Fig.~\ref{WP11_Ntot_sqrts_AA_pp_ee}, along with data from d+Au \cite{Bac04j},  p($\bar{p}$)+p, and e$^+$+e$^-$
annihilation into hadrons (the latter two compiled from references in \cite{Eid04}).  The d+Au value has also been divided
by the number of participant pairs.  The nucleus-nucleus data are for central
collisions.  However, this choice is inconsequential since, as will be
discussed in the following section, the total multiplicity per participant pair
appears to be approximately independent of centrality.

The various sets of data have very different trends.  The p+p (open
squares and crosses) and d+Au (open circle) data are consistently about 30\% below the
e$^+$+e$^-$ data, as shown in the lower panel where all of the data points
are divided by a fit to the e$^+$+e$^-$ data.  Starting at the lowest
energies, the A+A data rise much faster than both p+p and e$^+$+e$^-$ but then
the slope of the energy dependence changes and above $\sqrt{s_{_{NN}}}$
$\sim$20--30~GeV, the A+A data follow the trend of the e$^+$+e$^-$ data.  The
lower panel of the figure shows that these two sets agree to within 10\% over
a span of an order of magnitude in center-of-mass energy.

One proposed explanation for the difference between the p+p and e$^+$+e$^-$ data 
is that one must properly account for the ``leading particle effect''
which is present in hadron-hadron collisions, but not in e$^+$+e$^-$ annihilation.  
The distribution of protons
in $x_{_F}$ (see Appendix~\ref{SecAppB-2} for definition) for p+p collisions at
different energies was found to be approximately flat (with a spike at
$x_{_F}$=1 for elastic and diffractive events; a summary of these data can be found in
\cite{Ver02}).  One interpretation of these data is that a leading nucleon
typically carries away half of the beam energy.  In p+p collisions, the $x_{_F}$ of
the leading proton was found to directly anticorrelate with the particle
multiplicity, as if the leading particle simply removed energy that would
otherwise go into particle production \cite{Bre82,Bas80a,Bas80b}.  By
rescaling the center-of-mass energy for the p+p data by a factor of two (see open
diamonds in Fig.~\ref{WP11_Ntot_sqrts_AA_pp_ee}), one observes that the multiplicities of p+p
and e$^+$+e$^-$ reactions agree more closely over much of the energy range.

In contrast with the p+p data, which agree with the e$^+$+e$^-$ data over a
large energy range only after rescaling, there is reasonable agreement of the
total charged particle multiplicities between e$^+$+e$^-$ and A+A collisions
over $\sqrt{s}$ and $\sqrt{s_{_{NN}}}$ of about 20 to 200~GeV with no
rescaling.  At lower energies, one sees an apparent ``suppression'' of the A+A
multiplicity compared to both p+p and e$^+$+e$^-$.  This might be explained by
reference to the substantial baryon excess found in the particle yields at
these lower energies (e.g.\ the antiproton/proton ratio $\ll$1, see references
in \cite{Bac03a}).  The relatively larger number of baryons compared to pions
should tend to suppress the overall multiplicity, since the baryon chemical
potential reduces the entropy.  Essentially, the net baryons 
take up an increasing fraction of the available energy.  Additionally, the
overlap of the peak of the rapidity distributions of the net baryons and the
produced pions \cite{Bac01c} could result in increased pion absorption during the
evolution of the system.

The arguments made here suggest that the total multiplicity per participant pair is a universal
function of the available energy, irrespective of the colliding system \cite{Bac03e}.
All of the heavy ion data shown in Fig.~\ref{WP11_Ntot_sqrts_AA_pp_ee} are for 
central collisions, but as shown in Sec.~\ref{Sec4-B} the numbers remain 
constant over a broad range of impact parameter.  
This is a surprising result if p+p collisions are expected to be a
``reference system'', while the enhanced multiplicity in A+A is related to more
exotic physics.  Moreover, the prediction of the energy dependence of the e$^+$+e$^-$ multiplicity is
widely understood as a paradigmatic success of perturbative QCD \cite{Mue83}, 
while a broader range of processes are expected to contribute in heavy ion collisions.

This interpretation of the comparison of p+p and Au+Au systems is validated by
the $\sqrt{s_{_{NN}}}$=200~GeV d+Au results from PHOBOS \cite{Bac04i} shown in
Fig.~\ref{WP11_Ntot_sqrts_AA_pp_ee} for the most central collisions.  If it takes more than
one collision in order for all of the energy to be available for particle production, then one would expect the
participants in the deuteron to contribute approximately half the multiplicity
of an e$^+$+e$^-$ collision (i.e.\ with effective energy of $\sqrt{s}$ ), while
the participants in the gold nucleus would contribute half a p+p collision.
For a central d+Au collision, the ratio of gold to deuteron participants is
approximately 8, so the ``p+p-like'' collisions should dominate, making the
multiplicity closer to p+p, an expectation that is validated by the data.

It should be emphasized that this result applies mainly to the total
multiplicity and not necessarily to other details of particle production.  In other words,
this argument does not imply that A+A collisions are merely scaled up
e$^+$+e$^-$ annihilations.  The presence of elliptic flow and strangeness
enhancement, along with other observations, precludes this possibility.  Furthermore, it is not
argued that all observables in A+A collisions should be compared to similar
data from p+p at twice the center-of-mass energy.  Still, the similarities
between the total charged particle multiplicities of these various systems raise
the question of what are the decisive differences between the
larger and smaller systems.  Some insight may come from studying the role of
the size and shape of the collision volume, which will be addressed in later
sections.

While the physics scenario as stated is consistent with a broad range of
multiplicity data, it is complicated somewhat by the recent BRAHMS result on
the net baryon distribution, which is interpreted in terms of the net rapidity
loss of the incoming baryons \cite{Bea03}.  Although the measurements do not include the
bulk of the net baryons, the data can constrain the shape of the distributions
substantially.  The BRAHMS analysis finds that the average rapidity loss of the
net baryons in central Au+Au collisions at $\sqrt{s_{_{NN}}}$=200~GeV is
$\Delta y\sim$2~units \cite{Bea03}, which is consistent with values extracted
from p+A data at lower energy \cite{Bus84,Bus88}.  When translated into
``available'' energy, i.e.\ the total incoming energy minus the energy of the
net outgoing baryons, only about 75\% of the energy is left for particle production in
central A+A collisions.  It should be noted that this value is a lower limit
based on the assumption that the effects of longitudinal expansion can be
ignored.  If this reduced available energy is accounted for in the Au+Au data
as was done for p+p, the resulting data points in the bottom panel of
Fig.~\ref{WP11_Ntot_sqrts_AA_pp_ee} would increase by about 15\%.  This would imply that Au+Au
collisions are, in fact, able to convert the same amount of energy into a
slightly larger number of particles than are produced in e$^+$+e$^-$
annihilations at the same center-of-mass energy.  Given the systematic uncertainties in the various data sets, it
is difficult to determine which of these interpretations is correct.
Furthermore, given the current lack of understanding of the longitudinal
dynamics in RHIC collisions (see Sect.~\ref{Sec4-E}), the
validity of the assumption that all of the energy carried by the net baryons is
``unavailable'' for particle production is far from obvious.  What is
unambiguous is the surprisingly close correspondence of all systems despite the
common assumption that somewhat different physics dominates in each case.

In summary, the data show that the systematics of the total charged-particle
multiplicities are suggestive of a universal mechanism which affects ``bulk''
features of particle production in strongly-interacting systems.  The dominant
control variable in this picture appears to be the available or ``effective''
energy, per participant pair, which is apparently 50\% of $\sqrt{s_{_{NN}}}$ in a p+p or d+Au
collision, but appears to be a significantly larger fraction of
$\sqrt{s_{_{NN}}}$ in A+A and presumably all of $\sqrt{s}$ in e$^+$+e$^-$
reactions.  This may simply be related to the fact that typical participants
in an A+A collision are multiply struck when passing through the oncoming
nucleus.  A more complete description would involve a full explanation of the
nature and origin of the outgoing baryons in both nucleon-nucleon and
nucleus-nucleus collisions.  All of these issues thus require a more
comprehensive understanding of the early-time dynamics of the collision
process, including both the dynamics of baryon-number transport and entropy
production.

\subsection{Centrality dependence of total multiplicity} 
\label{Sec4-B}

One of the key tools for understanding particle production in high energy p+A
and A+A collisions is the study of the system-size dependence, either by
varying the size of the colliding nuclei or by classifying the collisions
according to centrality.  Variation of the collision centrality not only
changes the volume of the particle production region, but also the number of binary collisions per participant (see Appendix~\ref{SecAppB-3}
for more discussion of this topic).  In addition to changing the collision energy, varying centrality therefore provides
another handle, in principle, for changing the
balance of particle production between `soft' low-momentum processes and
point-like `hard' processes with large momentum transfer.

\begin{figure}[ht]
\begin{center}

\includegraphics[width=0.80\textwidth]{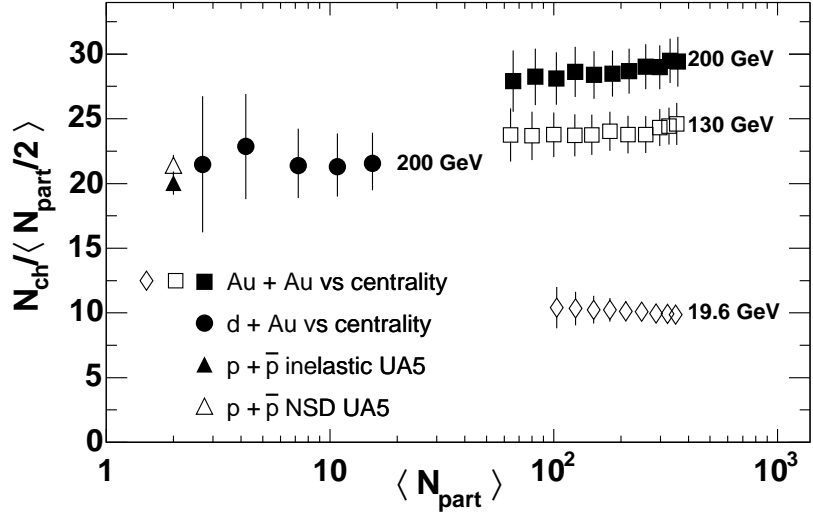}

\caption{ \label{WP12_Ntot_Npart_AuAu_19130200_dAu_pp}
Total integrated charged particle multiplicity per participant pair as a
function of number of participants. Data are shown for Au+Au collisions at
$\sqrt{s_{_{NN}}}$ of 19.6, 130 and 200~GeV \cite{Bac03e}, as well as d+Au 
\cite{Bac04i} and $\bar{p}+p$ at 200~GeV \cite{Aln87}. The vertical bars
include both statistical and systematic (90\% C.L.) uncertainties.}

\end{center}
\end{figure}

\begin{figure}[ht]
\begin{center}

\includegraphics[width=0.95\textwidth]{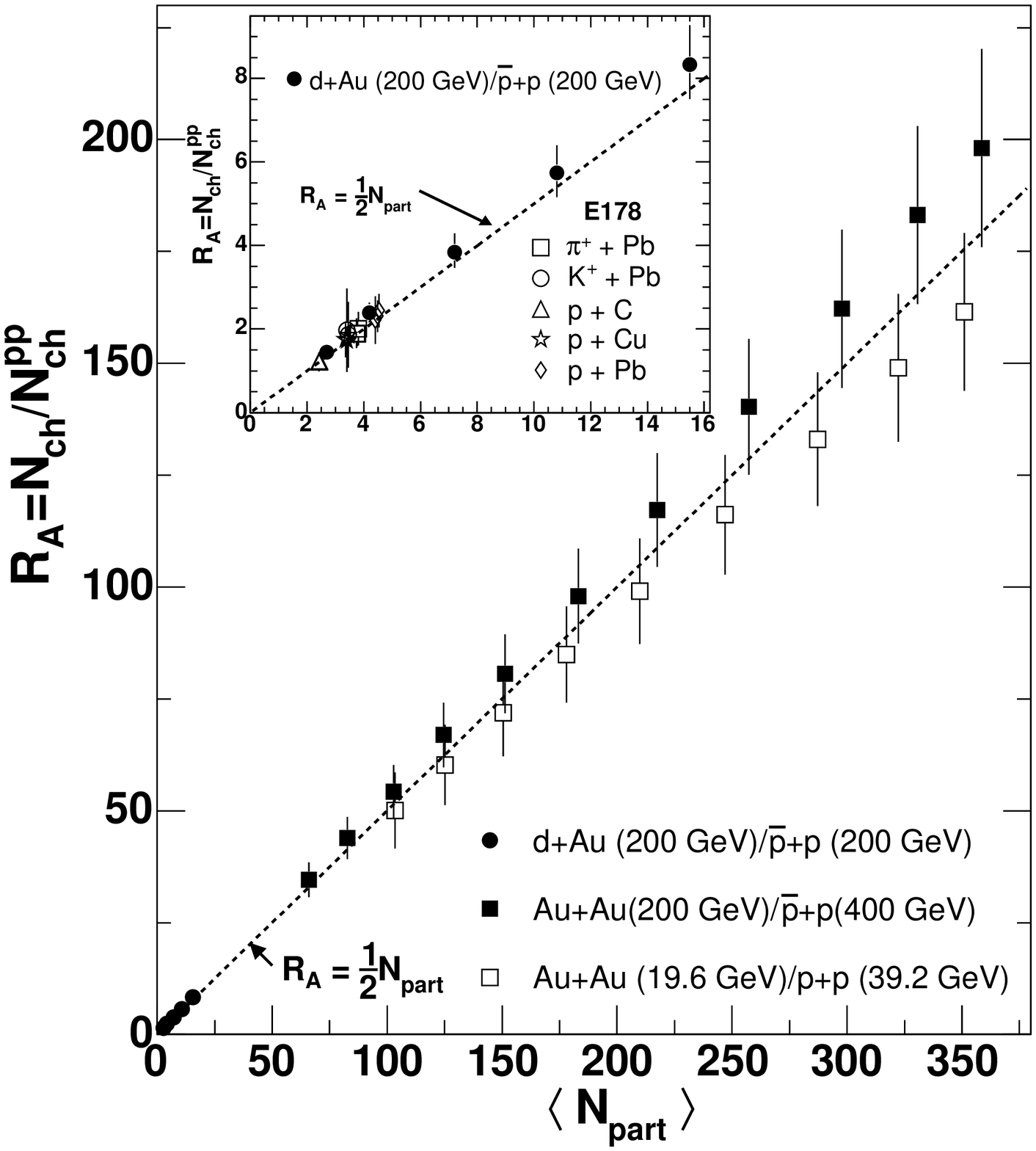}

\caption{ \label{WP13_NtotNpp_Npart_mA_dA_AA}
Ratios of total particle multiplicity data for a wide range of hadron-nucleus \cite{Eli80}, and nucleus-nucleus collisions
\cite{Bac04e,Bac04i} over  the multiplicity in proton(antiproton)-proton 
interactions \cite{Whi74,Whi76,Bog74,Aln87} are plotted
versus the number of participating nucleons.  The denominator for interactions
induced by mesons, protons, or deuterons is proton-proton data at the same
center-of-mass energy.  For Au+Au interactions, the denominator is 
proton(antiproton)-proton data at twice the center-of-mass energy.  The error bars
include both statistical and systematic effects.  Furthermore, they are
partially correlated due to common errors in $N_{ch}^{pp}$.  Note that all the
data fall on a common line with a slope of 1/2 (as expected since p+p has two
participants) and zero intercept.}

\end{center}
\end{figure}

One of the more striking features of total particle production in Au+Au
collisions at RHIC is the proportionality of the total charged-particle
multiplicity to  the number of participant pairs \cite{Bac03e}, as shown in
Fig.~\ref{WP12_Ntot_Npart_AuAu_19130200_dAu_pp} and compared to $\bar{p}$+p \cite{Aln87} and d+Au collisions \cite{Bac04i}.  The
figure also shows that the total charged particle multiplicity is proportional
to the number of participating nucleons in Au+Au collisions at all three
energies from $\sqrt{s_{_{NN}}}$=19.6 to 200~GeV.
The data suggest that the transition between p+p collisions and Au+Au is
probably not controlled simply by the number of participants, as even very
central d+Au collisions do not show any sign of trending up towards the level
of the Au+Au data.  As discussed in the preceding section, this aspect of the
total multiplicity is expected in the ``available energy'' {\it ansatz}, 
since the Au participants, which dominate the total number of participants in d+Au, are expected to
be more ``p+p-like''.

This topic represents one area where data for collisions
of lighter nuclei at RHIC could make an important contribution.  Extrapolation
of Au+Au analysis to very peripheral collisions inevitably suffers from
considerable systematic uncertainty in the number of participants.  Lessons
learned from analysis of lower energies and smaller systems such as d+Au are
currently being applied in an attempt to reduce those uncertainties.  However,
it is clear that data from lighter systems, currently being collected in Run V at
RHIC, will provide vital input to the interpretation of these results.

Further information about the centrality dependence is shown in
Fig.~\ref{WP13_NtotNpp_Npart_mA_dA_AA}, the inset of which shows a detailed comparison of the
PHOBOS d+Au results at $\sqrt{s_{_{NN}}}$=200~GeV \cite{Bac04i} with $\pi$+A,
K+A, and p+A for $\sqrt{s_{_{NN}}}\approx$10--20~GeV \cite{Eli80}.  In all
cases in the inset, the total charged particle multiplicity in hadron-nucleus
collisions is divided by the p+p multiplicity at the same collision energy.
Within the experimental uncertainty, the ratios all fall on the indicated line, demonstrating that the total charged particle multiplicity scales with the number of participant
pairs times the data for p+p at the same energy for all
hadron-nucleus systems, as was first recognized in earlier work \cite{Bus75,Bus77}.  
This feature of the data led to the ``wounded nucleon'' model of
Bia\l as, Bleszy\'{n}ski and Czy\.{z} \cite{Bia76}.
The range in $N_{part}$ over which this scaling is shown to apply is extended significantly by the PHOBOS charged particle multiplicity in
d+Au collisions versus centrality.

A similar analysis of Au+Au data for collisions at $\sqrt{s_{_{NN}}}$=19.6~GeV
and 200~GeV is shown in the main part of Fig.~\ref{WP13_NtotNpp_Npart_mA_dA_AA}
\cite{Bac04i}.  As for the hadron-nucleus data, the points fall along the line, exhibiting
scaling of the total multiplicity with the number of participant pairs, but in
this case multiplied by p($\bar{p}$)+p data at twice the center-of-mass energy \cite{Aln87,Whi74,Whi76,Bog74}.
A particularly striking feature, as discussed in Section~\ref{Sec4-A}, is the
fact that, for all these systems and energies, the total number of charged
particles is directly given by the number of participant pairs times the number
seen in p+p after accounting correctly for the energy carried away by the leading
baryon.

This continuation of the previously-observed approximate $N_{part}$ scaling, which is now seen to apply to all systems and over an expanded
range of energies from $\sqrt{s_{_{NN}}}$ below 10~GeV to the highest at
RHIC, represents one of the more surprising features of particle production at
RHIC.

\begin{figure}[ht]
\begin{center}

\includegraphics[width=0.75\textwidth]{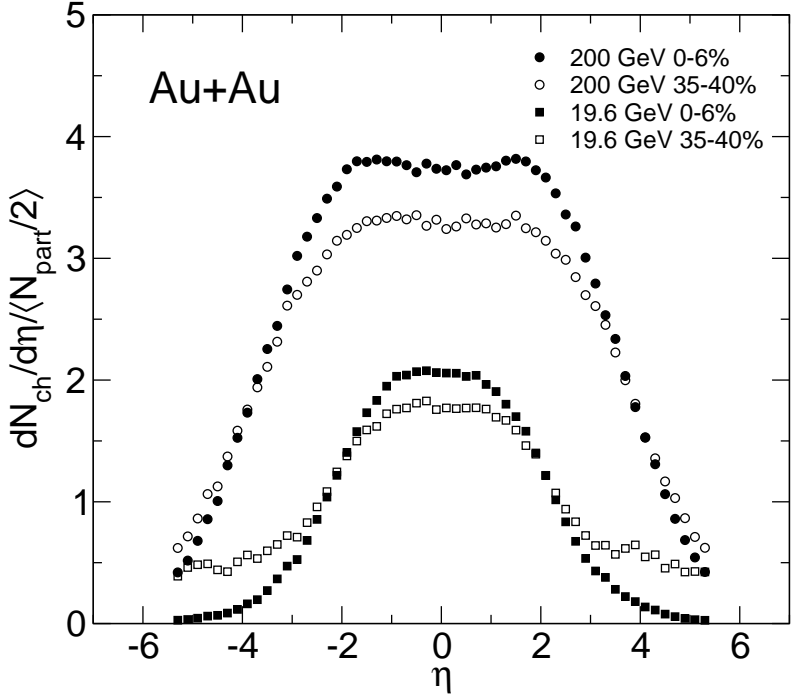}

\caption{ \label{WP14_dNdetaNpart_eta_AuAu_19_200}
Distributions of normalized pseudorapidity densities of charged particles
emitted in Au+Au collisions at two energies and two ranges of centrality \cite{Bac03c}.  The
data have been divided by the average number of pairs of participating nucleons
for each energy and centrality range.  The centrality is designated by the
fraction of the total inelastic cross section, with smaller numbers being more
central.  Systematic errors are omitted for clarity.  Statistical errors are smaller than the symbols.} 

\end{center}
\end{figure}

\begin{figure}[ht] 
\begin{center}

\includegraphics[width=0.70\textwidth]{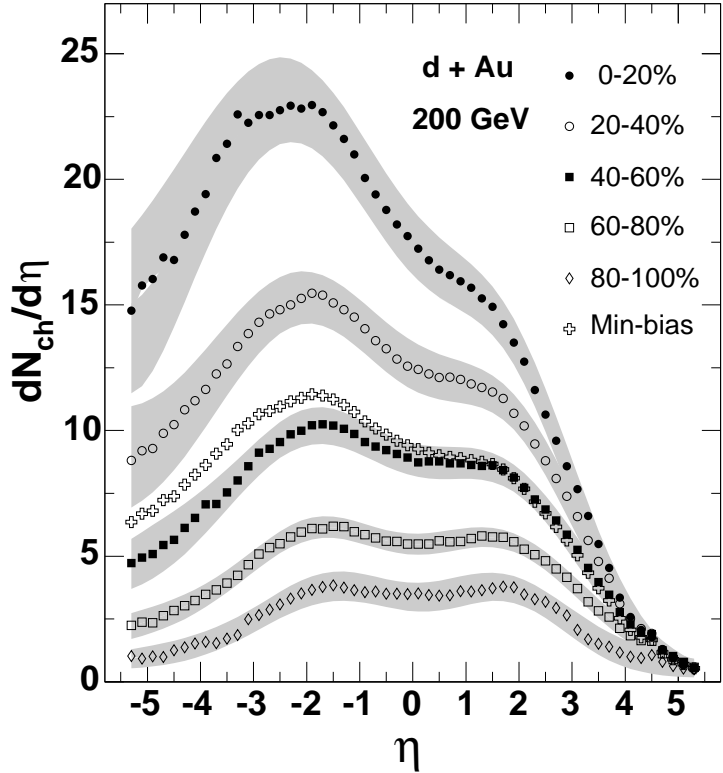}

\caption{ \label{WP15_dNdeta_eta_dAu_5cent}
Distributions of pseudorapidity densities of charged particles emitted in d+Au
collisions at $\sqrt{s_{_{NN}}}$=200~GeV for a variety of centralities \cite{Bac04i,Bac04j}. The
positive pseudorapidity direction is that of the deuteron.  The centrality is
designated by the fraction of the total inelastic cross section, with smaller
numbers being more central.  Grey bands indicate the systematic uncertainties
(90\% C.L.).}

\end{center}
\end{figure}

\subsection{Centrality dependence of pseudorapidity distributions}
\label{Sec4-C}

\begin{figure}[ht] 
\begin{center}

\includegraphics[width=0.70\textwidth]{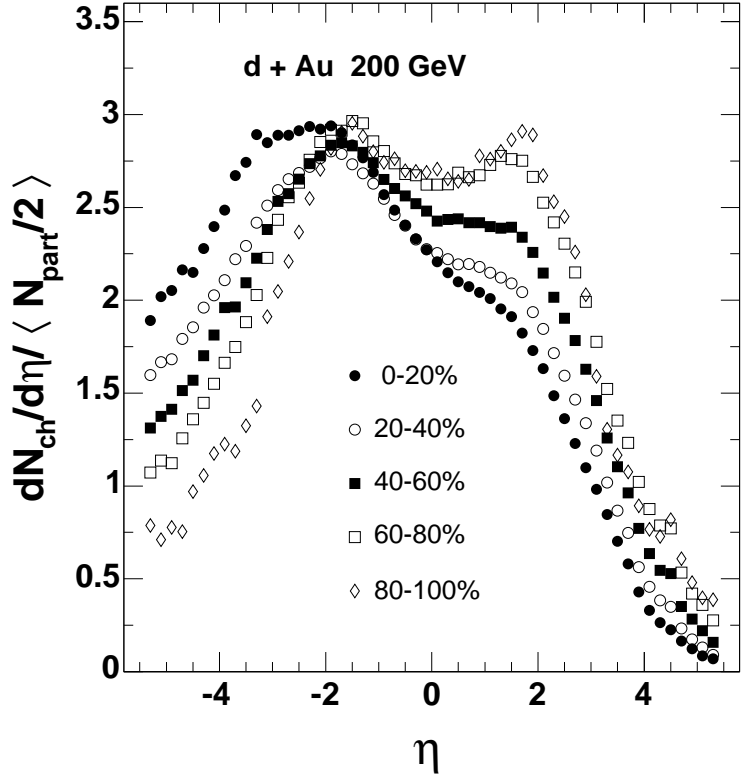}

\caption{ \label{WP16_dNdetaNpart_eta_dAu_5cent}
The data of Fig.~\ref{WP15_dNdeta_eta_dAu_5cent} are shown but in this case divided by the
average number of participant pairs in each centrality bin \cite{Bac04i}. Systematic errors are not shown.}

\end{center}
\end{figure}

It should be stressed that the universal $N_{part}$ scaling of the total number
of particles produced in Au+Au collisions does not result from rapidity
distributions whose shape is independent of centrality, or $N_{part}$.  The rapidity
distributions do depend on both centrality and on the nature of the colliding
systems, as is evident from Fig.~\ref{WP14_dNdetaNpart_eta_AuAu_19_200} for Au+Au \cite{Bac03c} and Figs.~\ref{WP15_dNdeta_eta_dAu_5cent}
and \ref{WP16_dNdetaNpart_eta_dAu_5cent} for d+Au \cite{Bac04i}.  However, the dependence of the shape on centrality, 
as first reported in \cite{Bac01b}, is very specific.

\begin{figure}[ht] 
\begin{center}

\includegraphics[width=0.85\textwidth]{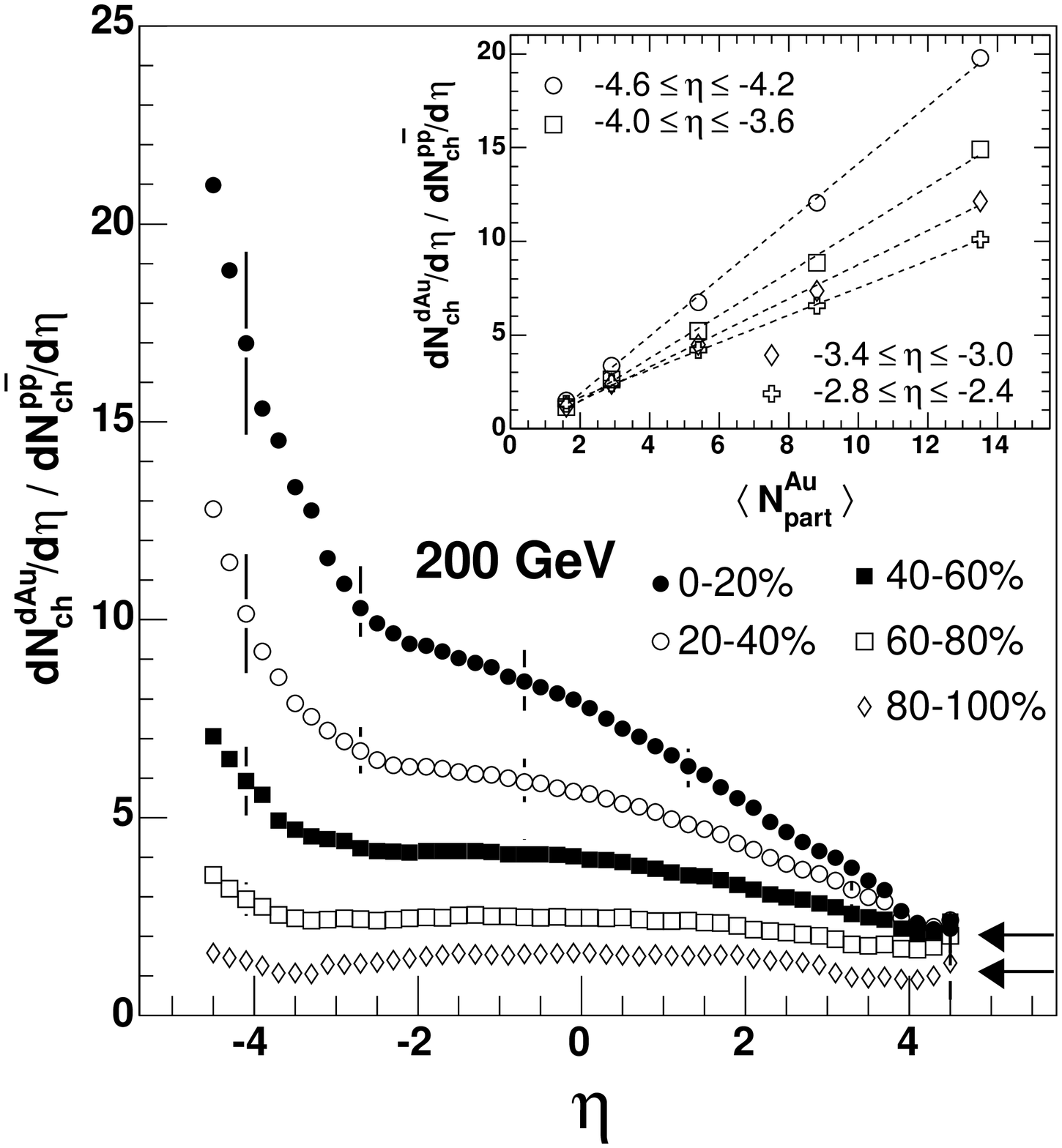}

\caption{ \label{WP17_dNdetaNpp_eta_dAu_5cent_inset} 
The main panel shows the distributions of pseudorapidity densities of charged
particles emitted in d+Au collisions with $\sqrt{s_{_{NN}}}$=200~GeV at
various centralities \cite{Bac04i} (see Fig.~\ref{WP15_dNdeta_eta_dAu_5cent}) divided by the
distribution for inelastic $\bar{p}+p$ collisions at the same energy
\cite{Aln87}.  The positive pseudorapidity direction is that of the deuteron.
Centralities are labeled by the fraction of total inelastic cross section in
each bin, with smaller numbers being more central.  The
lower and upper arrows on the right show the average number
of participants in the deuteron for the most peripheral (80--100\%) and most
central (0--20\%) bin, respectively.  The inset shows the values
averaged over several bins in negative pseudorapidity plotted versus the
average number of participants in the Au nucleus for the five centrality bins.}

\end{center}
\end{figure}

The Au+Au pseudorapidity distributions shown in Fig.~\ref{WP14_dNdetaNpart_eta_AuAu_19_200} appear to exhibit a sort of incompressibility in rapidity space.
Thus, a reduction in the number of particles at midrapidity is balanced by a similar
increase of the number of particles at high rapidities, with the total number
remaining constant.  Obviously, moving particles around in rapidity changes the
total longitudinal energy in the system.  If the total energy available for
produced particles depends only on the number of participants, energy must be
conserved by changes in the distribution of transverse momentum.

The centrality dependence of pseudorapidity distributions in asymmetric systems
can be studied using PHOBOS data for d+Au collisions as shown in
Figs.~\ref{WP15_dNdeta_eta_dAu_5cent} and \ref{WP16_dNdetaNpart_eta_dAu_5cent} \cite{Bac04j,Bac04i}.  With increasing centrality, an
increase in particle production (see Fig.~\ref{WP15_dNdeta_eta_dAu_5cent}) and a significant
change in shape of the distributions (see Fig.~\ref{WP16_dNdetaNpart_eta_dAu_5cent}) is observed. It
should be stressed that the appearance of a ``double-hump'' structure in the
d+Au distributions is primarily due to the effect of the Jacobian associated with the
transformation to $dN/d\eta$ from $dN/dy$ (see related discussion in Section~\ref{Sec4-E-3}).  Although the
shape changes in a non-trivial way, the integral of these distributions, when
extrapolated to full solid angle, is found to be proportional to the number of
participating nucleons, as was shown for many systems and energies in
Section~\ref{Sec4-B}.

The comparison of total particle multiplicity in d+Au and p+p can be extended
by studying the ratio $dN/d\eta$(d+Au)/$dN/d\eta$(p+p) as a function of pseudorapidity, as shown in
Fig.~\ref{WP17_dNdetaNpp_eta_dAu_5cent_inset} \cite{Bac04i,Aln87}.  The main panel of the figure shows this ratio 
for various d+Au centralities, as a function of
pseudorapidity.  The inset and the arrows at the lower right demonstrate that,
as was seen in p+A at lower energy \cite{Eli80,Bus77,Bri90,Dem84,Hal77}, the
data are consistent with a picture in which the density of produced particles which have a rapidity in the vicinity of the
incident deuteron (gold) is proportional to the number of deuteron (gold) participants.
The data suggest that the overall rapidity distribution, not
just the integral of the distribution, is strongly influenced by the collision
geometry.

In light of the discussion of particle production as a function of available
energy in Sect.~\ref{Sec4-A}, one might initially expect the ratio at
positive rapidity in Fig.~\ref{WP17_dNdetaNpp_eta_dAu_5cent_inset} to increase faster than the number of
deuteron participants.  This is because each deuteron participant interacts
with multiple Au participants and is therefore ``Au+Au-like'', while each Au
participant suffers far fewer collisions and is therefore ``p+p-like''.
Recall that the normalized multiplicity per participant pair in Au+Au collisions was higher than that in p+p
collisions at the same center-of-mass energy. 
However, it is important to keep in mind that the detailed
shape of the distribution, not just the relative height at the two ends, is a
complicated function of centrality.
For example, it has long been known that in p+A collisions, the yield of
all particles with rapidity within a unit or so of that of
the proton falls with increasing target mass \cite{Bar83}.
Thus, one should not expect conclusions from
integrated yields to apply simply to narrow fixed regions of pseudorapidity.

The longitudinal properties of particle production, and in particular the
dependence on center-of-mass energy, are discussed in more detail in
Sect.~\ref{Sec4-E}.

\subsection{Comparison of Au+Au and other systems} 
\label{Sec4-D}

\begin{figure}[ht] 
\begin{center}

\includegraphics[width=0.80\textwidth]{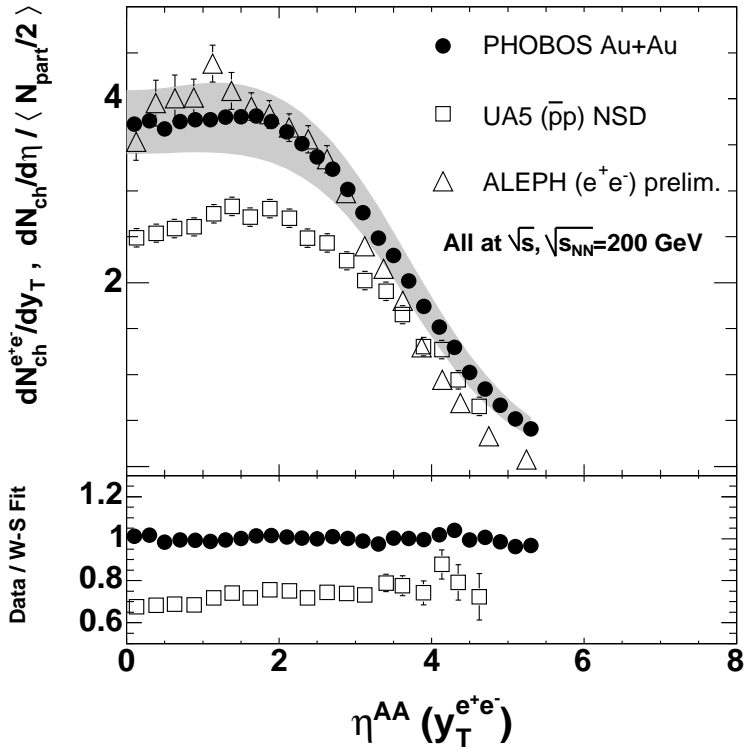}

\caption{ \label{WP18_dNdeta_eta_AA_pp_ee}
(Top panel) The $dN/dy_{_T}$ distribution for charged particles emitted in e$^+$+e$^-$
collisions \cite{Ste00} is compared to the $dN/d\eta$ distribution for charged particles
emitted in $\bar{p}$+p \cite{Aln86} and the normalized $dN/d\eta$ distribution for charged particles
emitted in the 3\% most central Au+Au collisions \cite{Bac03c}.  All three systems are at
$\sqrt{s_{_{NN}}}$ or $\sqrt{s}$ of 200~GeV. 
(Bottom panel) The Au+Au and p+p
data are both shown divided by a fit to the former \cite{Bac03e}.}

\end{center}
\end{figure}

Figure~\ref{WP11_Ntot_sqrts_AA_pp_ee} showed that the total charged particle multiplicities in the
e$^+$+e$^-$ and A+A systems are very similar at a given center-of-mass energy,
while those for p+p are somewhat smaller. To expand the comparison of these three very different systems, it is
interesting to consider the full distributions in pseudorapidity.  
However, this study is complicated by the fact that
the shapes of the Au+Au data vary dramatically with centrality (as is most
clearly evident in Fig.~\ref{WP14_dNdetaNpart_eta_AuAu_19_200}).  Figure~\ref{WP18_dNdeta_eta_AA_pp_ee} compares
$dN_{ch}/d\eta$ normalized by the number of participant pairs for the 3\% most
central Au+Au collisions \cite{Bac03c} to inelastic data for $\bar{p}$+p \cite{Aln86} and the distribution of
$dN_{ch}/dy_{_T}$ (see definition in Appendix~\ref{SecAppB-2}) in the e$^+$+e$^-$ data \cite{Ste00}, all at a $\sqrt{s_{_{NN}}}$ or $\sqrt{s}$ of 200~GeV \cite{Bac03e}.
The bottom panel of the figure demonstrates that the
lower total multiplicity seen in $\bar{p}$+p results from a pseudorapidity distribution
that is suppressed by roughly a constant factor over all emission angles. The figure shows
agreement in the overall rapidity distribution between A+A and e$^+$+e$^-$.
In comparing the two distributions, one should keep in mind the centrality 
dependence in the shape for Au+Au, as well as the difference between 
$dN/dy_{_T}$ and $dN/d\eta_{_T}$.  Studies using JETSET \cite{Sjo94} show that, 
for this data, the extracted $dN/dy_{_T}$ is about 10\% larger than 
$dN/d\eta_{_T}$ for $|y_{_T}|\sim$0 and about 10\% smaller than 
$dN/d\eta_{_T}$ for $|y_{_T}|\sim$4.  

\begin{figure}[ht] 
\begin{center}

\includegraphics[width=0.85\textwidth]{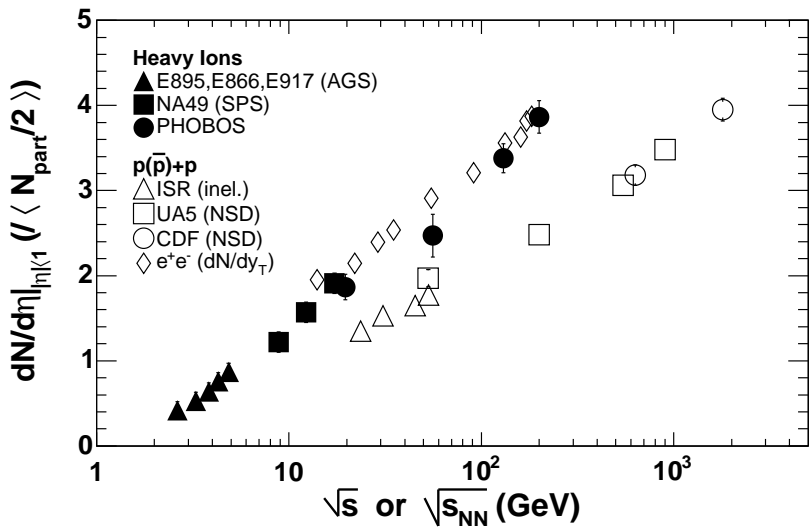}

\caption{ \label{WP19_dNdetaMid_sqrts_AA_pp_ee}
Pseudorapidity particle density near midrapidity as a function of energy for p($\bar{p}$)+p, A+A and e$^+$+e$^-$
reactions (where the e$^+$+e$^-$ density is $dN/dy_{_T}$, as explained in the text).
Data for p($\bar{p}$)+p and e$^+$+e$^-$ were extracted from results compiled in \cite{Eid04}.
Nucleus-nucleus data, shown for central collisions \cite{Bac03c,Bac00a,Bac02a,Bac02b,Bac02c,Bac04e,Afa02,Ant04,Ahl98a,Bac02e,Ahl00,Kla03,Dun99}, have been divided by the number of participating nucleon
pairs.  Note that midrapidity particle densities are not available for lower
energy p+p or e$^+$+e$^-$ collisions, in the latter case due to the lack of a well defined
jet structure.}  

\end{center}
\end{figure}

The similarity of the integrated multiplicity, as well as the shapes of the pseudorapidity distributions, for
e$^+$+e$^-$ and the most central Au+Au data suggests that there should be a
similarity in the evolution of the midrapidity density with collision energy, an
expectation that is verified by the data.  Figure~\ref{WP19_dNdetaMid_sqrts_AA_pp_ee} shows midrapidity particle density data from central heavy ion collisions \cite{Bac03c,Bac00a,Bac02a,Bac02b,Bac02c,Bac04e,Afa02,Ant04,Ahl98a,Bac02e,Ahl00,Kla03,Dun99} and from elementary collisions compiled from references in \cite{Eid04}.  This
additional close correspondence between the properties of central Au+Au and e$^+$+e$^-$
multiplicity data suggests that the agreement results from some underlying
feature of particle production, as opposed to being an accidental coincidence.
In particular, an understanding of why the shape of the pseudorapidity distribution for Au+Au collisions
approaches that of e$^+$+e$^-$ for more central interactions might
prove particularly enlightening.

The arguments presented in Sect.~\ref{Sec4-A} concerning total charged particle 
multiplicities should not be interpreted to imply that all observables in A+A 
will match those in p+p at a factor of two higher $\sqrt{s}$.  The midrapidity 
particle densities provide an instructive counterexample. Since the same total 
number of particles in p+p at a higher $\sqrt{s}$ are distributed over a 
broader range of pseudorapidity (see, for example, the top panel of 
Fig.~\ref{WP21_dNdeta_etaMNSybeam_pp_ee}), a factor of two shift in the p+p center-of-mass energy 
obviously cannot result in midrapidity densities equal to those measured in 
A+A.  An examination of Fig.~\ref{WP19_dNdetaMid_sqrts_AA_pp_ee} reveals that the data confirm this expectation.

\begin{figure}[ht] 
\begin{center}

\includegraphics[width=0.85\textwidth]{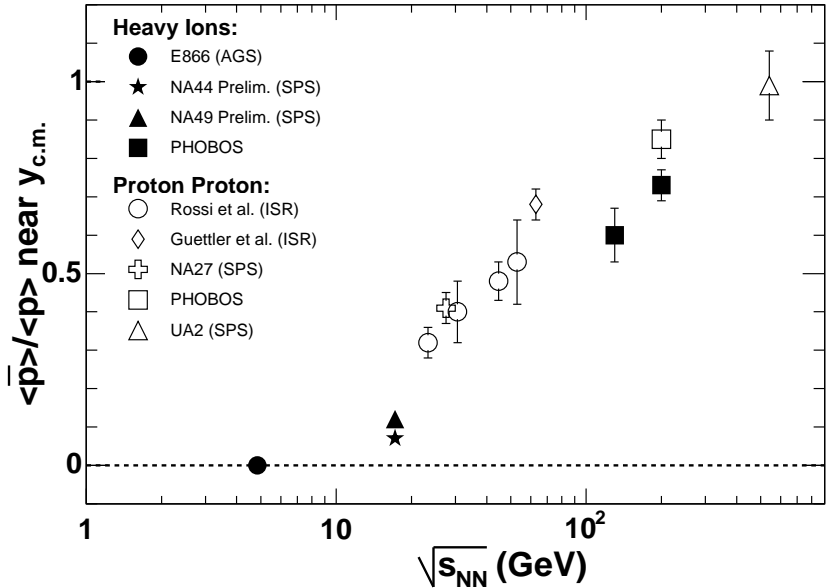}

\caption{ \label{WP20_pbarp_sqrts_AA_pp}
Antiproton to proton ratios near midrapidity as a function of
$\sqrt{s_{_{NN}}}$ for p($\bar{p}$)+p collisions (open symbols) \cite{Bac04l,Ros75,Gue76,Ban83,Agu91} and central A+A
collisions (filled symbols) \cite{Bac01a,Bac03a,Bea96,Bac99,Ahl00,Ahl99,Ahl98b}. Error bars include both statistical and systematic errors.}

\end{center}
\end{figure}

A less trivial counterexample is illustrated  in Fig.~\ref{WP20_pbarp_sqrts_AA_pp} which shows 
ratios of the yields of antiprotons over protons emitted near midrapidity in 
p($\bar{p}$)+p, as measured by PHOBOS at RHIC \cite{Bac04l} and experiments at other energies \cite{Ros75,Gue76,Ban83,Agu91}, and in A+A collisions \cite{Ahl00,Bac01a,Bac03a,Bea96,Bac99,Ahl99,Ahl98b} as a function of $\sqrt{s_{_{NN}}}$ .  The 
ratios for d+Au at $\sqrt{s_{_{NN}}}$=200~GeV \cite{Bac04b} (discussed in Sect.~\ref{Sec2-B-1}
and shown in Fig.~\ref{WP5_pbarp_nu_dAu_models}) are consistent with the value shown on the 
figure for p+p.  As discussed in Section~\ref{Sec2-B}, the relevant physics
for understanding this ratio involves the interplay of baryon transport and
antibaryon-baryon pair creation.  In this case, in contrast to the situation
for particle multiplicities, it is clear that the ratios for the nucleus-nucleus
data are comparable to those in nucleon-nucleon collisions at significantly 
{\em lower} center-of-mass energies.  Although this result may not be unexpected
given the larger baryon rapidity loss in A+A as compared to p+p, it serves to 
illustrate the importance of a systematic study to unravel the dynamical 
differences between the simpler and more complicated systems.

Finally, the extraction of nuclear modification factors, $R_{AA}$, requires the 
explicit use of a p+p reference spectrum.  The conventional choice is to use
minimum bias data from inelastic interactions of p+p at the same collision
energy, and all of the PHOBOS analyses have adhered to this standard.  On the
other hand, it was shown in Fig.~\ref{WP19_dNdetaMid_sqrts_AA_pp_ee} of this section and
Fig.~\ref{WP11_Ntot_sqrts_AA_pp_ee} of Sect.~\ref{Sec4-A} that the charged particle multiplicity
per participant (both at midrapidity and integrated over all solid angle) is
larger in A+A than in p+p at the same energy.  At $\sqrt{s}$ values of 200~GeV
and above, it is known that the $p_{_T}$ spectra in p+p events with higher than
average total multiplicity fall off less steeply than those for minimum bias
events \cite{Arn82,Alb90,Boc96}.  It should be stressed that we do not claim
that an alternative p+p reference spectra is in any way inherently more
appropriate.  However, since the physics that determines the shapes of the
transverse momentum spectra in p+p and A+A is not fully understood, such an
alternative comparison could prove instructive.  Therefore, one should keep
these ambiguities in mind when interpreting data for the $R$-factors, particularly
the specific value of the factors at large transverse momentum.

It should be noted that, although the relative yield at low and high $p_{_T}$
changes with multiplicity in p+p collisions, there is evidence that the change in shape is
relatively small above $p_{_T}\sim$2~GeV/c \cite{Boc96}. In addition, the
question of what p+p reference spectrum to use does not affect modification
factors such as $R_{PC}^{N_{part}}$ which directly compare A+A at different
centralities.  Therefore, any possible ambiguities in nuclear modification
factors due to the variation of the $p_{_T}$ distribution with multiplicity in
p+p do not significantly impact any of the conclusions presented in this
paper.

Of course, for very peripheral A+A collisions, all observables must evolve to 
match those in p+p (or, to be exact, the appropriate mix of p+p, p+n, n+p, and n+n) 
collisions at the same $\sqrt{s}$.  The current PHOBOS analysis of Au+Au 
collisions typically spans a range of impact parameters corresponding to a 
variation in the average number of participants in each centrality bin of more 
than a factor of 5--6, i.e.\ from roughly 60 up to 350 or more.  One remarkable 
aspect of this broad data set is that, over this range, the total particle 
multiplicity deviates very little from its central value when suitably 
normalized by the number of participants (see Fig.~\ref{WP12_Ntot_Npart_AuAu_19130200_dAu_pp}).  The 
normalized pseudorapidity density near midrapidity does vary and is tending 
towards the p+p value but is still far above it for the most peripheral 
collisions studied to date (see discussion in Sect.~\ref{Sec4-F}). The shape 
and magnitude of the transverse momentum distributions also vary but only 
slightly and they show little sign of tending towards the p+p distribution 
(see Fig.~\ref{WP8_RAA_pT_AuAu_64_200}).  One can speculate that these deviations between 
peripheral Au+Au and p+p collisions might result from the fact that the number 
of collisions per participant (or the fraction of the participants that are 
multiply struck) rises extremely rapidly with decreasing impact parameter for 
these most grazing collisions (See Appendix~\ref{SecAppB-1} and 
Fig.~\ref{WP36_Ncoll_Npart_b_Npart}). 

In summary, comparisons of data for A+A and more elementary systems reveal an
intriguing array of similarities and differences.  Clearly, it is not possible
to describe A+A collisions as trivial combinations of any other simpler
systems.  Rather than assuming that a single data set, such as p+p data at the
same $\sqrt{s_{_{NN}}}$, can serve as an ideal ``reference'' set for
interpreting the complete dynamics of A+A interactions, the properties of a
variety of systems should be studied over a range of energies and centralities
to elucidate the similarities and differences among them.  Such a study will lead to
a more complete understanding of the salient features of the underlying
physics, especially how the characteristics of the exciting regime of high
energy density created in central Au+Au collisions at RHIC energies relate to
those for other types of interactions.

\subsection{Extended longitudinal scaling}
\label{Sec4-E}

This section describes several features of the pseudorapidity dependence of
observables in a variety of systems.  In particular, the distributions of particle yield and elliptic flow are found to be 
largely independent of center-of-mass energy over a broad region of pseudorapidity 
when shifted by $y_{beam}$ and thereby effectively viewed in the rest frame of
one of the colliding particles.  In addition, no evidence is found for a broad 
region near midrapidity displaying the characteristic constant value of observables expected for a 
boost-invariant scenario.

\subsubsection{Longitudinal dependence of particle production: Elementary systems}
\label{Sec4-E-1}

Before considering the energy dependence of pseudorapidity distributions in
heavy ion collisions, it is instructive to review the extensive 
literature devoted to interpretations of, and expectations for, such
distributions in simpler systems.  A very general picture of elementary
hadron-hadron collisions emerged in the late 1960's, consisting of two sources
of particle production.
This concept led to
the prediction of two types of scaling laws for the distributions of final
state particles in the regions of the longitudinal momentum space which are either near to or far from the colliding partners. 

Particles near beam and target rapidity were thought to be governed by the
``limiting fragmentation hypothesis'' \cite{Ben69}. In this model, the 
momentum distribution of 
particles of species ``i'' in the rest frame
of one of the original colliding hadrons (commonly denoted with a prime to
distinguish it from the center-of-mass frame), $E_i d^3N_i/dp^{\prime 3}$, or equivalently
$d^3N_i/p_{_T}dy^{\prime} dp_{_T}d\phi$, 
becomes energy-independent at high enough collision energy. The central concept
is that the 
``projectile'' hadron, when seen in the frame of the ``target'', 
is Lorentz-contracted into a very narrow strongly-interacting pancake which
passes through the target. This interaction leaves behind a complicated 
excited state whose properties do not depend in detail on the energy or even 
identity of the projectile, and which
then ``fragments'' into a final state distribution of
particles, $E_i d^3N_i/dp^{\prime 3}$. 
It was generally assumed that this process produced particles
primarily in a restricted window of rapidity around $y^{\prime}$=0, possibly even
leading to a complete lack of particles at midrapidity in a very high energy
hadron-hadron collision \cite{Cho70}.

\begin{figure}[ht]
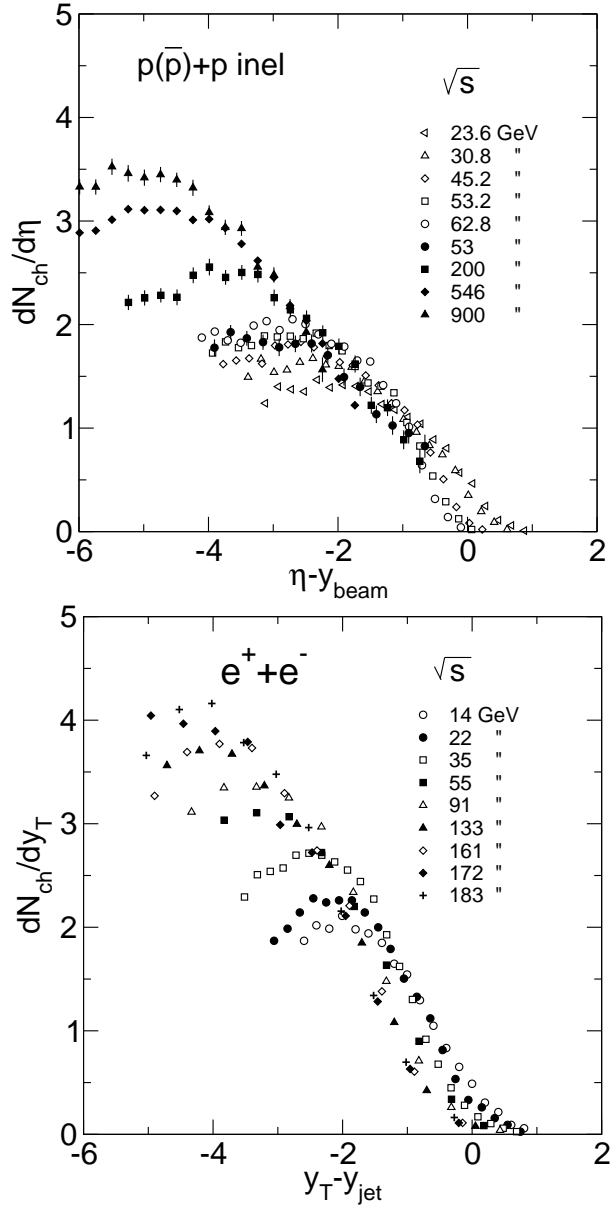
 
\begin{center}

\includegraphics[width=0.57\textwidth]{WP21a_dNdeta_etaMNSybeam_pp_ee.eps}
\break\hfill
\includegraphics[width=0.57\textwidth]{WP21b_dNdeta_etaMNSybeam_pp_ee.eps}

\caption{ \label{WP21_dNdeta_etaMNSybeam_pp_ee}  
(Top panel) Distributions of pseudorapidity density of charged particles emitted
in p($\bar{p}$)+p collisions at a range of energies versus the variable
$\eta-y_{beam}$ \cite{Tho77,Aln86}. (Bottom panel) Similar data for particles emitted along the
jet axis in an e$^+$+e$^-$ collision versus the variable $y_{_T}-y_{jet}$,
defined in Appendix~\ref{SecAppB-2} \cite{Abr99}. In both cases, when effectively viewed in the ``target''
rest frame, these collisions exhibit longitudinal scaling (energy
independence).}

\end{center}
\end{figure}

In contrast, particles near midrapidity in the center-of-mass frame were
expected to form a rapidity plateau with a constant $dN/dy$, independent of
energy and the nature of the hadrons in the initial collision \cite{Fey69,Fey72}. 
Similarly, in heavy ion collisions, 
a boost-invariant central plateau where ``the initial conditions \ldots are invariant with
respect to [longitudinal] Lorentz transformations''
(i.e.\ 
observables are independent of $y$) was predicted \cite{Bjo83}.
Furthermore, the extent of this boost-invariant region was expected to grow
with energy. 

For elementary collisions such as p+p, and even e$^+$+e$^-$, this general
picture failed completely.
Instead, the extended
longitudinal scaling, seen 
in the form of $x_{_F}$ scaling,
pointed the way to 
the current view in terms of QCD, modeled for instance in the widely used Pythia code \cite{Sjo01}.
This formulation
generalized the concept of ``fragmentation'', which ``describes the way the 
creation of new quark-antiquark pairs can break up a high-mass system into
lower-mass ones, ultimately hadrons'' \cite{Sjo03}.  It should be noted that
energy independence, or scaling, in $E_i d^3N/dp^{\prime 3}$ (i.e.\ full ``limiting
fragmentation'') implies scaling of both $dN/d{y^{\prime}}$ and $dN/dx_{_F}$. 

Figure~\ref{WP21_dNdeta_etaMNSybeam_pp_ee} shows $dN/d{\eta^{\prime}}$ for p($\bar{p}$)+p collisions
\cite{Aln86,Tho77} and $dN/d(y_{_T}$-y$_{jet}$) for e$^+$+e$^-$ collisions
\cite{Abr99} (see Appendix~\ref{SecAppB-2} for definitions). 
Lorentz
boosts of pseudorapidity, $\eta$, are not as trivial as those of rapidity, but
${\eta^{\prime}} \equiv \eta - y_{beam}$ (or $\eta+ y_{beam}$) approximates
${y^{\prime}}$. Furthermore, as noted above, the limiting fragmentation concept implies 
scaling in the full distribution, $E_i d^3N_i/dp^{\prime 3}$.  Since ${\eta^{\prime}}$ is just a
function of (${y^{\prime}}$, $p_{_T}$, $m_i$), scaling in $dN/d{\eta^{\prime}}$ is also implied directly.
For these elementary
systems, instead of a growing boost-invariant plateau, an extended version of
limiting fragmentation is found, which leads to longitudinal scaling (energy 
independence) over more than four units of rapidity,
extending nearly to midrapidity. 
The entire system can be described in terms of either string
``fragmentation'' or in terms of a parton cascade, leading naturally to extended longitudinal 
scaling.

\begin{figure}[ht] 
\begin{center}

\includegraphics[width=0.95\textwidth]{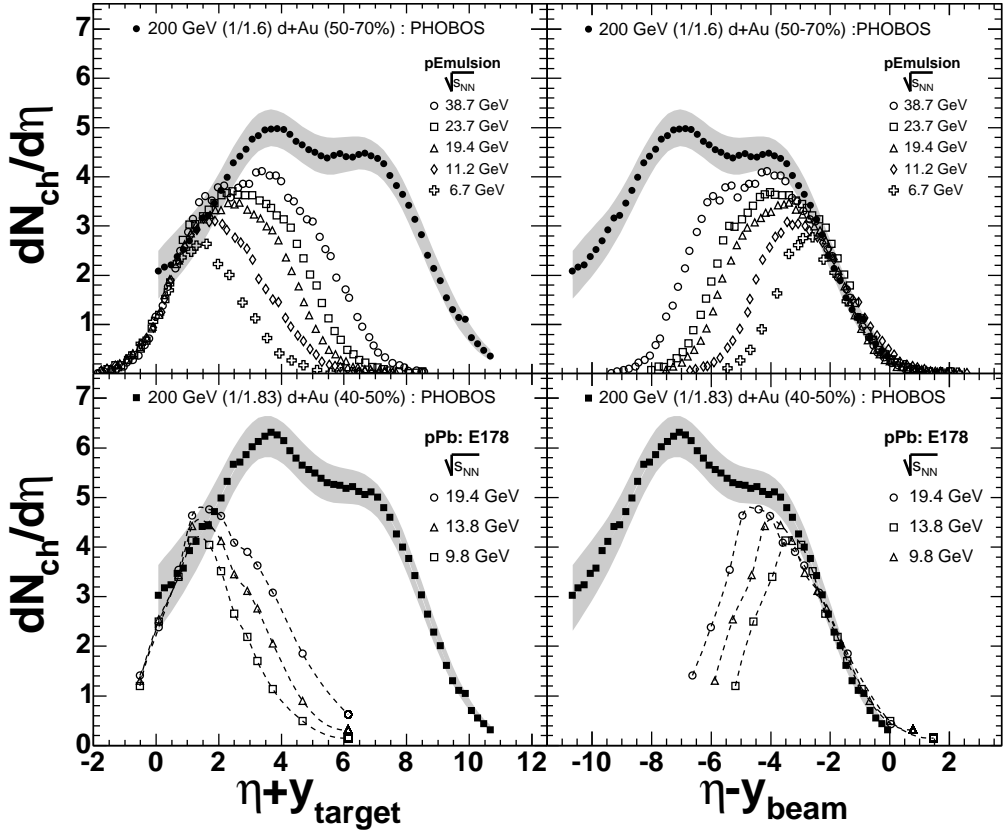}

\caption{ \label{WP22_dNdeta_etaPMybeam_dAu_pA}
A compilation of distributions of pseudorapidity densities of charged particles
emitted in p+A and d+A collisions at a variety of energies\cite{Bac04i,Eli80,Ott78,Fre87}.  Grey tracks are included in the distributions shown for emulsion data.  
The data are plotted versus the variables
$\eta+y_{target}$ and $\eta-y_{beam}$ calculated using the rapidity of the
larger (left panels) or smaller (right panels) of the colliding species.  Note
that the data at all energies and at both ends of
the pseudorapidity range follow common curves.}

\end{center}
\end{figure}

\subsubsection{Longitudinal dependence of particle production: d+A and p+A}
\label{Sec4-E-2}

In the case of asymmetric systems, the concept of extended longitudinal scaling can be
explored separately in the rest frame of the two projectiles. 
Such studies, applied to hadron-nucleus collisions, were of particular interest 
in the 1970's \cite{Bus77}.  The specific question was whether the region
of rapidity in which the particle yield is A-dependent expands with 
increasing collision energy \cite{And76,Got74,Nik81}.  Many models predicted that an extended 
A-dependent region, indicative of long-range order, should not occur.  
Instead, only a localized region near the rapidity of the larger 
collision partner would be affected by the target mass, and further, the
height and width of this region was expected to be independent of, or at most
weakly dependent on, beam energy.  One prediction of these expectations was that
the integrated yield in p+A would approach the value observed in p+p at high
beam energies, since the small A-dependent region would become increasingly
unimportant \cite{Sto75}.  Instead, to the surprise of many people, a broad A-dependent region
was observed, displaying characteristics very similar to the extended 
longitudinal scaling observed in simpler systems 
\cite{Bus88,Eli80,Bri90,Hal77,Ott78,Fre87}.

Pseudorapidity distributions from PHOBOS can be used to extend these studies
to d+A collisions at RHIC energies.  In Fig.~\ref{WP22_dNdeta_etaPMybeam_dAu_pA}, a
compilation of pseudorapidity density data for proton+(nuclear emulsion)
\cite{Ott78,Fre87} and p+Pb \cite{Eli80} at various energies is shown, together
with PHOBOS data for d+Au at $\sqrt{s_{_{NN}}}$=200~GeV \cite{Bac04i}, with the
centrality and normalization for the d+Au results chosen appropriately.  To be
more specific, the d+Au pseudorapidity densities are divided by the number of
participating nucleons in the deuteron (by definition this would be unity for
p+A). Furthermore, the d+Au centrality bin was selected such that the ratio of
the number of participating nucleons in the Au nucleus to the number in the
deuteron was equal to the number of participating nucleons from the lead or
emulsion in p+A.  This latter quantity is commonly denoted $\bar{\nu}$, the average
number of collisions per participant in the smaller projectile (see definitions
in Appendix~\ref{SecAppB-1}).  
Fig.~\ref{WP22_dNdeta_etaPMybeam_dAu_pA} clearly demonstrates that extended longitudinal scaling 
also is manifested in d+A collisions at RHIC energies.  

\subsubsection{Longitudinal dependence of particle production: Au+Au at RHIC}
\label{Sec4-E-3}

\begin{figure}[ht]
\begin{center}

\includegraphics[width=0.65\textwidth]{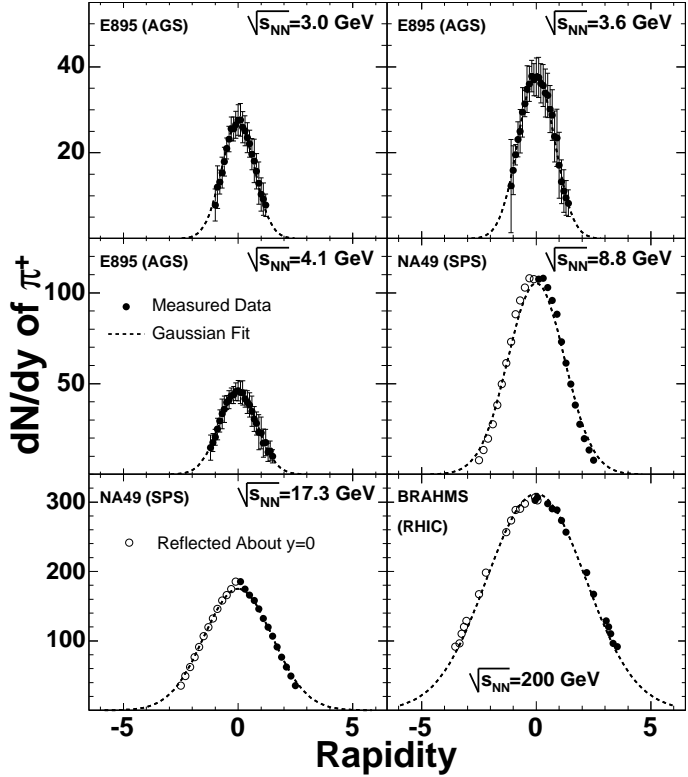}

\caption{ \label{WP23_dNdy_y_AA_6sqrts}
 Rapidity densities of positive pions emitted in central collisions of Au+Au
(AGS and RHIC) \cite{Kla03,Bea04}
and Pb+Pb (SPS) \cite{Afa02}
at a variety of beam
energies.  Note that, in contrast to Fig.~\ref{WP1_dNdeta_eta_AuAu_19_130_200}, yields in rapidity
space are well represented by Gaussians with no evidence for a broad
midrapidity plateau.}

\end{center}
\end{figure}
  
The uniquely broad pseudorapidity coverage of the PHOBOS detector allows
similar studies to be performed for heavy ion collisions at RHIC energies.  At
first the pseudorapidity distributions themselves, shown in
Fig.~\ref{WP1_dNdeta_eta_AuAu_19_130_200}, suggest that $dN_{ch}/d\eta$ may develop a small
boost-invariant central plateau, but these plots are misleading for this
purpose. Pseudorapidity is known to distort the rapidity distribution for
production angles near $0^\circ$ and $90^\circ$.  Demonstrating this point, the
rapidity distributions of positive pions measured by BRAHMS \cite{Bea04}, as well as
similar data at lower energies \cite{Afa02,Kla03}, are all well represented by Gaussian fits, as
shown in Fig.~\ref{WP23_dNdy_y_AA_6sqrts}.  In short, there are no indications of the
existence of a broad boost-invariant central plateau in the final particle
distributions.

\begin{figure}[t]
\begin{center}

\includegraphics[width=0.95\textwidth]{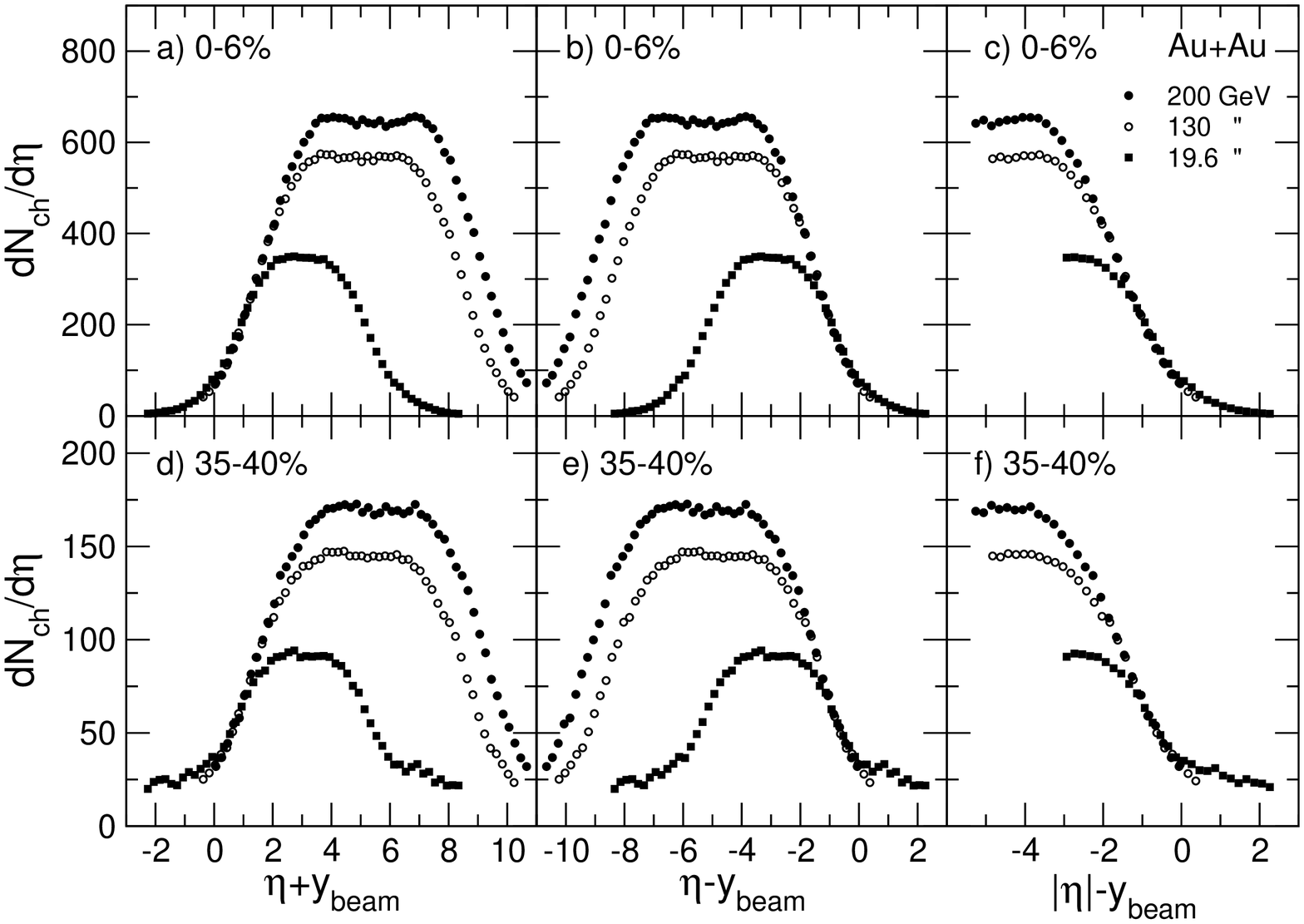}

\caption{ \label{WP24_dNdeta_etaPMybeam_AA_19_130_200}
Distributions of pseudorapidity densities of charged particles emitted in 
Au+Au collisions at three energies and two centrality ranges \cite{Bac03c} 
are plotted versus 
${\eta^{\prime}} \equiv \eta - y_{beam}$ 
(or $\eta +y_{beam}$).  In the far right panel, data for positive and negative 
$\eta$ have been averaged to generate data versus $|\eta| - y_{beam}$.
Systematic errors (identical to those on Fig.~\ref{WP1_dNdeta_eta_AuAu_19_130_200}) are not shown and statistical errors are smaller than the symbols.
Note that the data from all three energies follow
a common curve.}

\end{center}
\end{figure}

In Fig.~\ref{WP24_dNdeta_etaPMybeam_AA_19_130_200}, the data shown in Fig.~\ref{WP1_dNdeta_eta_AuAu_19_130_200} are effectively shifted to
the rest frame of one of the gold nuclei \cite{Bac03c}.  
The data at both centralities
show an extended scaling with the longitudinal velocity in the rest frame of
one of the projectiles, identical behavior to that seen in simpler systems (see,
for example, \cite{Whi74,Whi76,Aln87,Hal77}).  Similar behavior in nucleus-nucleus collisions over a
narrower range in $\eta^{\prime}$ was first observed by BRAHMS \cite{Bea01,Bea02}.

Figure~\ref{WP24_dNdeta_etaPMybeam_AA_19_130_200} illustrates one example of how  the scaling behaviors 
can be used 
to  
infer the properties of particle production
which lie outside the experimental acceptance at large collision energies.  
If one accepts the assumption that
the ${\eta^{\prime}}$ distributions at all energies are identical in the region
corresponding to larger $\eta$, the data from lower energies can be used to
constrain the extrapolation of the higher energy data to the full solid angle.
In addition, it should be noted that the corrections to the PHOBOS
multiplicity data depend strongly on emission angle of the particles and also
are significantly asymmetric between positive and negative pseudorapidities.
The latter effect results primarily from the offset of the PHOBOS magnet from
the center of the interaction region (see Fig.~\ref{WP33_PhobosDet_big}).  The good
agreement seen when comparing particles emitted at different angles and for
both signs of pseudorapidity indicates the robustness of the analysis procedure,
as well as providing interesting physics insight.
 
Fig.~\ref{WP24_dNdeta_etaPMybeam_AA_19_130_200} illustrates the observation that longitudinal scaling 
holds over an even more extended range of pseudorapidity in these seemingly
complex high energy A+A collisions at RHIC.  
Based on the pseudorapidity distribution (and, as will be discussed in
following sections, elliptic flow and perhaps even HBT), no evidence is seen
in any hadron-hadron or ion-ion collisions for two energy independent
fragmentation regions separated by a boost invariant central plateau which 
grows in extent with increasing collision energy.  Thus, the expectation from
the boost-invariant description of the energy evolution of rapidity
distributions is not valid for heavy ion collisions either.  In fact, there is
no boost invariant central plateau and, instead, the rapidity distribution
appears to be dominated by two broad ``fragmentation-like'' regions, whose extent increases 
with energy.  We call this effect ``extended longitudinal scaling''.

\subsubsection{Longitudinal dependence of elliptic flow: Au+Au at RHIC}
\label{Sec4-E-4}

\begin{figure}[ht] 
\begin{center}

\includegraphics[width=0.75\textwidth]{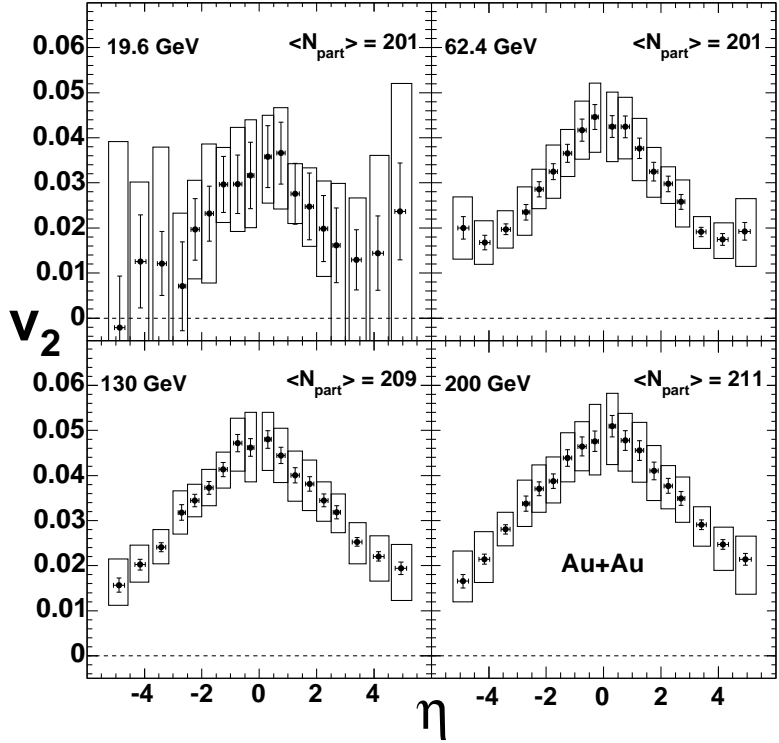}

\caption{ \label{WP25_v2_eta_AA_4sqrts}
Pseudorapidity dependence of elliptic flow of charged particles for the 40\%
most central collisions of Au+Au (average number of participating nucleons
indicated) at a variety of beam energies \cite{Bac04g}.  Note the linear fall-off at higher 
$|\eta|$ and the lack of evidence for a constant value over a broad
midrapidity region.  Boxes indicate systematic uncertainties (90\% C.L.).}

\end{center}
\end{figure}

\begin{figure}[ht] 
\begin{center}

\includegraphics[width=0.95\textwidth]{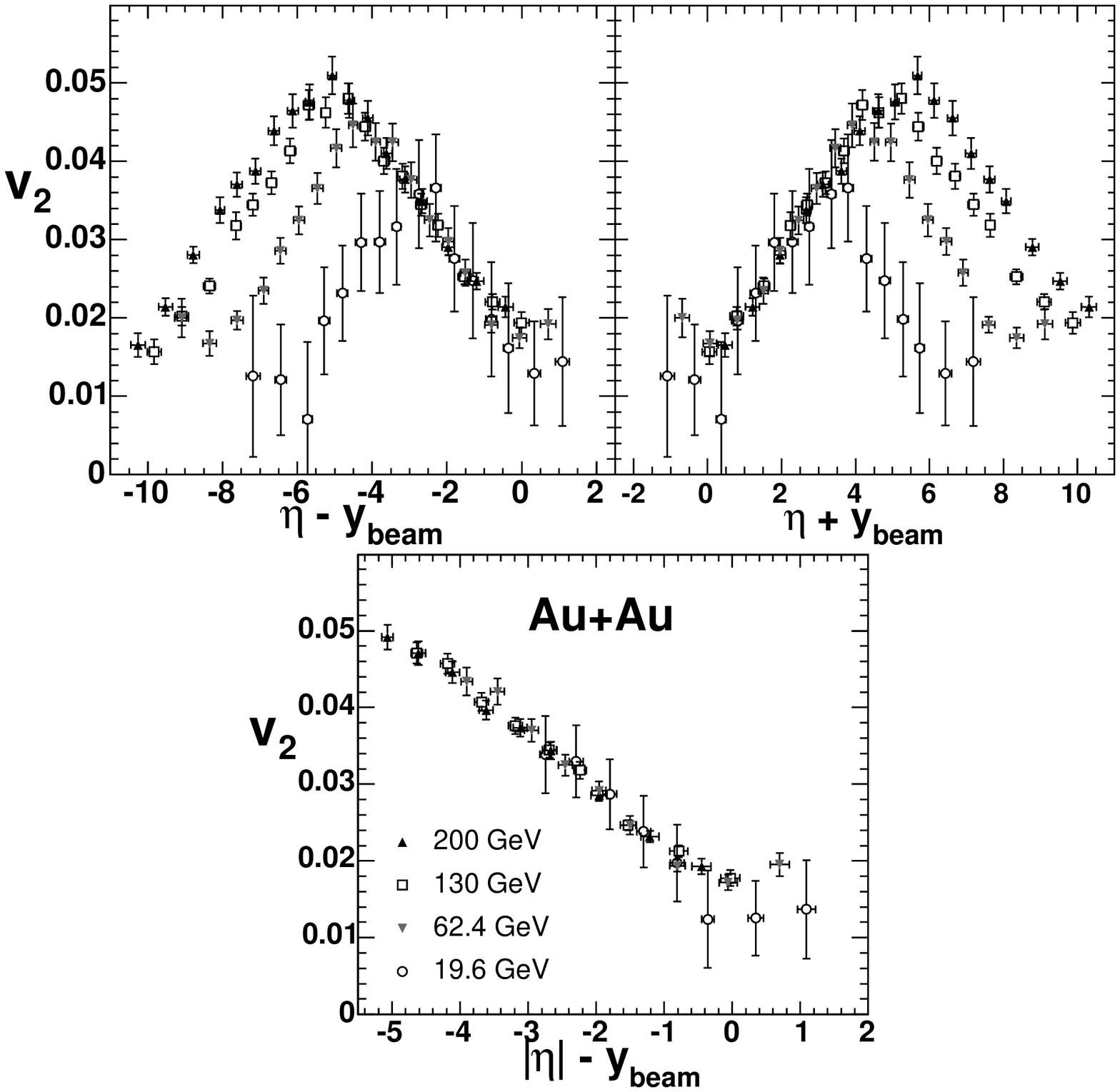}

\caption{ \label{WP26_v2_etaPMybeam_AA_4sqrts}
The flow data of Fig.~\ref{WP25_v2_eta_AA_4sqrts} are shown in the top left (top right) panel
versus the variable ${\eta^{\prime}}=\eta-y_{beam}$
(${\eta^{\prime}}=\eta+y_{beam}$) \cite{Bac04g}.  In the bottom panel, data at positive and
negative pseudorapidity have been averaged to give $v_2$ as a function of
$|\eta|$.  These results were then plotted versus the variable
${\eta^{\prime}}=|\eta|-y_{beam}$.  As for the particle densities shown in
Fig.~\ref{WP24_dNdeta_etaPMybeam_AA_19_130_200}, the flow data at all energies follow a common curve.  In
the case of flow, this curve holds over the entire range from beam or target
to midrapidity.}

\end{center}
\end{figure}

In addition to the pseudorapidity distributions of yields of produced particles,
longitudinal scaling can also be seen in the elliptic flow of particles produced in heavy ion collisions. As discussed in Section~\ref{Sec3}, the elliptic flow parameter, $v_2$,
provides a sensitive probe of the properties in the early stages of the
collision, one of which is the presence or absence of boost-invariance. Boost
invariant ``initial conditions'' (i.e.\ right after the collision) should lead
to a boost-invariant $v_2(y)$.  Kinematic effects result in a difference
between $v_2(y)$ and $v_2(\eta)$, but the changes are small ($<$10\% at 200~GeV to
$<$20\% at 19.6~GeV)\cite{Kol01b,Bac04g}. The small magnitudes of these differences
mean that they do not affect the conclusions discussed here and that a
boost-invariant scenario (in rapidity) should also result in elliptic flow which
is approximately flat over a large region of pseudorapidity.  In
Fig.~\ref{WP25_v2_eta_AA_4sqrts}, the pseudorapidity dependence of the elliptic flow
parameter, $v_2$, is shown for semi-central Au+Au events at various energies \cite{Bac04g}.
Clearly, no boost invariant central plateau is seen.  Thus, there are no
indications of the existence of a broad boost-invariant central plateau in the
final particle distributions or in the state formed shortly after the collision,
as reflected by $v_2$.

In Fig.~\ref{WP26_v2_etaPMybeam_AA_4sqrts}, the elliptic flow data from Fig.~\ref{WP25_v2_eta_AA_4sqrts} are
replotted effectively in the rest frame of one of the gold nuclei.  Once again the
phenomenon of extended longitudinal scaling is revealed, this time for $v_2$ \cite{Bac04g}.  As
discussed above, there is a small modification of the shape if $v_2$ is
plotted versus rapidity instead of $\eta$ but this change does not significantly
impact the comparison of different energies.  There appears to be a single
universal curve governing the elliptic flow as a function of ${\eta^{\prime}}$
over a broad range down to midrapidity at each energy studied.  This extended longitudinal
scaling behavior of elliptic flow in Fig.~\ref{WP26_v2_etaPMybeam_AA_4sqrts} has further
implications since elliptic flow builds up early in the collision.  Therefore,
the dependence on the location in $\eta^{\prime}$ space must reflect
the conditions very shortly after the collision, and then these early
conditions lead to the measured elliptic flow.

\subsubsection{Longitudinal dependence: Lessons from HBT}
\label{Sec4-E-5}

Particle interferometry, in the form of Hanbury-Brown Twiss (HBT) correlations
\cite{Han54,Han56}, provides an extra, although much more indirect, test of
the ideas of boost-invariance in heavy ion collisions. Since pions are bosons,
they constructively interfere when they are near to each other in phase
space. Correlation measurements in momentum space can therefore reveal the
source size in position space. In particular, HBT correlations are sensitive to
the spatiotemporal distributions of particles at thermal freeze-out (i.e.\ the
point of the last elastic interactions).  
See \cite{Tom02} for a recent review.
Appendix~\ref{SecAppB-3} contains
more details including a description of the source parameterizations. Most
theoretical studies of HBT assume ideal (i.e. non-viscous) hydrodynamics and a boost-invariant source which exhibits
longitudinal Hubble flow ($z=v_zt$, where $z$ and $v_z$ are the longitudinal position and
velocity, respectively). These assumptions simplify the 
coupled differential equations 
and allow the
use of 2D transverse expansion overlaid on the boost-invariant longitudinal
expansion (a scenario often called 2+1D hydrodynamics).  While this
basic hydrodynamic picture was roughly successful in describing some aspects of
the elliptic flow (see Figs.~\ref{WP6_v2_Npart_AuAu_130_200} and \ref{WP7_v2_pT_AuAu_200}), these models have
failed to describe the HBT data from RHIC \cite{Ris96,Hir02,Sof02}.
 
\begin{figure}[ht] 
\begin{center}

\includegraphics[width=0.75\textwidth]{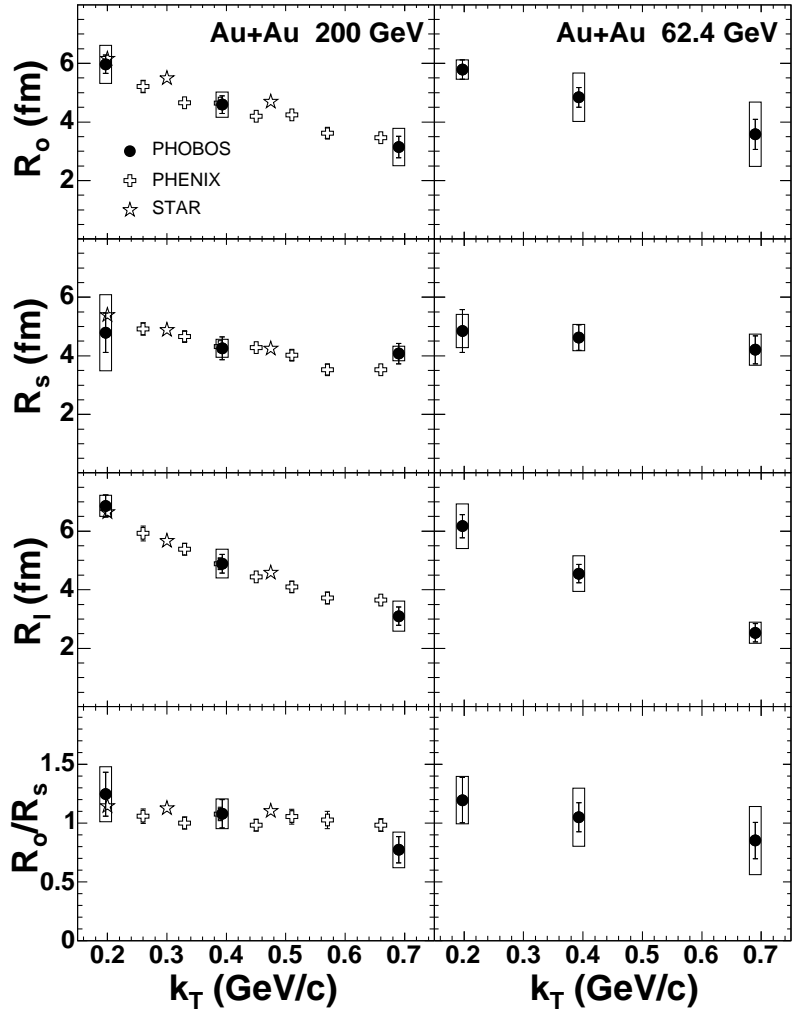}

\caption{ \label{WP27_HBT_RBP_kT_AA_2sqrts}
Bertsch-Pratt parameters $R_o$, $R_s$, and $R_l$, and the ratio $R_o/R_s$ for $\pi^-\pi^-$
pairs emitted in central collisions of Au+Au at $\sqrt{s_{_{NN}}}$ of 200~GeV
(left panels) and 62.4~GeV (right panels) as a function of pair transverse
momentum k$_{_T}$ \cite{Bac04k}. For comparison, data from STAR \cite{Ada04d} (open
stars) and PHENIX \cite{Adl04b} (open crosses) are presented at
$\sqrt{s_{_{NN}}}$=200~GeV.  PHOBOS systematic errors are shown as boxes;
systematic errors from STAR and PHENIX are not shown.}

\end{center}
\end{figure}

The influence of a possible new phase on HBT measurements has a long history \cite{Pra86}.
Under the assumptions of boost-invariant
hydrodynamics, the $R_o/R_s$ ratio should be large if a long-lived source is
formed and should typically be larger than $\sqrt{2}$ in any case.  
Figure~\ref{WP27_HBT_RBP_kT_AA_2sqrts} shows the results of fits using the Bertsch-Pratt
parameterization, along with the $R_o/R_s$ ratio from $\sqrt{s_{_{NN}}}$=62.4
and 200~GeV Au+Au collisions \cite{Bac04k} (see Appendix~\ref{SecAppB-3} for definitions). 
The data at 200~GeV are compared to the results of other RHIC experiments \cite{Ada04d,Adl04b}.
In contrast to expectations, the ratio of $R_o/R_s$ appears to be close to unity in heavy ion
collisions. Similar results were found in heavy ion collisions at lower
energies (see references in \cite{Bac04k}). The smallness of both $R_o/R_s$ and $R_l$ has come
to be known as the ``HBT puzzle''. It has been postulated that relaxing the
assumption of boost-invariance \cite{Bak03,Ren04}, or allowing non-zero
viscosity~\cite{Tea03}, may resolve this discrepancy.

\begin{figure}[t]
\begin{center}

\includegraphics[width=0.70\textwidth]{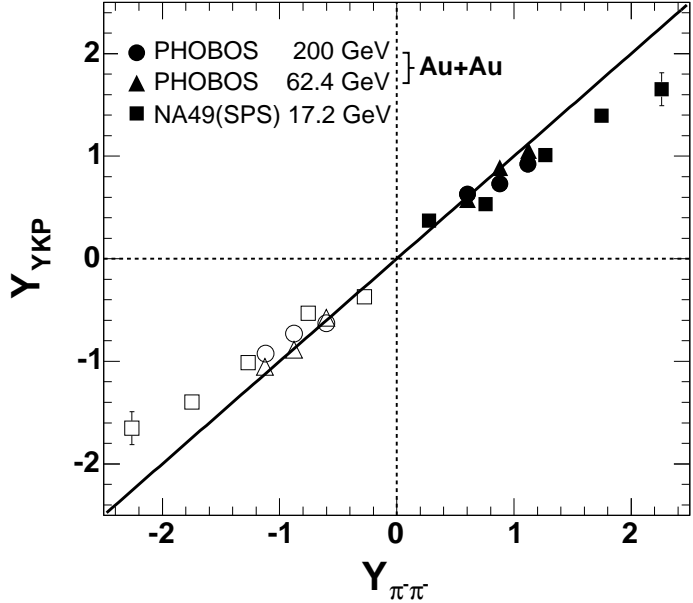}

\caption{ \label{WP28_HBT_YYKP_Ypi_AA_3sqrts}
The rapidity source parameters for $\pi^-\pi^-$ pairs emitted in central Au+Au
collisions at RHIC \cite{Bac04k} and Pb+Pb collisions at the SPS \cite{App98}.  This element of the
Yano-Koonin-Podgoretskii parameterization specifies
the rapidity (in the nucleus-nucleus center-of-mass system) of the source,
$Y_{YKP}$, from which the pions were emitted.  The abscissa of this plot is
the average rapidity of the pions themselves.  The filled symbols are the
measured data, the open symbols have been reflected about midrapidity.  The
line with a slope of 1 is drawn to guide the eye.}

\end{center}
\end{figure}

The detailed nature of the longitudinal properties of particle production can
also be explored by HBT measurements, in this case in a very direct way as
shown in Fig.~\ref{WP28_HBT_YYKP_Ypi_AA_3sqrts}. The data show the average rapidity of the
source of the pions (derived from the source velocity in the
Yano-Koonin-Podgoretskii parameterization) as a function of the rapidity of the
pions themselves \cite{Bac04k}.  A clear systematic trend is observed, and again the results
are very similar to what was found at the SPS \cite{App98}.  Under the simple assumption of
all pions being emitted from a single source located at the center of mass, the
ordinate of all points would be equal to zero.  If, instead, the system consisted
of a series of independent sources at different rapidities (i.e.\ a strong
longitudinal position-momentum correlation) the points would fall on the line.
The ``locality'' revealed by HBT studies of pion correlations in rapidity 
space suggests that the longitudinal distribution of particle properties is 
established very early, with the subsequent  evolution and freezeout having only 
short range correlations in rapidity.

\subsubsection{Extended longitudinal scaling: Summary}
\label{Sec4-E-6}

To summarize this section, the data demonstrate that extended longitudinal
scaling, reminiscent of ``limiting fragmentation'' over a broad region of 
longitudinal momentum, seems to be a dominant feature of particle production
for all colliding systems.  Based on all of the data, no evidence is seen in
any hadron-hadron or ion-ion collisions for two energy independent
fragmentation regions separated by a boost invariant central plateau which 
grows in extent with increasing collision energy.  The lack of a broad
boost-invariant central plateau is seen in both the final particle
distributions and in the state formed shortly after the collision as reflected
by $v_2$.  It is difficult to reconcile this with the common assumption that
particle production at midrapidity results from different physics than that in
the fragmentation region, particularly at the higher energies.  Furthermore, 
the similarity of the longitudinal scaling of both particle densities and
elliptic flow suggests the possibility of some direct connection between the
two, implying that the final particle multiplicities also result from the
properties of the very early evolution.

A good way to appreciate the significance of these results is to consider what
would be observed in the detectors if a collider could operate its two
beams at different energies.  For simplicity, the conventional RHIC
designation for the two counter-rotating beams, namely ``blue'' and
``yellow'', will be used.  If the energy of the blue beam was set to a
rapidity of 2, for example, the results show that, as the rapidity of the
yellow beam was increased up to a little beyond 2, the particle density and
elliptic flow seen in the detectors covering the blue beam hemisphere would
show a gradual increase and then reach a limiting value.  With the blue beam
fixed at a rapidity of 2, the particle density would not increase beyond this
limiting value on the blue beam side even if the yellow beam was set to infinite
rapidity.  The only way to further increase the particle density or elliptic
flow in the blue beam hemisphere would be to increase the energy of the blue
beam.

In p+p collisions, extended longitudinal scaling was understood to be a
consequence of $x_F$ scaling in string fragmentation (or, equivalently, in 
parton cascades).  No similar, widely accepted, explanation exists for the observation
of this behavior in the more complex p+A, d+A, and A+A collisions.

\subsection{Factorization of energy and centrality dependence} 
\label{Sec4-F}

The previous sections have described separately the dependencies of a variety
of observables on energy and centrality.  These independent discussions may
have obscured the remarkable extent to which these two dependencies factorize.
This section will describe several aspects of PHOBOS data which display this phenomenon.

\begin{figure}[t]
\begin{center}

\includegraphics[width=0.80\textwidth]{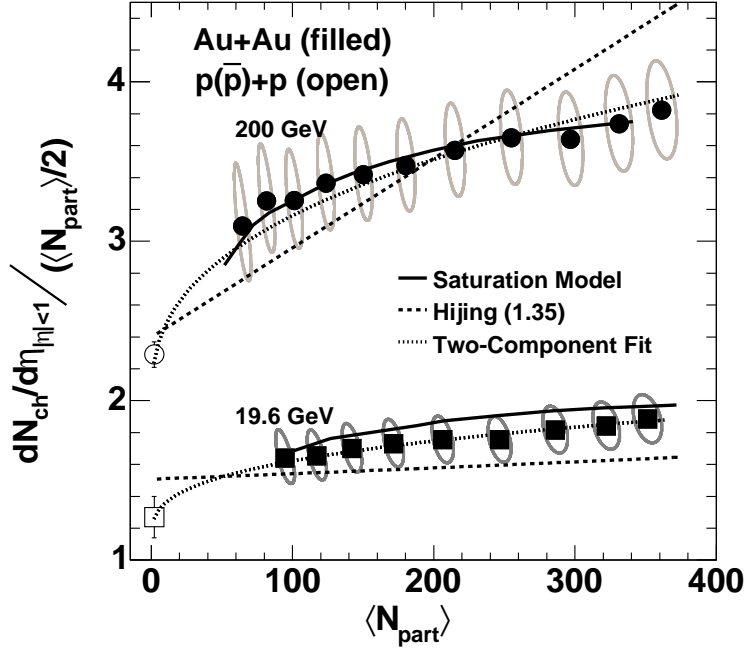}

\caption{ \label{WP29_dNdetaMidNpart_Npart_AA_19_200}
Pseudorapidity density of charged particles emitted near midrapidity divided
by the number of participant pairs as a function of the number of
participants. Data are shown for Au+Au at collision energies of 19.6 and 200~GeV \cite{Bac04e}.  
Data for p($\bar{p}$)+p \cite{Aln86,Ans89,Tho77} measured at 200~GeV and an interpolated value at
19.6~GeV are shown as open symbols.  
The grey ellipses show the 90\% C.L. systematic errors.
The results of two models \cite{Gyu94,Kha01a,Kha01c} and one parameterized fit \cite{Kha01b} are shown for comparison.}

\end{center}
\end{figure}

One simple example of factorization was revealed by the PHOBOS measurements of
the total charged particle multiplicity divided by the number of pairs of
participating nucleons in Au+Au collisions at three energies, from 19.6 to
200~GeV (see Fig.~\ref{WP12_Ntot_Npart_AuAu_19130200_dAu_pp}).  The data for the different
energies are separated by a factor that is constant as a function of centrality.
In other words, the centrality and energy dependence of the yield per
participant in Au+Au collisions factorize over the range of the two control
variables.  In this case, the factorization occurs trivially, as the total
charged particle yield per participant is centrality-independent at all
energies. Whether this factorization is a fundamental property of particle
production in Au+Au collisions can be tested by studying the yields per
participant more differentially in pseudorapidity and transverse momentum.

In Fig.~\ref{WP29_dNdetaMidNpart_Npart_AA_19_200}, the pseudorapidity density of charged particles per participant pair near
midrapidity is shown as a
function of centrality for collision energies of 19.6~GeV and 200~GeV \cite{Bac04e}. 
Data for $\bar{p}$+p collisions at 200~GeV and an interpolated value at 19.6~GeV are also plotted \cite{Aln86,Tho77,Ans89}.
Over
the centrality range shown here, the normalized yield at midrapidity increases
by approximately 25\% from mid-peripheral to central collisions.  Early theoretical
explanations attributed this increase to the contribution of the hard component
of particle production, which would grow with the relative increase in the
number of binary nucleon-nucleon collisions in more central events. As an
example of such a superposition of soft and hard particle production, the
results of a HIJING calculation \cite{Gyu94} are shown as dashed lines. The model shows an
increase in the yield per participant pair, although steeper than that seen in
the higher energy data.

\begin{figure}[ht]
\begin{center}

\includegraphics[width=0.80\textwidth]{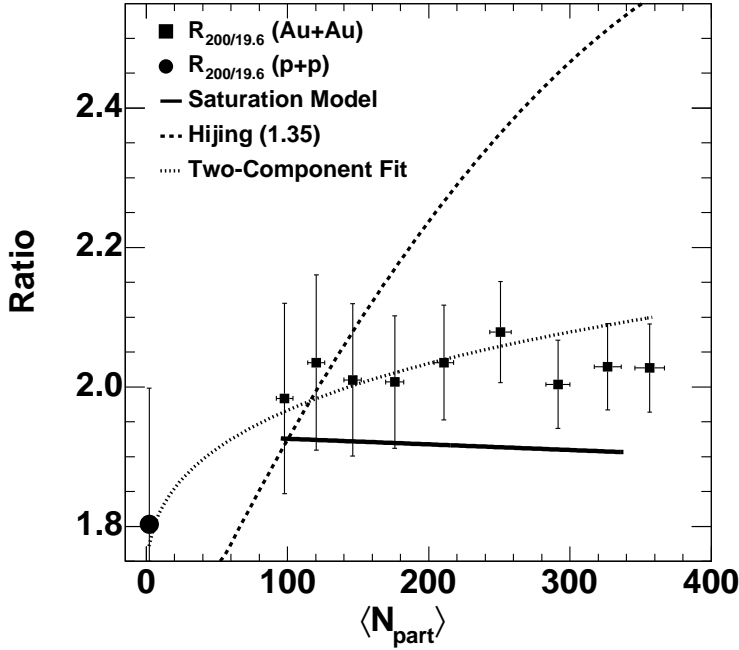}

\caption{ \label{WP30_dNdetaMidNpartRatio_Npart_AA_19_200}
Ratio of the pseudorapidity densities of charged particles emitted near
midrapidity for Au+Au at 200~GeV over 19.6~GeV as a function of the number of
participants \cite{Bac04e}.  The closed circle shows the ratio for collisions of protons.  The error bars include both statistical and 1-$\sigma$ systematic errors. 
The ratios for the same two models and one fit shown in
Fig.~\ref{WP29_dNdetaMidNpart_Npart_AA_19_200} are displayed for reference. }

\end{center}
\end{figure}

However, this explanation is challenged by the detailed study of the energy
dependence of midrapidity particle yields shown in Fig.~\ref{WP30_dNdetaMidNpartRatio_Npart_AA_19_200}, where
the centrality dependence of the ratio of the data for 200 over 19.6~GeV is
plotted \cite{Bac04e}. Within the experimental uncertainty, this ratio is independent of
centrality, whereas the contribution from hard processes would be expected to
show a large increase over this collision energy range. This is illustrated by
the HIJING prediction for this ratio (shown as a dashed line), which completely
fails to capture the factorization of energy and centrality dependence for the
midrapidity yield per participant.  A similar result was found earlier (over 
a smaller span in beam energy) using the centrality dependence of normalized 
midrapidity yields from Au+Au at $\sqrt{s_{_{NN}}}$=130~GeV 
\cite{Bac02a,Bac02c}.

The results of an attempt to investigate the interplay of hard and soft
scattering without invoking a complicated model are shown as dotted
lines.  In this case, a very simplistic two component fit \cite{Kha01b} was
performed to separately extract the fractions of the particle yield which
scaled with the number of participants (soft scattering) and the number of collisions (hard scattering).  A reasonably good fit
to the data is found but the fitted parameters suggest that, within the
uncertainties, there would be an identical contribution from hard scattering at both
beam energies, a result which is totally unexpected for minijet dominated physics.

Also shown in Figs.~\ref{WP29_dNdetaMidNpart_Npart_AA_19_200} and \ref{WP30_dNdetaMidNpartRatio_Npart_AA_19_200} is the result of a saturation model
calculation \cite{Kha01a,Kha01c}.  This model, which, as mentioned in
Section~\ref{Sec2-A}, yields a reasonably good match to the energy evolution of
particle yields at RHIC energies, also does a much better job of describing the
centrality evolution than the HIJING model.
     
\begin{figure}[ht] 
\begin{center}

\includegraphics[width=0.95\textwidth]{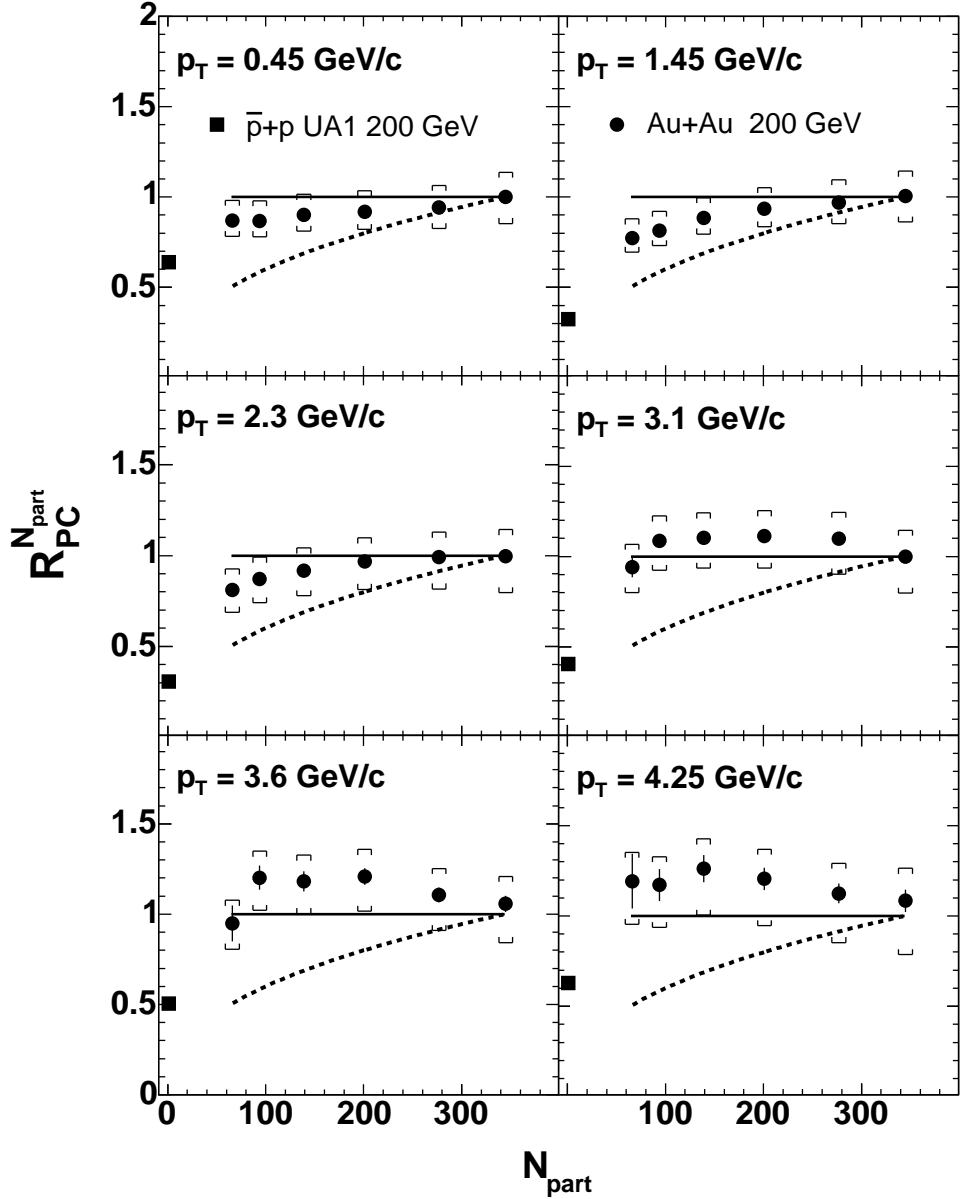}

\caption{ \label{WP31_RPCNpart_Npart_AA_6pT}
Particle yield normalized by the number of participant pairs and then divided
by a fit to the central data (see definitions in
Appendix~\ref{SecAppB-3}) as a function of centrality for Au+Au collisions at
$\sqrt{s_{_{NN}}}$=200~GeV, for six transverse momentum ranges \cite{Bac04a}.  Bars  and
brackets show statistical and systematic uncertainties, respectively.  The
solid (dashed) line shows the expectation for $N_{part}$ ($N_{coll}$) scaling
from peripheral to central collisions.  Squares show data for p+p collisions
from UA1 \cite{Alb90} with the same normalization factor.}

\end{center}
\end{figure}

Another example of non-trivial centrality dependence that is energy independent
was shown by the pseudorapidity distributions in Figs.~\ref{WP14_dNdetaNpart_eta_AuAu_19_200} and
\ref{WP24_dNdeta_etaPMybeam_AA_19_130_200}.  The former showed that the shape of the distributions differed
significantly as a function of centrality.  The latter demonstrated that the
distributions at different beam energies were found to line up when plotted in
the approximate rest frame of one of the incoming nuclei, i.e.\ using the
variable ${\eta^{\prime}}\equiv\eta-y_{beam}$.  Thus, the shape evolution with
centrality is independent of beam energy over a very broad range in
$\eta^{\prime}$.

Additional evidence for factorization is provided by the transverse momentum
distributions briefly mentioned in Sect.~\ref{Sec3}.  In the absence of
medium effects, one would expect that the volume scaling (i.e.\ proportionality
to $N_{part}$) observed for the bulk production of hadrons turns into scaling
with the number of binary collisions ($N_{coll}$) when measuring reaction
products of point-like hard processes.  This transition should be visible when
studying particle production as a function of transverse momentum.  However, as
is now known (see Fig.~\ref{WP8_RAA_pT_AuAu_64_200}), particle production at large transverse
momenta seems to be significantly modified in the presence of the medium in heavy ion
collisions.  The strength of this modification is more clearly illustrated in
Fig.~\ref{WP31_RPCNpart_Npart_AA_6pT} which shows the nuclear modification factor for charged
hadrons in six bins of $p_{_T}$ as a function of $N_{part}$ \cite{Bac04a}.  In the figure,
yields at a given transverse momentum in collisions of varying centrality were
normalized by the number of participant pairs and then divided by a fit to the
same quantity in central data (see Appendix~\ref{SecAppB-3} for definition).  Data for p+p collisions from UA1
\cite{Alb90} are shown with the same normalization factor.  It is striking to
see that the medium modification results in charged particle yields that, over
the centrality range studied here, more closely scale with $N_{part}$ than with
the number of binary collisions, even for transverse momenta above 4~GeV/c.

\begin{figure}[ht]
\begin{center}

\hspace{-0.5cm}
\includegraphics[width=1.02\textwidth]{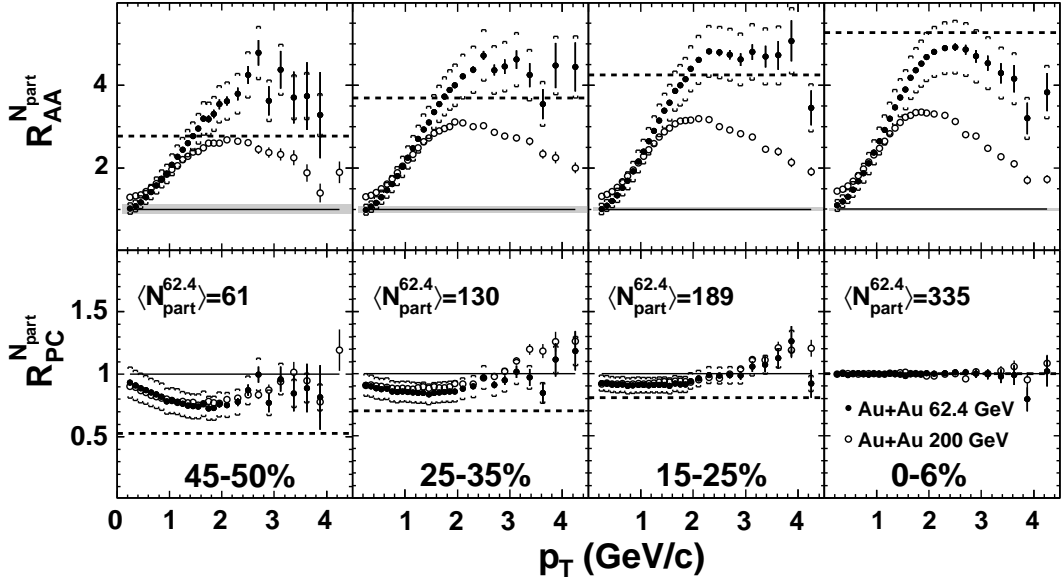}

\caption{ \label{WP32_RPCNPart_RAANpart_pT_AA_8panel}
Nuclear modification factors versus transverse momentum for Au+Au at two beam
energies and a variety of centralities \cite{Bac04d} calculated using two different reference
distributions: (top row) $N_{part}$/2 times p+p yields \cite{Bre95,Dri82,Alb90}, or (bottom row) the ratio of
$N_{part}$ times a fit to the distribution for central Au+Au.  Filled
symbols are for $\sqrt{s_{_{NN}}}$=62.4~GeV, open symbols are for 200~GeV.
Bars and brackets show
statistical and systematic uncertainties, respectively.  
The grey bands in the top row show the systematic error in the overall scale due to $N_{part}$.  
Centralities are labeled by the fraction of total inelastic cross section in
each bin, with smaller numbers being more central and the number of
participants at the lower energy are indicated.  
The solid (dashed) line shows the expectation for $N_{part}$ ($N_{coll}$) scaling
(See discussion in Appendix~\ref{SecAppB-3}).  
Note the small variations with centrality in
both the magnitude and shape of the ratios calculated using $N_{part}$ and also
that $R$-factors normalized using central Au+Au data (bottom row) are identical at
the two beam energies.}

\end{center}
\end{figure}

The observation of $N_{part}$ scaling at high transverse momentum suggests
that the medium is almost completely ``black'' or ``absorbing'' to 
produced fast particles.  This conclusion follows if one assumes $N_{coll}$
scaling of the primary production throughout the entire volume of the
collision zone followed by complete absorption except on the surface.  The
volume to surface ratio (proportional to the nuclear radius $R$ or equivalently $A^{1/3}$) has a centrality dependence that is
similar to the dependence for the ratio of the number of collisions to the number of
participants.  However, since the centrality dependence of particle production
is seen to be very similar at all transverse momenta, it is also possible that
the usual simplistic assumption of participant dominance at low $p_{_T}$
evolving into collision dominance at higher values needs to be reconsidered.

Data from the most recent RHIC run have been used to study the evolution of
the transverse momentum distributions as a function of both collision centrality and
energy.  The measurements were performed near midrapidity at
collision energies of 62.4~GeV and 200~GeV \cite{Bac04d}.  In Fig.~\ref{WP32_RPCNPart_RAANpart_pT_AA_8panel},
particle production as a function of centrality and $p_{_T}$ is shown for
these two energies in terms of $R_{AA}^{N_{part}}$ and $R_{PC}^{N_{part}}$ (Ref.~\cite{Bac04d} shows additional centrality bins).  As defined in Appendix~\ref{SecAppB-3},
$R_{AA}^{N_{part}}$ shows the variation in the yield per participant pair
relative to p+p collisions \cite{Alb90,Bre95,Dri82} (upper row of Fig.~\ref{WP32_RPCNPart_RAANpart_pT_AA_8panel}) and
$R_{PC}^{N_{part}}$ shows the variation in yield per participant pair relative
to central Au+Au collisions (bottom row).

As discussed earlier, the range in $p_{_T}$ from a few hundred~MeV/c to more than 4
GeV/c is assumed to cover very different regimes of particle production, from
soft coherent processes to independent binary scattering. Over the collision energy range
from 62.4 to 200~GeV, overall particle production in p+p increases by less than
a factor 2, whereas the yield at $p_{_T}$=4~GeV/c increases by an order of
magnitude.  This clearly shows the change in the balance of lower and higher
transverse momenta particles, which presumably reflects the different energy
dependencies of soft and hard particle production in p+p collisions over this
energy range. For central Au+Au collisions however, the ratio of the yields
between 200~GeV and 62.4~GeV at $p_{_T}$=4~GeV/c is only about
 4 (with a factor of 1.6 increase in the $p_{_T}$-integrated
multiplicity), i.e.\ the huge increase in the yield of high $p_{_T}$ particles
in p+p is not reflected in Au+Au.

The top row of Fig.~\ref{WP32_RPCNPart_RAANpart_pT_AA_8panel} clearly demonstrates that the overall shape and
magnitude of $R_{AA}^{N_{part}}$ depend strongly on beam energy
and, to a lesser extent, also on centrality.  In particular, at both energies
the yield per participant at any given $p_{_T}$ changes by less than 25\% over
the centrality range from 60 to 340 participants, with an even smaller
variation at the highest $p_{_T}$.  Even more surprisingly, the comparison in
terms of $R_{PC}^{N_{part}}$ in the bottom row of the figure shows that the
remaining variation of the yield per participant pair is the same for both
energies over the full $p_{_T}$ and centrality range. This means that the
energy and centrality dependences of particle production also factorize 
over this entire range in energy, centrality, and $p_{_T}$. This is particularly striking, as the
factorization therefore covers both the bulk particle production at low
$p_{_T}$, as well as rare particle production at intermediate and high
$p_{_T}$, believed to be governed by different particle production
mechanisms. In particular, at intermediate $p_{_T}$ above 1~GeV, particle
production is thought to be influenced by the effects of radial hydrodynamic
flow, the $p_{_T}$ broadening due to initial and final state multiple
scattering (``Cronin effect''), the balance between `soft' and `hard' particle production,
parton recombination and fragmentation, and the in-medium energy loss of fast 
partons. All of these contributions to the overall particle yields are expected
to show distinctly different centrality and energy dependencies at different
$p_{_T}$, yet the overall result is a factorization of energy and centrality
dependence at all $p_{_T}$ within the experimental uncertainty.  

The observed factorization in the energy and centrality dependencies of transverse
momentum spectra,
combined with similar observations for total and midrapidity yields as well as the rapidity distributions, strongly
suggests that the data reflect the dominant influence of yet-to-be-explained
overall global constraints in the particle production mechanism in A+A
collisions.

\section{Conclusion} 
\label{Sec5}

PHOBOS data and results from the other RHIC experiments, combined with very
general arguments which are either model independent or depend on fairly simple
model assumptions, lead to a number of significant conclusions.

In central Au+Au collisions at RHIC energies, a very high energy density medium
is formed.  Conservative estimates of the energy density at the time of first
thermalization yield a number in excess of 3~GeV/fm$^3$, and the actual density could be significantly larger.  This is far greater than hadronic densities and
so it is inappropriate to describe such a medium in terms of simple hadronic
degrees of freedom.  Unlike the weakly interacting QGP expected
by a large part of the community before RHIC turn-on, the constituents of the
produced medium were found to experience a significant level of
interactions.  If this medium is a new form of QCD matter,
as one would expect from lattice gauge calculations for such a high energy
density system, the transition to the new state does not appear to produce any
signs of discontinuities in any of the observables that have been studied.  To
the precision of the measurements, all quantities evolve smoothly with energy,
centrality, and rapidity.  
Although it does not provide strong evidence against other possibilities, this 
feature of the data is consistent with the results of recent lattice QCD calculations  
which suggest that the transition from this novel high energy density
medium to a hadronic gas is a crossover.

An equally interesting result was the discovery that much of the data can be
expressed in terms of simple scaling behaviors.  In particular, the data
clearly demonstrate that proportionality to the number of participating nucleons, $N_{part}$, is a
key concept which describes much of the phenomenology.  Further, the total
particle yields per participant from different systems are close to identical when compared at
the same available energy; the longitudinal velocity dependences of elliptic flow and
particle yield are energy independent over a very broad range, when effectively viewed in the rest frame of one of the
colliding nuclei; and many
characteristics of the produced particles factorize to a surprising degree into
separate dependences on centrality and beam energy.  

All of these observations
point to the importance of the geometry of the initial state and the very early
evolution of the colliding system in determining many of the properties of the
final observables.  Future data at RHIC, most especially collisions of lighter
nuclei, as well as higher energy nucleus-nucleus data from the LHC, will help
to further evaluate the range of validity of these scaling behaviors.  It is
possible that models which describe the initial state in terms of parton saturation will
play a role in explaining some or all of these scaling properties, but such an
identification is not yet clear.  What is clear is that these simple scaling
features will constitute an integral component or essential test of models which attempt
to describe the heavy ion collision data at  ultrarelativistic energies.
These unifying features may, in fact, provide some of the most significant inputs
to aid the understanding of QCD matter in the region of the phase diagram where a
very high energy density medium is created.

\begin{ack}

The PHOBOS collaboration would like to express our gratitude to the RHIC
management and construction personnel and to the operations staff of the BNL
Collider-Accelerator Department for their hard work and truly spectacular
success.  Without the broad range of colliding species and energies, and
continually increasing luminosities, provided by RHIC over a very short span of
time, the PHOBOS physics program would not have been nearly so rich.
We would like to express special thanks to 
S.Basilev,
B.D.Bates,
R.Baum,
A.Bia\l as,
M.Ceglia,
Y.-H.Chang,
A.E.Chen,
T.Coghen,
C.Conner,
J.Corbo,
W.Czy\.{z},
B.D\c{a}browski,
M.Despet,
P.Fita,
J.Fitch,
M.Friedl,
K.Ga\l uszka,
R.Ganz,
\linebreak
J.Godlewski,
C.Gomes,
E.Griesmayer,
J.Halik,
P.Haridas,
A.Hayes,
D.Hicks,
W.Kita,
J.Kotu\l a,
H.Kraner,
C.Law,
M.Lemler,
J.Ligocki,
J.Micha\l owski,
\linebreak
J.M\"ulmenst\"adt,
M.Neal,
M.Nguyen,
A.Noell,
M.Patel,
M.Plesko,
M.Rafelski,
M.Rbeiz,
D.Ross,
J.Ryan,
A.Sanzgiri,
J.Scaduto,
J.Shea,
J.Sinacore,
\linebreak
S.G.Steadman,
M.Stodulski,
Z.Stopa,
A.Straczek,
M.Strek,
K.Surowiecka,
\linebreak
R.Teng,
B.Wadsworth,
K.Zalewski, and
P.\.{Z}ychowski without whose efforts the PHOBOS experiment would not have 
been possible. 
We would like to thank \mbox{Krishna} Rajagopal for many valuable suggestions and
enlightening discussions.
We acknowledge the generous support of the Collider-Accelerator Department
(including RHIC project personnel) and Chemistry Department at BNL.  We
thank Fermilab and CERN for help in silicon detector assembly.  We thank the
MIT School of Science and LNS for financial support.   
This work was partially supported by U.S. DOE grants 
DE-AC02-98CH10886,
DE-FG02-93ER40802, 
DE-FC02-94ER40818,  
DE-FG02-94ER40865, 
DE-FG02-99ER41099, and
W-31-109-ENG-38, by U.S. 
NSF grants 9603486, 
0072204,            
and 0245011,        
by Polish KBN grant 1-P03B-062-27(2004-2007), and
by NSC of Taiwan Contract NSC 89-2112-M-008-024.

\end{ack}

\vspace{1.5in}

\appendix

\section{The PHOBOS detector}
\label{SecAppA}

\begin{figure}[ht]
\begin{center}

\includegraphics[width=0.95\textwidth]{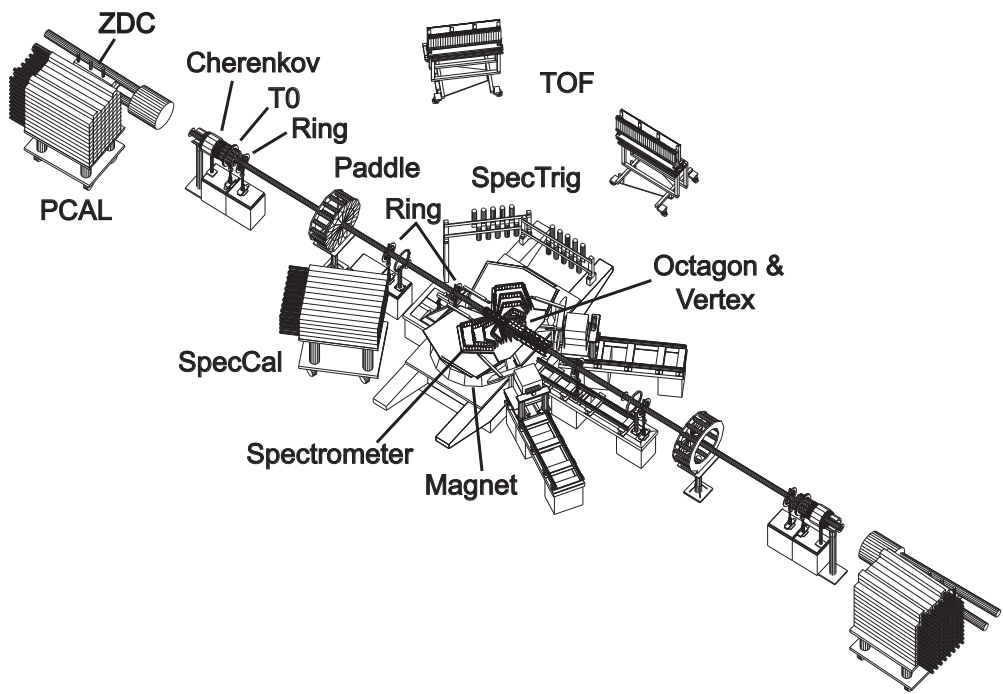}

\caption{ \label{WP33_PhobosDet_big}
The layout of the PHOBOS detector during the RHIC run in early 2004.  The
beams collide at a point just to the right of the double-dipole magnet, the
top of which is not shown.  The PCAL and ZDC calorimeters 
are drawn to scale but are located about 3 times farther from the interaction
point than shown.}

\end{center}
\end{figure}

\begin{figure}[ht] 
\begin{center}

\includegraphics[width=0.75\textwidth]{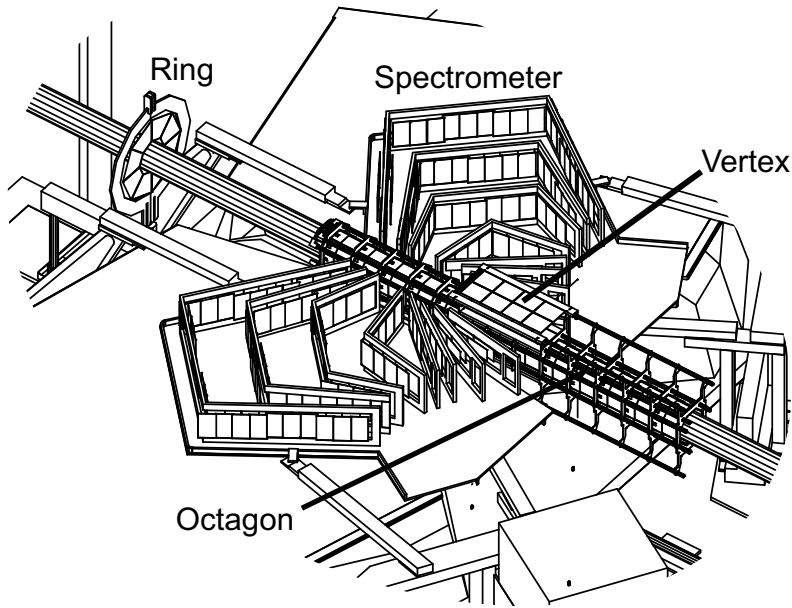}

\caption{ \label{WP34_PhobosDet_center}
The elements of the PHOBOS detector in the vicinity of the beam collision
point.}

\end{center}
\end{figure}

The PHOBOS experimental setup is composed of three major sub-systems: a charged
particle multiplicity detector covering almost the entire solid angle, a two
arm magnetic spectrometer with particle identification capability, and a suite
of detectors used for triggering and centrality determination.  More details
can be found in \cite{Bac03b}.  The active elements of the multiplicity
detector and tracking detectors in the spectrometer are constructed entirely of
highly segmented Si wafers with individual readout of the energy deposited in
each pad \cite{Bac00b,Bac99a,Bac98}.  The layout of the experiment during the 2004 run
is shown in Fig.~\ref{WP33_PhobosDet_big}.  An enlarged view of the region around the
beam collision point is shown in Fig.~\ref{WP34_PhobosDet_center}.  Table~\ref{TabApp11}
lists the colliding systems, center-of-mass energies, and data samples
collected by PHOBOS during the first four RHIC runs.
 
\begin{table}
\begin{center}
\begin{tabular}{|c|c|c|c|c|c|c|} \hline
RHIC& Colliding & $\sqrt{s_{_{NN}}}$ & Beam & Dates of PHOBOS & Total \\
Run & System & & Rapidity & Data Taking & Events (M) \\ \hline
 & Au+Au &   55.87~GeV &4.094& 6/13/00--6/16/00 & 1.8*\\ 
\raisebox{1.5ex}[0pt]{1}& Au+Au &  130.4~GeV & 4.942 & 8/15/00--9/4/00 & 4.3\\ 
\hline
 & Au+Au &  130.4~GeV &4.942& 7/8/01 & 0.044\\ 
 & Au+Au & 200.0~GeV &5.370& 7/20/01--11/24/01 & 34\\ 
\raisebox{1.5ex}[0pt]{2}& Au+Au &  19.59~GeV &3.044& 11/25/01--11/26/01 & 0.76*\\
 & p+p  &   200.0~GeV &5.362& 12/28/01--1/25/02 & 23\\ \hline
 & d+Au &  200.7~GeV &5.370& 1/6/03--3/23/03 & 146\\ 
\raisebox{1.5ex}[0pt]{3}& p+p  &  200.0~GeV &5.362& 4/13/03--5/24/03 & 50\\ \hline
 & Au+Au  & 200.0~GeV &5.370& 1/5/04--3/24/04 & 215\\ 
4 & Au+Au &  62.40~GeV &4.205& 3/24/04--4/2/04 & 22\\ 
 & p+p  &  200.0~GeV &5.362& 4/18/04--5/14/04 & 28\\  \hline

\end{tabular}
\caption{
\label{TabApp11}
Summary of data collected by PHOBOS during the first four RHIC runs.  Note that event totals
given in the last column represent the number summed over the entire variety of
triggering conditions, including minimum-bias events, interactions occurring in
a restricted
range of the collision vertex, collisions selected to be more central or more
peripheral, and collisions satisfying the high-$p_{_T}$ spectrometer
trigger. Note that triggers for the Au+Au runs at 19.6 and 56~GeV (marked with *) had very
loose requirements on timing with the result that only a relatively small
fraction of the events were usable in the currently published analysis. 
}
\end{center}
\end{table}

The Si pad detectors used to measure multiplicity consist of a single layer
covering almost the entire $4\pi$ solid angle.  These detectors
measure the total number of charged particles emitted in the collisions, as well
as detailed information about their distribution in azimuthal and polar angle
(or equivalently pseudorapidity, $\eta$).  The Si modules are mounted onto a
centrally located octagonal frame (Octagon) covering $|\eta|\leq 3.2$, as
well as three annular frames (Rings) on either side of the collision vertex, extending the coverage out to $|\eta|\leq 5.4$.

The Si modules forming the arms of the spectrometer are mounted on  eight frames.
Depending on the trajectory, charged particles traverse between 13 and 16
layers of Si as they pass through the spectrometer.  The first layer is only 
10~cm from the nominal interaction vertex. 
The magnet pole tips are arranged to produce almost no magnetic
field in the vicinity of the first  six layers.  The field then rises rapidly to a
roughly constant value of $\sim$2 Tesla for the remaining layers.  The Si wafers are
finely segmented to provide 3-dimensional space points used in the track
finding.  The solid angle covered depends on the vertex location along the beam
direction and extends over about 3/4 of a unit of $\eta$ for any given vertex location, with a total coverage of roughly $0<\eta<2$
.  Each arm covers
approximately 0.1 radians in azimuth for particles that traverse all of the
layers.  The momentum resolution is close to 1\% for particles with momenta near 0.5~GeV/c and rises about
1\% for each additional 3~GeV/c.  

Particle identification is provided using two
techniques.  Charged particle energy loss is measured in each Si layer.  Combining this
information with the momentum from the tracking can separate pions from kaons
out to about 700~MeV/c and pions from protons out to about 1.2~GeV/c.  Additional
particle identification is provided by two Time-of-Flight (TOF) walls, each
consisting of 120 plastic scintillator slats.  Before the start of the 2003
RHIC run, these walls were moved farther from the interaction point, extending
particle identification capability out to momenta roughly 2-3 times that achievable
using energy loss in the silicon detector.  In their new locations, the TOF walls cover
roughly half the azimuthal acceptance of the spectrometer.  

Before the 2004
run, a small hadronic calorimeter (SpecCal) was installed behind one of the
spectrometer arms.  Consisting of 50 lead/scintillator modules, each 10~cm
square by about 120~cm long, this detector can be used to measure the energy of
high momentum particles traversing part of the spectrometer acceptance.

The primary event trigger for all colliding systems was provided by two sets
of 16 plastic scintillator slats (Paddles) covering $3.2<|\eta| <4.5$.  Imposing an upper limit on the time difference between the signals in the two arrays eliminated
most beam-gas interactions and provided a rough selection of collision vertex
locations along the beam line.  To enhance the data sample of useful events, a
more precise measure of vertex location was generated using two arrays of 10
\v{C}erenkov counters (T0s).  This was necessary because the range of vertex
positions for which the multiplicity and tracking detectors have reasonable acceptance is considerably
shorter than that created by the overlap of the colliding beam bunches.  For different colliding
systems, the T0 detectors could be moved to different locations along the beam
line in order to optimize the efficiency of the vertex determination while
minimizing the number of events with multiple particles traversing a single counter.  A more precise
vertex location is found off-line using signals from the Vertex detector,
which is composed of two sets of two layers each of Si modules.  With high
segmentation along the beam direction, correlating hits in the inner and outer
layer can be used to determine the vertex along the beam line to an accuracy
of better than 0.4 mm.  This detector also determines the height of the beam 
but with limited resolution.  The vertical position and horizontal position 
perpendicular to the beam can be found using tracks from the spectrometer.

Colliding systems such as p+p or
d+Au, which produce smaller numbers of particles, have fewer events with tracks traversing the spectrometer.  The
spectrometer trigger uses an additional array of scintillator slats (SpecTrig)
mounted between the tracking detectors and the TOF walls.  Coincidences
between the SpecTrig and TOF hit slats, combined with the vertex location from
T0, were used online to select events containing a high momentum track in the
acceptance of both the spectrometer and the TOF.

The Zero-Degree-Calorimeters (ZDC) have a cross-sectional area of 10$\times$12~cm$^2$
centered on the direction of the beam and are located about 18 m from the
nominal interaction point.  Particles hitting these detectors must first
traverse the initial RHIC accelerator magnet which separates the two
counter-circulating beams.  Therefore, the ZDC signal results almost
exclusively from spectator neutrons which are not bound in
nuclear fragments and whose transverse momentum remains close to zero after
the interaction.  Due to the response time of this detector, partly resulting
from its long distance from the collision point, it was not possible to use
ZDC signals in the primary event trigger for the bulk of the physics data.
However, this device was used on-line in special runs to check triggering
efficiency for the other detectors and also off-line in studies of centrality
determination.

Similar to the ZDC, the Proton Calorimeters (PCAL) are located behind the
first accelerator magnets, but in this case next to the outer edge of one of the beam pipes.  The
magnets bend spectator protons to an angle of more than twice that of the beam
particles so these protons will exit the beam pipe and shower in the PCAL.  As
with the ZDC, only individual protons, as opposed to those bound in clusters,
can be detected.  The PCAL is particularly useful for studies of d+Au
collisions.  On the side of the outgoing deuteron, the combination of PCAL
and ZDC signals can be used to divide the event sample into p+Au, n+Au,
and d+Au subsets, i.e.\ events in which only one or both of the incoming
nucleons interacted.  On the side of the outgoing Au nucleus, the PCAL is
primarily sensitive to protons knocked out of the Au, which is a measure of the
total number of collisions suffered by the interacting nucleons in the
deuteron.

\section{Definitions of terms}
\label{SecAppB}

In this section, detailed definitions are given for the important event and particle
characterization parameters, as well as a number of the critical observables
used in the physics analysis.

\subsection{Event characterization}
\label{SecAppB-1}

In interpreting data from heavy ion collisions, the primary event
characterization parameters are the energy of the collision and the overlap of
the two nuclei at the moment when they interact, commonly referred to as
centrality.  In order to compare fixed target, colliding beam, symmetric, and
asymmetric systems all on a common footing, the collision energy is defined
using the center-of-mass energy available when a single nucleon from one
projectile collides with a single nucleon from the other projectile, ignoring Fermi motion.  The
standard notation for this quantity is $\sqrt{s_{_{NN}}}$, referred to as the
nucleon-nucleon center-of-mass energy.  For symmetric colliding beams, each of which has
the same energy per nucleon, $\sqrt{s_{_{NN}}}$ is simply twice that value
and the nucleon-nucleon frame is also the lab frame. When colliding deuterons 
and gold at RHIC, both beams were run at the same relativistic $\gamma$ (and 
therefore the same rapidity) as the gold beams in the 200~GeV Au+Au collisions.
The mass difference caused by the binding energy is responsible for the fact 
that the d+Au collisions are slightly asymmetric in the lab frame.  The 
deuteron has a total energy of 100.7~GeV/nucleon, only 0.7\% larger than the gold 
beam value of 100.0~GeV/nucleon. Consequently, the nucleon-nucleon frame does 
not coincide with the lab frame, but the shift in rapidity is only +0.004
units.  For collisions of p+p, in contrast, the relativistic $\gamma$ (and hence the rapidity) were
adjusted in order to compensate for the small mass
difference and, thereby, to achieve the same $\sqrt{s_{_{NN}}}$ of
200~GeV as for the highest energy Au+Au collisions.  At
RHIC, data have been taken for a wide range of $\sqrt{s_{_{NN}}}$ (see
Table~\ref{TabApp11}) ranging from a value close to the maximum achieved at the
SPS up to a value more than 10 times larger.

\begin{figure}[ht]
\begin{center}

\includegraphics[width=0.95\textwidth]{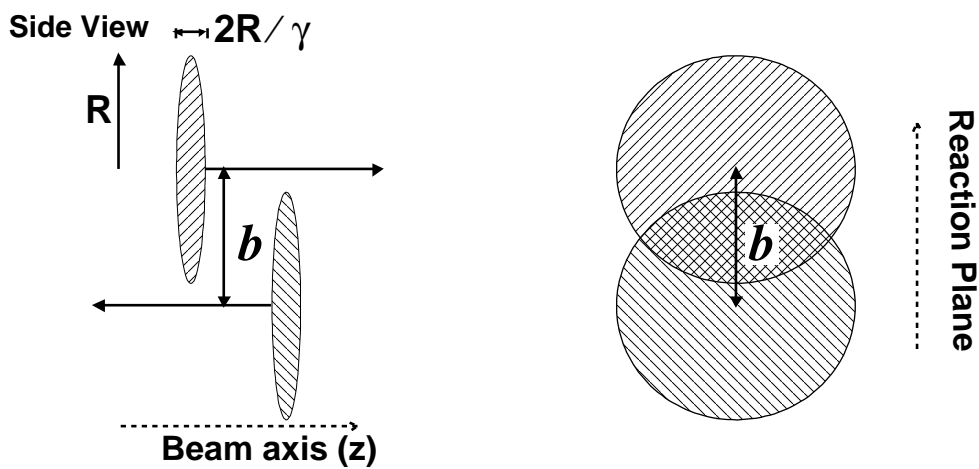}

\caption{ \label{WP35_collide_diagram}
(Left panel) A side view in the nucleon-nucleon center-of-mass frame of two relativistic heavy ions colliding.  (Right
panel) A view along the beam axis, where the cross-hatched almond-like
overlap region is indicated.  The reaction plane for a particular collision is
the plane defined by the impact parameter, $b$, and the beam axis (z).}

\end{center}
\end{figure}

A direct measure of the collision geometry is given by the impact parameter,
$b$, which is the transverse distance between the centers of the colliding
heavy ions.  It is defined such that $b=0$ for central collisions, see
Fig.~\ref{WP35_collide_diagram}.

In most physics analyses of heavy ion collision data at highly relativistic energies,
the impact parameter is not considered particularly useful in characterizing
the important influence of geometry on the outcome of a given interaction.
Instead, two parameters which quantify the critical distinctions are used:
namely the number of participating nucleons, $N_{part}$, and the number of
binary nucleon-nucleon collisions, $N_{coll}$.  In defining these variables,
two important assumptions are made.  First, since the collision duration at
such high energies is very short compared to the typical time-scale for nuclear
rearrangement or movement of nucleons within the nucleus, it is assumed that
only the nucleons in the overlap region (the cross-hatched area in the right
panel of Fig.~\ref{WP35_collide_diagram}) experience any substantial interactions (i.e.\
participate) in the collision.  Second, the collisions suffered by a given nucleon as it
traverses the other nucleus may not be distinct sequential events, and thus it may be most meaningful to simply count
the total number of collisions.  For observables such as elliptic flow
which are sensitive to the shape of the initial overlap region, a third
parameter, namely the spatial asymmetry of this region derived from the impact parameter and
the radii of the colliding nuclei, can be used. 
 
\begin{figure}[ht]
\begin{center}

\includegraphics[width=0.65\textwidth]{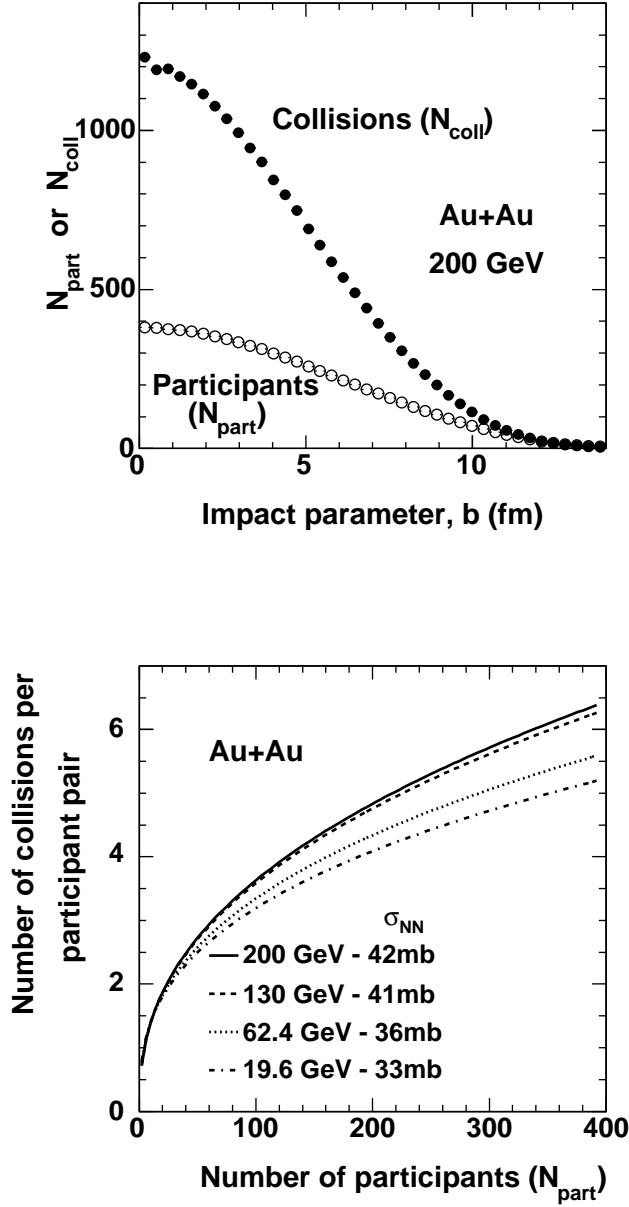}

\caption{ \label{WP36_Ncoll_Npart_b_Npart}
(Top panel) $N_{part}$ and $N_{coll}$ vs. impact parameter, $b$, for
Au+Au collisions at $\sqrt{s_{_{NN}}}$=200~GeV.  (Bottom panel) 
The average number of collisions, $N_{coll}$, divided by the average number of
participant pairs versus $N_{part}$ for Au+Au at a variety of beam energies.
See text for discussion.}

\end{center}
\end{figure}

In determining the number of participating nucleons, or equivalently the number
of nucleons which interact, only those which are struck by nucleons from the
other nucleus (as opposed to ones which were hit only in secondary scatterings)
are counted.  This is the same quantity as ``wounded nucleons'' introduced by
Bia\l as, Bleszy\'{n}ski and Czy\.{z} \cite{Bia76}.  
In some publications, the notation $N_{wound}$ is used for what is herein 
referred to as $N_{part}$ and the notation $N_{part}$ includes nucleons 
suffering secondary scatterings.  When comparing PHOBOS data with results from
other experiments, care has been taken to use the appropriate values.
$N_{part}$ depends on the collision
geometry and is typically calculated using a Glauber model of the collision.
The key ingredients in this calculation are (1) nucleons are distributed
according to a nucleon density function (e.g. Woods-Saxon), (2) nucleons in each
nucleus travel in straight lines through the colliding system, and (3)
nucleons interact according to the inelastic cross section, $\sigma_{_{NN}}$,
as measured in proton-proton collisions.  For the energies at RHIC, the values
assumed for $\sigma_{_{NN}}$ were 33, 36, 41, and 42~mb for
$\sqrt{s_{_{NN}}}$=19.6, 62.4, 130, and 200~GeV, respectively.  
In all cases, the nucleons were assumed to be hard spheres distributed
according to a Wood-Saxon functional form of 
\[ P(R)=R^2\left( 1+e^{\frac{(R-r_0)}{a}}\right)^{-1}, \]
where $r_0$=6.38~fm and $a$=0.535~fm for all energies.
The open
circles in the top panel of Fig.~\ref{WP36_Ncoll_Npart_b_Npart} show an example of the results of such a
model calculation relating $N_{part}$ and impact parameter for Au+Au collisions
at one of the RHIC energies.  The number of participants is usually assumed to
have a strong influence on the bulk properties of particle production but it is
shown in the physics sections of this paper that $N_{part}$ (or $N_{part}/2$)
provides a convenient benchmark to study the effects of the collision geometry
on many measured experimental quantities.

As introduced above, $N_{coll}$ denotes the number of binary nucleon-nucleon
collisions in a heavy ion reaction.  As in the calculation of $N_{part}$, only
primary collisions, i.e.\ those occurring along the straight-line trajectory of
nucleons through the opposing nucleus, are counted.  This quantity can also be
calculated in a Glauber model, with typical results being shown as closed
circles in the top panel of Fig.~\ref{WP36_Ncoll_Npart_b_Npart}.  The yield from hard scattering
(i.e.\ large momentum transfer) processes is expected to scale as $N_{coll}$.
For symmetric A+A collisions, simple geometrical arguments imply that
$N_{coll}$ would scale as roughly $A^{4/3}$.  Thus, for collisions of more than
two participants, the number of binary nucleon-nucleon collisions is larger than
the number of participants, with the difference increasing dramatically for
smaller impact parameters.

One possibly important aspect of centrality in heavy ion collisions which goes
beyond the simple increase in the number of participants or collisions is shown
in the bottom panel of Fig.~\ref{WP36_Ncoll_Npart_b_Npart}.  There, the number of collisions
is divided by the number of participating pairs to derive the average number of
collisions suffered by each participant.  A similar parameter, typically
denoted $\bar{\nu}$ and calculated from $\bar{\nu}=(A\sigma_{pp})/ \sigma_{pA}$ where the
$\sigma$'s are inelastic cross sections, is commonly used to characterize
centrality or target dependences of observables in p+A collisions \cite{Bus77}.
In nucleus-nucleus collisions, the calculated average number of collisions per
participant varies by a large factor as a function of centrality and also has
some dependence on energy due to the varying nucleon-nucleon cross section.

\subsection{Particle characterization}
\label{SecAppB-2}

In describing the trajectories of particles emitted in heavy ion collisions, a
distinction is typically made between longitudinal (i.e.\ along the beam direction) and
transverse motion.  The former may reflect some remnant of the original motion
of the beam while the latter is largely generated in the interaction.  The
physics variable typically associated with the longitudinal motion is rapidity,
denoted $y$ and defined as
$y=\frac{1}{2} \ln((E+p_{_\parallel})/(E-p_{_\parallel})) = \ln((E+p_{_\parallel})/m_{_T})$ with
$E$ and $p_{_\parallel}$ being the total energy and the component of the
particle's momentum along the beam, respectively, and $m_{_T}$ being the
transverse mass defined below.  Rapidity has the important property of being additive in
Lorentz transformations from one reference frame to another which differ by
velocity along the beam.  Thus, the shape of the distribution of any quantity
plotted versus rapidity is the same in any such frame.  Unfortunately, it is
frequently difficult to experimentally determine the particle identification,
or in some cases even the momentum itself, necessary to calculate rapidity.  In
such instances, it is common to replace rapidity with pseudorapidity, denoted
$\eta$ and defined as $\eta= -\ln(\tan(\theta/2))$, where $\theta$ is the polar
angle to the beam axis.  For particles whose total momentum is large compared
to their mass, i.e.\ for particles whose velocity is close to the speed of
light ($\beta=v/c\approx$1), the two measures are close to identical, except for polar angles very close to zero.  Since the
produced particles are typically dominated by pions whose transverse momentum
alone averages a few hundred MeV/c or more, the use of pseudorapidity is a
quite reasonable approximation.  A variable frequently used in elementary
collisions is the Feynman $x_{_F}$ variable given by the ratio of the momentum along
the beam to the maximum possible value, $x_{_F}=p_{_{\|}}/p_{_{\|}max}$.

Another aspect of the distributions as a function of longitudinal velocity that
proves to be very interesting is the comparison of distributions at a variety
of beam energies but viewed in the rest frame of one of the projectile
particles.  For distributions as a function of rapidity, this can be done
exactly and trivially by simply subtracting the rapidity of the beam from the
rapidity of each particle.  In the case of pseudorapidity distributions, the
transformation is not exact but a reasonably close approximation is found using
the shifted pseudorapidity, denoted $\eta^{\prime}$ and defined as $\eta^{\prime} =
\eta - y_{beam}$, where $\eta$ is the pseudorapidity of a particle and
$y_{beam}$ is the beam rapidity.  The quantity $y_{beam}$, which is given by
$\frac{1}{2}\ln((E+p)/(E-p))=\ln((E+p)/M)$ with $E$, $p$, and $M$ being the energy,
momentum, and mass of the beam, respectively, is 
given in Table~\ref{TabApp11} for the various colliding systems and energies.
Fermi motion of 300~MeV/c would spread the nucleons out 
by typically $\approx$0.3~units in rapidity.

The transverse motion is most often characterized using simply the component of
the momentum, denoted $p_{_T}$, that is perpendicular to the beam axis.
Occasionally, the so-called transverse mass, $m_{_T}=\sqrt{p_{_T}^2 + m_0^2}$,
is used where $m_0$ is the rest mass of the particle.  The use of this more
complicated variable is motivated by its appearance as the natural scaling
parameter for particles emitted by a thermal source.  It can also be used to
combine energy, transverse momentum, and rapidity of a particle via the 
identity $E=m_{_T}\cosh(y)$.

The various particle characterization variables can be related using the
following identities:
\[ p_{_{\|}}=m_{_T}\sinh(y)=p_{_T}\sinh(\eta).\]
For relativistic beam energies, $p_{_{\|}}\approx \left(\sqrt{s}/2\right)x_{_F}$ 
and for $y$ larger than about 1--2, $\sinh(y)\approx \left(e^y\right)/2$ so that: 
\[ \eta^{\prime}\equiv\eta - y_{beam}\approx \ln(x_{_F})-\ln\left (\frac{p_{_T}}{M}\right)\] 
\[y^{\prime}\equiv y-y_{beam}\approx\ln(x_{_F})-\ln\left (\frac{m_{_T}}{M}\right), \]
where $M$ is the nucleon mass.

In the case of jets emitted in e$^+$+e$^-$ annihilation, the motions of
individual particles along and transverse to the beam are not the most interesting
quantities.  Instead, distributions are characterized by the trajectories of
particles relative to the jet direction, the so-called thrust axis.  Since data
exist most frequently in the form of unidentified charged particles, the motion
along the thrust axis is traditionally defined using $y_{_T}$, the rapidity
calculated using the momentum parallel to the jet direction and assuming the
pion mass.  The required shift to compare different beam energies in a common
frame, as was done for $y^{\prime}$ or $\eta^{\prime}$, is not intuitively obvious.
In this paper, the somewhat arbitrary choice was made to replace $y_{beam}$ in
the formulas above with $y_{jet}$ which is the rapidity calculated using the
center-of-mass energy combined with the assumption of the proton mass.
Therefore, the same shift was used in both e$^+$+e$^-$ and p+p at the same
$\sqrt{s}$.

\subsection{Notation for observables}
\label{SecAppB-3}

The most basic observable characterizing particle production is the total
number of particles emitted.  Two experimental hurdles complicate the
extraction of this number from the data.  The first is that only charged
particles are easily detected.  Although assumptions can be made concerning the
ratio of charged and neutral particles, the multiplicity data is almost always
presented in terms of the number of charged particles.  Adjustments for the
number of unobserved neutrals is typically only done when needed in a specific
calculation, for example in the discussion of the energy density presented in
Sect.~\ref{Sec2-A}.  The notation $N_{ch}$(A+B) is used to denote the total
charged particle yield, integrated over all solid angle, in collisions of
species A with species B.  To date, PHOBOS has measured 
$N_{ch}$(d+Au) at a variety of centralities for one center-of-mass energy and
$N_{ch}$(Au+Au) over a broad range of both centrality and beam energy.  Note
that in all cases the multiplicity is defined to be ``primary'', i.e.\ those
particles emitted in the initial interaction.  Corrections are applied to the
data to remove all other ``secondary'' particles, which are created in weak or electromagnetic decays
of primary particles and interactions of primary particles with material in the
detector.  The second complication in extracting total numbers is that no
detector can be fully hermetic, i.e.\ capable of detecting every single
particle emitted.  As a result, it is always necessary to measure distributions
of particles and extrapolate into the unmeasured regions.  

Because the PHOBOS multiplicity detector measures only the emission angle of
charged particles, the extracted distribution is the number of charged
particles per unit pseudorapidity, denoted $dN_{ch}/d\eta$.  The experimental
layout is designed to minimize the amount of material between the collision
vertex and the active elements and, therefore, the cut-off in transverse momentum
is low and the losses of particles with low $p_{_T}$ are small.  The correction
for secondary particles which are added to the total by decays or interactions
in the material is typically much larger than the correction for particles
that are lost.  In addition, the very broad coverage in $\eta$ provided by the
PHOBOS setup results in a relatively small extrapolation for particles emitted
at small angles with respect to the beam.  Thus, PHOBOS can provide information
about $dN_{ch}/d\eta$ and $N_{ch}$ which are unique at RHIC. As mentioned
above, it is also interesting to study particle distributions shifted into the
rest frame of one of the projectiles.  The shifted distribution,
$dN_{ch}/d{\eta^{\prime}}$, can be used as a measure of the charged particle
pseudorapidity density as effectively viewed in the rest frame of one of the colliding
nuclei, although one should keep in mind that such a shift is, in principle, associated with a small distortion of the distributions.

As discussed in the main body of this paper, the particle density is
highest near $y$ or $\eta$ of zero and, therefore, it is generally assumed that
the potential for creation of any new state of matter is also highest in that
region.  As a result, the properties of observables ``near midrapidity'' are
of particular interest.  For the midrapidity multiplicity distribution, the
range chosen is $\pm$1 unit in $\eta$ so the pseudorapidity distribution is
averaged over this range to generate $dN_{ch}/d\eta\rfloor_{|\eta|\le1}$.

In cases where the momentum and angle of the particles are measured,
distributions in both transverse momentum and rapidity (or pseudorapidity in
cases without particle identification) can be generated.  The transverse
distributions are commonly presented in a form which is Lorentz invariant
given by $Ed^3\sigma/dp^3$, with $E$ and $p$ being the total energy and vector
momentum of the particle, respectively.  Since the interesting quantity is
typically the number of particles in a given event, i.e.\ the distribution that
integrates to give multiplicity, this is more commonly expressed as invariant
yield $Ed^3N/dp^3$.  When integrating over all orientations of the reaction
plane, azimuthal symmetry can be assumed and the differential momentum volume
can be expressed in cylindrical coordinates as $dp^3\rightarrow2\pi p_{_T} dp_{_T}
dp_{_\parallel}$.  Furthermore, the component of the momentum parallel to the
beam can be transformed using $dp_{_\parallel}=Edy$ where $y$ is the rapidity,
resulting in the final form $d^2N/2\pi p_{_T}dp_{_T}dy$.  When using transverse
mass, the transformation is trivial since $p_{_T}dp_{_T}= m_{_T}dm_{_T}$ and
only the horizontal axis changes in the distributions.  In cases without
particle identification, rapidity is approximated by pseudorapidity, yielding
$d^2N/2\pi p_{_T}dp_{_T}d\eta$.

When comparing transverse momentum distributions for more complicated systems
to data from proton-proton collisions, one could simply take the ratio of the
two distributions as a function of $p_{_T}$ to study the change in magnitude or
shape.  This ratio is called the nuclear modification factor since it is a
measure of the modification of the properties of the emitted particles
resulting from the presence of the nucleus in the interaction.  In order to
test specific theories of how the yield should scale, the standard procedure is
to normalize the  A+A (or, equivalently, scale the p+p) data by some factor.  The resulting ratio comparing
collisions of species A with species B to p+p is typically denoted $R_{AB}$
defined as
\[R_{AB}=\frac{1}{Norm} \frac{dN_{A+B}/dp_{_T}}{dN_{p+p}/dp_{_T}}=\frac{1}{N_{coll}} \frac {dN_{A+B}/dp_{_T}}{dN_{p+p}/dp_{_T}}. \]
The most common normalization, and the
one usually indicated by the simple notations $R_{AA}$, $R_{dAu}$, etc.,\ is
$N_{coll}$ as shown in the rightmost formula above.  This arises from the interest
in studying the behavior of high transverse momentum particles and the belief
that the yield from such ``hard'' processes should scale with the number of binary
nucleon-nucleon collisions.  Analysis by the PHOBOS collaboration has
demonstrated that the number of pairs of participating nucleons is often the more appropriate
scaling variable.  To avoid confusion, ratios using this latter normalization
are denoted 
\[ R_{AB}^{N_{part}}= \frac{1}{N_{part}/2} \frac{dN_{A+B}/dp_{_T}}{dN_{p+p}/dp_{_T}}. \]
Note that a p+p collision has one pair of participants.  
This normalization will
be generically referred to as the number of participant pairs even in
asymmetric collisions.

It is frequently of interest to study the evolution of the shape and magnitude
of these distributions as a function of centrality for nucleus-nucleus
collisions.  The most direct display of this evolution involves dividing data
from one centrality bin by that from a different bin.  In this case,
both distributions need to be suitably normalized.  The notation PC (CP) is
used for ratios where peripheral (central) data is divided by central
(peripheral).  The PHOBOS collaboration has recently advocated the use of
$R_{PC}$ since different experiments have different reach in centrality and the
central data typically have significantly smaller statistical and systematic
errors.  In keeping with the convention described above, the definitions with
the different normalizations are
\[ R_{PC}=\frac{N_{coll}^{central}}{N_{coll}^{periph}}\frac{dN_{A+B}^{periph}/dp_{_T}}{dN_{A+B}^{central}/dp_{_T}}\] 
and 
\[R_{PC}^{N_{part}} = \frac{N_{part}^{central}}{N_{part}^{periph}}\frac{dN_{A+B}^{periph}/dp_{_T}}{dN_{A+B}^{central}/dp_{_T}}. \]
Note that the practical application of these definitions typically uses a fit to the
distribution that appears in the denominator in order to avoid propagating statistical
point-to-point fluctuations.

In the case of pure $N_{coll}$ scaling, $R_{AB}$ and $R_{PC}$ would be unity while
$R_{AB}^{N_{part}}$ and $R_{PC}^{N_{part}}$ would be unity for perfect $N_{part}$ 
scaling.  The variation of $R_{AB}$ for $N_{part}$ scaling (see, for example, 
Fig.~\ref{WP8_RAA_pT_AuAu_64_200}) or the variation of $R_{PC}^{N_{part}}$ and 
$R_{AB}^{N_{part}}$ for $N_{coll}$ scaling (see 
Figs.~\ref{WP31_RPCNpart_Npart_AA_6pT} and \ref{WP32_RPCNPart_RAANpart_pT_AA_8panel})
depends on the ratio of $N_{coll}$ to $N_{part}$.  Careful examination of the numbers
in Tables~\ref{TabNpartNcoll62} and \ref{TabNpartNcoll200} in Appendix~\ref{SecAppC-1} 
reveals that, for a given centrality, this ratio depends slightly on beam energy.  
When comparing data at 62.4 and 200~GeV, the difference is never more than 15\%.  For
clarity, the dashed lines in Figs.~\ref{WP8_RAA_pT_AuAu_64_200} and 
\ref{WP32_RPCNPart_RAANpart_pT_AA_8panel} show only the value for the lower 
beam energy.

Using an event-by-event measurement of the orientation of the two colliding
nuclei, the study of particle distributions can be extended to include a third
coordinate, namely the azimuthal angle.  In relativistic heavy ion collisions,
the generic terms ``directed flow'' and ``elliptic flow'' are used for the
measurement of anisotropy in the azimuthal distributions of particles relative
to the reaction plane.  The reaction plane for a particular collision is the
plane defined by the impact parameter and the beam axis ($b$ and $z$ in
Fig.~\ref{WP35_collide_diagram}).  In flow analyses, the distribution of particles in the
azimuthal angle, $\phi$, (always taken relative to the reaction plane for a
particular collision) is measured and expressed in terms of a Fourier
expansion, $dN_{ch}/d\phi = N_0(1 +2v_1cos(\phi) + 2v_2cos(2\phi)+\ldots)$.
The amplitude of the first $\phi$-dependent term, $v_1$, is called directed
flow.  Elliptic flow is the name given to the amplitude of the second term of
the Fourier expansion, $v_2$.  This latter anisotropy in the form of a
variation in particle yield in momentum space results primarily from the
non-spherical shape in position space of the initial collision volume (see the
cross-hatched region in the right panel of Fig.~\ref{WP35_collide_diagram}).

Moving beyond single particle distributions, additional information can be
obtained by studying the correlations of particles.  In heavy ion collisions,
the most common multi-particle observable studied is the HBT correlation, named
for Hanbury-Brown and Twiss who pioneered an analogous technique for studying
the size of objects in astronomy \cite{Han54,Han56}.  The procedure is most often applied to pairs
of like-sign pions and depends on the quantum mechanical connection between
separation in coordinate and momentum space for identical particles.  The data
are presented as the ratio of the distribution of pairs in some
relative-momentum variable divided by a distribution which matches the correct
occupancy of the two-particle phase space but which does not contain the
effects of the two-particle correlation.  This normalization is obtained by
pairing particles found in different events which have been matched for
centrality and other event-characterization variables.  The resulting
correlation functions can be fit using a variety of parameterizations of the
source distributions.   From such parameterizations, information about the spatiotemporal extent
of the emission source can be extracted.  One commonly used system is the
so-called Bertsch-Pratt coordinates \cite{Pra86,Ber89,Hei96}.  For a given
pair of identical particles with average momentum $k$, the coordinates are:
longitudinal ($R_l$) along the beam direction ($z$), outwards ($R_o$) in the
($z$, $k$) plane perpendicular to $z$, and sidewards ($R_s$) perpendicular to the
other two directions.  The Yano-Koonin-Podgoretskii parameterization also
includes spatial parameters for the longitudinal and transverse sizes of the
source, as well as parameters describing the duration and longitudinal velocity
of the source \cite{Yan78,Pod83}.

\section {Techniques for determining centrality} 
\label{SecAppC}

\begin{figure}[ht] 
\begin{center}

\includegraphics[width=0.60\textwidth]{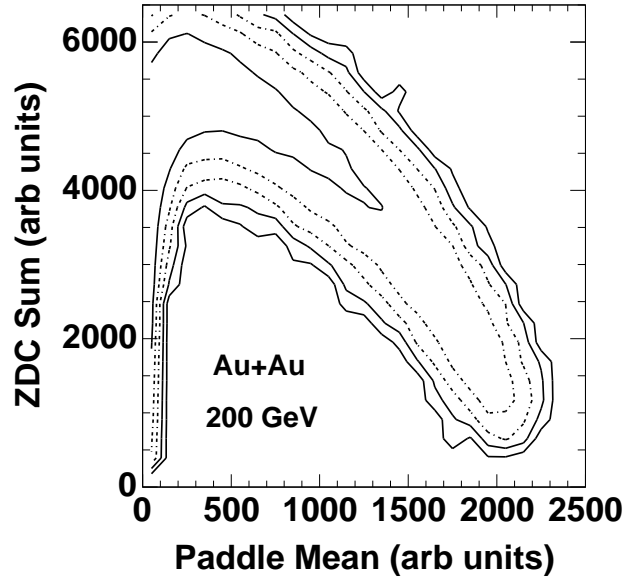}

\caption{ \label{WP37_ZDC_Paddle}
Correlation between spectator neutrons measured in the PHOBOS ZDC's (ZDC Sum)
and charged particle multiplicity measured in the Paddle counters (Paddle
Mean) for $\sqrt{s_{_{NN}}}$=200~GeV Au+Au collisions.  The contours are logarithmic
with a factor of 4 in yield between adjacent levels.}

\end{center}
\end{figure}

As briefly discussed in Appendix~\ref{SecAppB}, determining the centrality of a
heavy ion collision is extremely important for event characterization.  Knowing
the centrality provides a geometrical scale for use in any studies of the
underlying collision dynamics and affords the possibility of a more meaningful
comparison to ``baseline'' data from elementary proton or electron collisions.
The primary event centrality in PHOBOS is determined by utilization of signals
from the Paddle scintillator counters, as well as the Octagon and Ring silicon
detectors, all of which are sensitive to charged particle multiplicities in
various regions of pseudorapidity.  These signals, through bins in the percentage
of total cross section, provide a measure of centrality.  The validity of this
technique is based on the experimental observation of a strong correlation
between the charged particle multiplicity signals in, for example, the Paddle
scintillator counters and neutral beam ``remnants'' (spectator neutrons) as measured in
the Zero-Degree-Calorimeters (ZDCs).  This correlation is shown in
Fig.~\ref{WP37_ZDC_Paddle} for $\sqrt{s_{_{NN}}}$=200~GeV Au+Au collisions at
PHOBOS.

The specific methods developed within PHOBOS to determine centrality depend on
both the collision species (Au+Au versus d+Au) and the collision energy.
The technique is to associate an experimentally measured signal to a well-defined
centrality related variable, such as the number of participating nucleons,
$N_{part}$.  For this technique to be meaningful, a monotonic relation must
exist between the multiplicity signals in the chosen region of pseudorapidity and
$N_{part}$.  This assumption is justified by the experimental correlation shown
in Fig.~\ref{WP37_ZDC_Paddle} (the remnant neutrons are anti-correlated with
$N_{part}$ for the 50\% most central collisions).  Additional evidence for the validity of this technique has been
obtained using extensive Monte Carlo (MC) studies using event generators (such as
HIJING, AMPT, RQMD, and Venus) and a full GEANT simulation of the PHOBOS
detector.  An outline of some of these techniques follows.

\subsection{Centrality determination in Au+Au collisions} 
\label{SecAppC-1}

There are four main considerations that must be addressed in the course of
determining the event centrality: the event selection,
detection efficiency, choice of pseudorapidity region to utilize, and the event
generator simulations to extract $N_{part}$.

\begin{figure}[ht] 
\begin{center}

\includegraphics[width=0.75\textwidth]{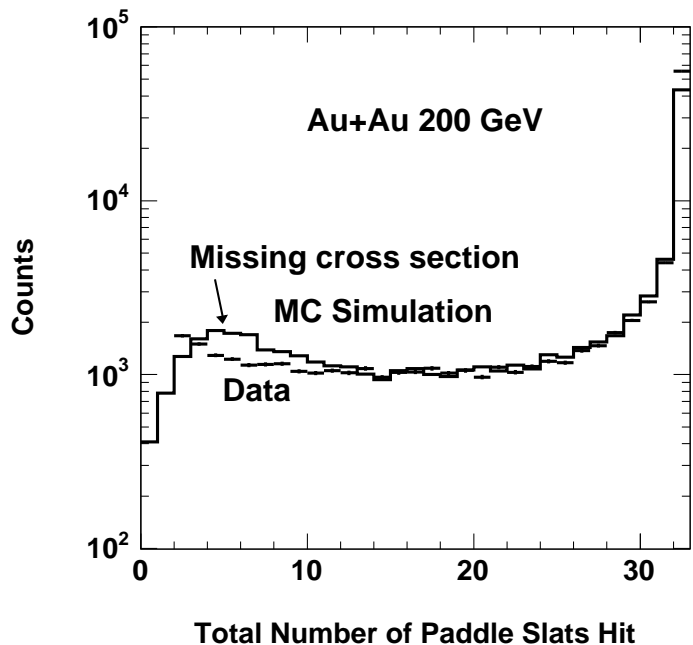}

\caption{ \label{WP38_TotPaddleHit_Data_MC}
Illustration of the detection efficiency determination in Au+Au collisions
using a comparison between Monte Carlo (MC) simulation and data for the number
of Paddle slats hit.  Data are shown for Au+Au collisions at $\sqrt{s_{_{NN}}}$=200~GeV. The
same technique was used for $\sqrt{s_{_{NN}}}$=62.4 and 130~GeV Au+Au collisions.}

\end{center}
\end{figure}

The initial event selection must cleanly identify and separate true Au+Au
collisions from numerous background sources, such as beam-gas interactions, while
simultaneously providing the smallest possible bias on the resulting data set.
In PHOBOS this was accomplished by using a combination of energy and time
signals from the Paddle counters and the ZDCs.  A selection of events with less than 4~ns 
time-difference between the two Paddle signals was combined with cuts on the ZDC
individual and summed timing signals.  Additional logic ensured no loss of very central
events that have a high Paddle signal and correspondingly few numbers of
spectator neutrons available to hit the ZDCs.  This selection provided a basic
``valid collision'' definition for Au+Au collisions at $\sqrt{s_{_{NN}}}$=62.4,
130 and 200~GeV.  For the lowest energy Au+Au collision of
$\sqrt{s_{_{NN}}}$=19.6~GeV, the ZDC timing requirement had to be modified due to the
substantially reduced efficiency for detection of the lower energy neutrons.

The detector efficiency was determined for two ``minimum-bias'' trigger
configurations of at least one (two) hits in each scintillator
Paddle counter array.  For both configurations, a
loss of peripheral events had to be accounted for before bins in percentage of total
cross section could be correctly fashioned.  The fraction of lost peripheral
events was determined using comparisons of the total number of Paddle slats hit
in both data and the full MC simulations (see Fig.~\ref{WP38_TotPaddleHit_Data_MC}).  This
analysis yielded a total detection efficiency of 97\% and 88\% for the two trigger
configurations, respectively, for Au+Au collisions at
$\sqrt{s_{_{NN}}}$=200~GeV.

\begin{figure}[t]
\begin{center}

\includegraphics[width=0.85\textwidth]{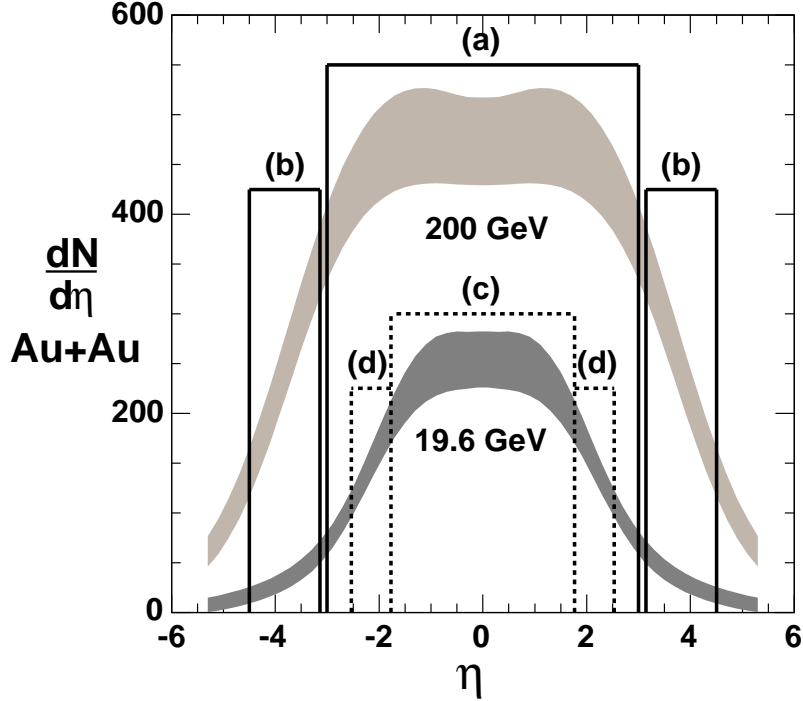}

\caption{ \label{WP39_dNdeta_eta_TrigRegions}
Pseudorapidity density distributions from $\sqrt{s_{_{NN}}}$=200 (light, top
band) and 19.6 (dark, bottom band)~GeV Au+Au collisions, for the most central
25\% of the cross section \cite{Bac03c}.  The boxed areas (a--d) illustrate the
separate regions in pseudorapidity used in the centrality
determination for each collision energy. Region (b) illustrates the
pseudorapidity coverage of the Paddle scintillator counters, and the other
regions were developed for centrality determination using the Octagon silicon
detector.}

\end{center}
\end{figure}

Using the collision event selection criteria outlined above and the
deduced trigger detection efficiency, the next task is to find an appropriate
experimental quantity for use in determining the event centrality.  For Au+Au
collisions, a consistent centrality determination was found to be relatively
independent of the choice of detector (and hence the pseudorapidity limits), as long as the chosen
region contained substantial particle multiplicity.  The signal from the Paddle counters, with a pseudorapidity coverage of
$3.2<|\eta|<4.5$ (region (b) of Fig.~\ref{WP39_dNdeta_eta_TrigRegions}), worked well as a
centrality measure for collision energies of $\sqrt{s_{_{NN}}}$=62.4, 130 and
200~GeV.

For the lowest energy of 19.6~GeV, new pseudorapidity regions had to be chosen due
to a reduction in the monotonicity between the multiplicity signals in the
Paddle counters and both the number of spectator neutrons seen in the ZDCs and
$N_{part}$, as determined from MC simulations.  In addition, the Paddles are
traversed by significantly fewer particles at 19.6 than 200~GeV (see dark grey
band in Fig.~\ref{WP39_dNdeta_eta_TrigRegions} or the bottom panel of Fig.~\ref{WP1_dNdeta_eta_AuAu_19_130_200}) and,
consequently, a different pseudorapidity region had to be chosen.   In order to create a
centrality measure at 19.6~GeV similar to that obtained from the Paddles at
200~GeV, the Paddle pseudorapidity range was scaled down to a smaller region by
the ratio of beam rapidities $y^{beam}_{19.6} / y^{beam}_{200}=0.563$ (region
(d) of Fig.~\ref{WP39_dNdeta_eta_TrigRegions}).  The resulting $\eta$ region,
$1.8\leq|\eta|\leq 2.5$, lies wholly within the Octagon silicon detector
coverage of $|\eta|\leq 3.2$ for collisions which occur within $\pm$10~cm
of the nominal vertex position.  Thus, the charged particle multiplicity measured in the region (d) was used as a centrality measure
 for 19.6~GeV and allowed for  a direct centrality comparison to the
original Paddle-based method at 200~GeV.  Additional centrality measures were
developed at both energies, in pseudorapidity regions close to midrapidity, which used the multiplicity
signals of charged particles traversing the Octagon in the pseudorapidity
regions (a) and (c), where region (c) is scaled by a factor of 0.563 compared to region (a).  This technique of matching centrality regions allowed direct
comparisons of midrapidity and away from midrapidity centrality
determinations across a factor of ten in collision energy.  Also, both
pseudorapidity regions have been found to have very different rates of particle
production and intrinsic dependences on $N_{_{part}}$.  By utilizing these two
independent regions, the assumption that the centrality measure and $N_{_{part}}$ have
to be only monotonic and not necessarily linear can be explicitly tested.  An
insignificant difference was found when analyzing data with both methods, at
both energies \cite{Bac04e}.

\begin{figure}[ht] 
\begin{center}
\includegraphics[width=0.95\textwidth]{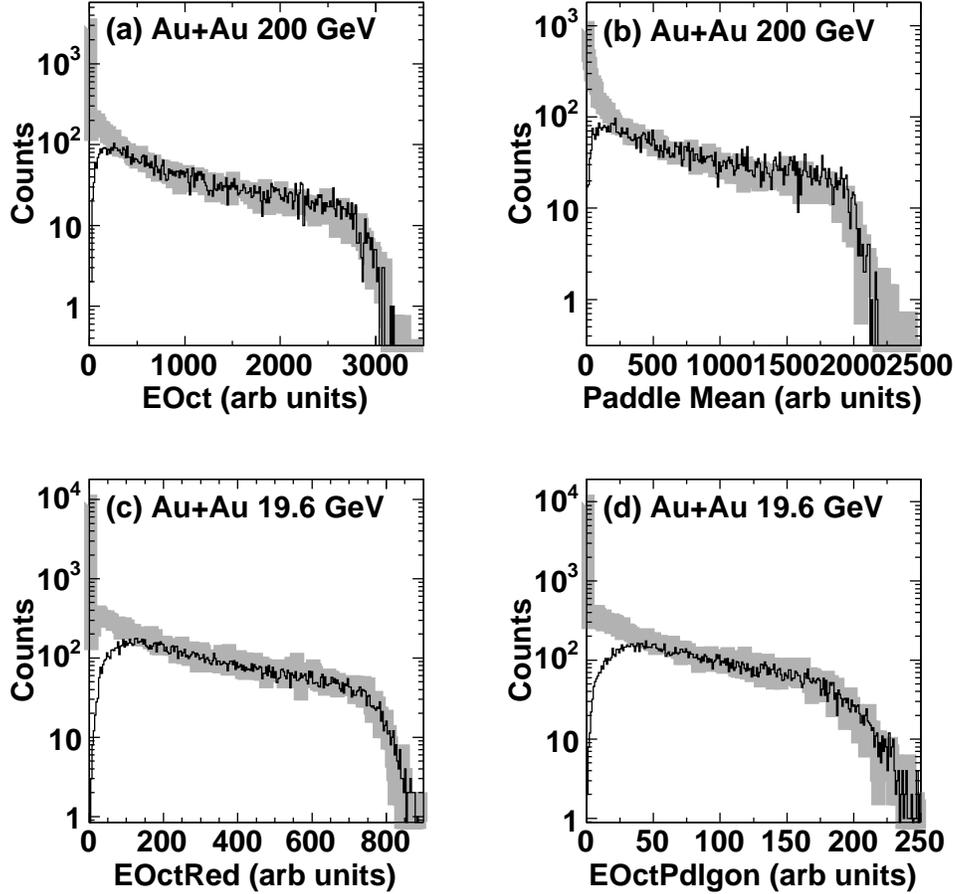}

\caption{ \label{WP40_Eoct_etc_Data_MC}
Charged particle multiplicity signal distributions measured in the four 
pseudorapidity regions (boxes labeled (a--d) in
Fig.~\ref{WP39_dNdeta_eta_TrigRegions}) used in the centrality determination.
Black histograms are data and the grey distributions are MC simulations for Au+Au at two energies.  All
data are shown for a restricted collision vertex of $| z |\leq 10$~cm,
and thus have an additional inefficiency for low multiplicities as evident
from the figures where the data falls below the (unbiased) MC simulations for
peripheral events.}

\end{center}
\end{figure}
 
\begin{figure}[ht] 
\begin{center}

\includegraphics[width=0.95\textwidth]{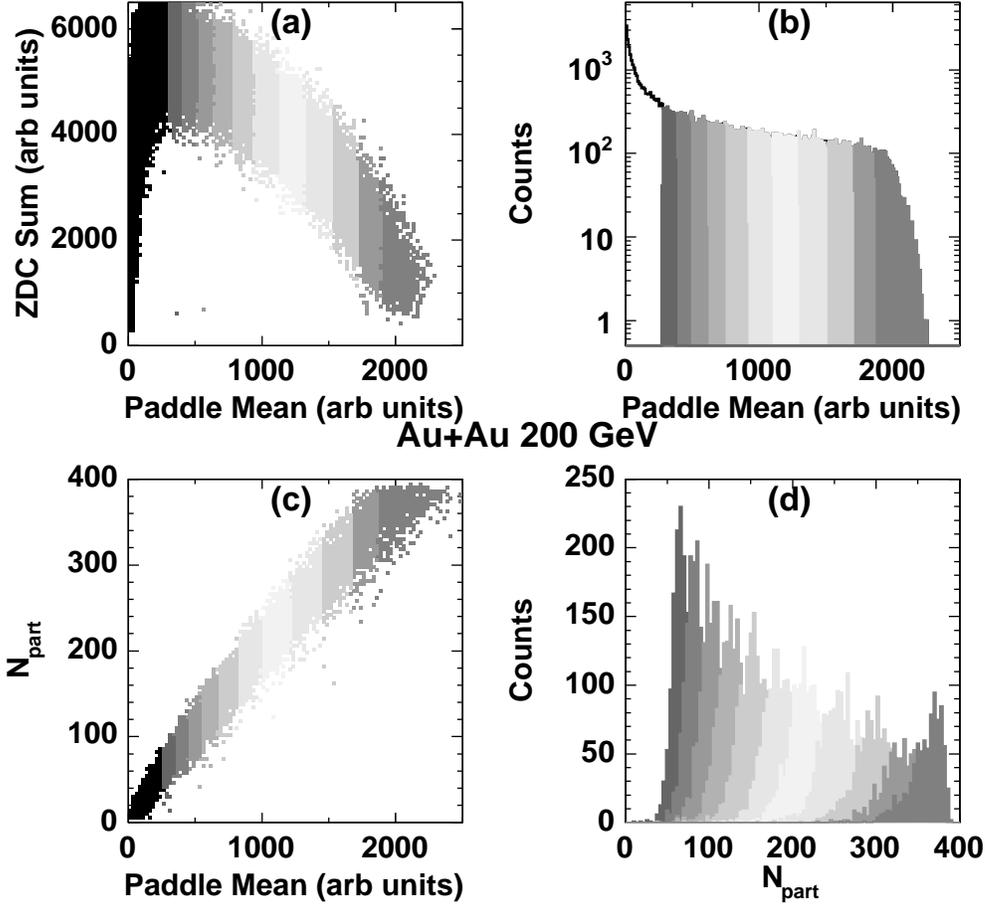}

\caption{ \label{WP41_Trigger_ZDC_Paddle_Npart}
Illustration of how the centrality of a heavy ion Au+Au collision is defined (results for $\sqrt{s_{_{NN}}}$=200~GeV are shown).
Only the top 50\% of cross section, where there is 100\% detection and vertex
reconstruction efficiency, is used. Panel (a) shows the experimental
correlation between the charged particle multiplicity signals in the Paddle
counters (Paddle Mean) and the signals in the ZDC's from spectator neutrons.
The shaded bands represent bins in percentile of cross section cut on the
Paddle Mean signal.  Panel (b) is a projection of (a) onto the Paddle Mean
axis.  Panel (c) shows a corresponding MC calculation where a monotonic
relation is observed between the Paddle Mean signal and $N_{part}$, the number
of participating nucleons.  From this correlation, the average $N_{part}$ (see
panel (d)) can be extracted for each bin in percentile of cross section.}

\end{center}
\end{figure}

Use of the Octagon silicon detector signals as a centrality measure introduces
an additional complication not present for the Paddle counters.  The precise
vertex position of each event is required for the merging and angle correction
of valid hits in the Octagon.  PHOBOS has developed several techniques to
determine the primary collision vertex, including use of the Vertex detector
and straight-line tracks in the first six planes of the Spectrometer.
However, due to the requirement of any of these valid vertices, the resulting
data set is not only biased by the intrinsic trigger efficiency, but also by
the vertex reconstruction efficiency.  Additional inefficiencies are
introduced for low multiplicity events and this fact is the primary reason
that PHOBOS has, thus far, only published data for the top 50\% of cross
section for Au+Au data, where there are no such inefficiencies.  Despite
these additional complications, exploiting the Octagon detector signals as a
centrality measure greatly expands the available solid angle for centrality
determination.  As shown in Fig.~\ref{WP40_Eoct_etc_Data_MC}, a good match between the
data and MC simulations in all regions of pseudorapidity shown in
Fig.~\ref{WP39_dNdeta_eta_TrigRegions} gives confidence in the validity of the procedure.

Once the choice of pseudorapidity region for the centrality
determination is made and the corresponding efficiency is determined, the
resulting multiplicity related distribution can be divided into percentile of
total cross section bins, as illustrated in Fig.~\ref{WP41_Trigger_ZDC_Paddle_Npart}, panels (a)
and (b).  Comprehensive MC simulations of these signals, that include Glauber
model calculations of the collision geometry, allow the estimation of
$N_{part}$ for a cross section bin, as illustrated in
Fig.~\ref{WP41_Trigger_ZDC_Paddle_Npart}, panels (c) and (d).  The most central collisions
(b$\sim$0, see Fig.~\ref{WP35_collide_diagram}) will have the largest number of
participants with the obvious upper limit of $N_{part}$=394 for a
``perfectly central'' Au+Au collision, where all nucleons interact.

Systematic uncertainties on the extracted values of $N_{part}$ were
determined with MC simulations that included possible errors in the
overall detection efficiency and also utilized different types of event
generators.  The uncertainty on the deduced $N_{part}$ increased from
$\sim$3\% for central collisions to $\sim$9\% for mid-peripheral.

\begin{table}
\begin{center}
\begin{tabular}{|r|r|r|r|r|r|r|r|} \hline
Energy & $\sigma_{_{NN}}$ & A & syst & stat & B & syst & stat\\
\hline
19.6 & 33 & 0.310 & 0.013 & 0.001 & 1.356 & 0.007 & 0.001 \\
62.4 & 36 & 0.296 & 0.012 & 0.001 & 1.376 & 0.007 & 0.001 \\
130 & 41 & 0.274 & 0.016 & 0.001 & 1.408 & 0.010 & 0.001 \\
200 & 42 & 0.271 & 0.016 & 0.001 & 1.413 & 0.010 & 0.001 \\
\hline
\end{tabular}
\caption{
\label{TabGlaub}
List of the nucleon-nucleon cross section used for the four Au+Au energies
followed by the parameters of the power law fit to $N_{coll}$ vs $N_{part}$,  
$N_{coll}=A\times \left(N_{part}\right)^{B}$, along with their systematic and statistical errors.  The systematic errors 
between the two parameters are highly correlated.}
\end{center}
\end{table}

In principle, $N_{coll}$ could be extracted from the same elaborate simulation 
procedure used for $N_{part}$.  In practice, however, three issues arise.  
First, the ratio of $N_{coll}$ over $N_{part}$ varies dramatically with centrality
(see bottom panel of Fig.~\ref{WP36_Ncoll_Npart_b_Npart}), but the
experimental observables used in the centrality determination, 
when normalized by $N_{part}$, depend only weakly on centrality
(see, for example, Figs.~\ref{WP12_Ntot_Npart_AuAu_19130200_dAu_pp} and 
\ref{WP29_dNdetaMidNpart_Npart_AA_19_200}).  Secondly, while the relationship between 
$N_{coll}$ and $N_{part}$ is very sensitive to the assumed nucleon-nucleon 
cross-section, the correspondence between the observables and $N_{part}$ is 
relatively insensitive to such changes.  In contrast, factors which strongly 
impact the extraction of $N_{part}$, such as the overall detection efficiency 
and the detailed properties of the produced particles, have no influence on the 
correspondence between $N_{coll}$ and $N_{part}$.  For these reasons, it was 
found more effective to determine the values and systematic uncertainties for 
$N_{coll}$ from the derived values of $N_{part}$ by using 
a parameterization of the results of a Glauber calculation (see
Appendix~\ref{SecAppB-1} and Fig.~\ref{WP36_Ncoll_Npart_b_Npart}).
Specifically, the results of the simulation are fit to a power law of the form
$N_{coll}=A\times \left(N_{part}\right)^{B}$ with the parameters given in
Table~\ref{TabGlaub}.  The systematic errors of the fit procedure are
determined by fitting different ranges of the data.  Since the two parameters
are highly anti-correlated (the normalized correlation coefficient ranges from 
-0.997 to -0.998), the changes observed in the different fits, and hence the deduced 
systematic errors, are highly correlated.
This functional form works
well down to values of $N_{part}\sim 20-30$ but begins to deviate by
$\sim$10--20\% for smaller values.  Tables~\ref{TabNpartNcoll62} and
~\ref{TabNpartNcoll200} summarize the values of $N_{part}$ and $N_{coll}$ and
their systematic uncertainties for the centrality bins used for Au+Au at
$\sqrt{s_{_{NN}}}$=62.4 and 200~GeV.

It should be noted that additional physics considerations may impact the
extracted value of $N_{coll}$.  As one example, the results of a
straightforward Glauber simulation can be compared to the output of the HIJING
code for Au+Au at $\sqrt{s_{_{NN}}}$=200~GeV.  At all impact parameters, the
numbers for $N_{part}$ are equal in the two cases to within 10 particles or
less (with the HIJING value consistently higher).  In contrast, $N_{coll}$ from
HIJING is found to be lower by roughly 10\% (resulting from the particular 
implementation of nuclear shadowing in the code \cite{Gyu94}), with a slight 
increase in the difference for more central collisions.

\begin{table}
\begin{center}
\begin{tabular}{|r|r|r|r|r|r|r|r|} \hline
Bin & $N_{part}$ & syst & $N_{coll}$ & syst 1 & syst 2 & syst 3 & syst T\\
\hline
45-50\% & 61 & 7 & 85 & 13 & 3 & 4 & 14\\
35-45\% & 86 & 9 & 136 & 20 & 4 & 7 & 22\\
25-35\% & 130 & 10 & 240 & 26 & 7 & 12 & 29\\
15-25\% & 189 & 9 & 402 & 27 & 12 & 20 & 36\\
6-15\% & 266 & 9 & 643 & 30 & 19 & 32 & 48\\
0-6\% & 335 & 11 & 883 & 40 & 26 & 44 & 65\\
\hline
\end{tabular}
\caption{
\label{TabNpartNcoll62}
List of centrality parameters extracted for each of the fractional cross
section bins used in the analysis of Au+Au at $\sqrt{s_{_{NN}}}$=62.4~GeV.
Bins are labeled by the percentage of the total inelastic cross section with
smaller numbers being more central.  The systematic error in $N_{part}$ is found
as described in the text.  There are three components of the systematic error
in $N_{coll}$: 1) the propagation of the uncertainty in $N_{part}$ through the
power law function, 2) the value from the systematic uncertainty in the fit,
3) an estimate of the systematic uncertainty in the Glauber model itself, 
and T) total found by summing contributions in quadrature.}
\end{center}
\end{table}

\begin{table}
\begin{center}
\begin{tabular}{|r|r|r|r|r|r|r|r|} \hline
Bin & $N_{part}$ & syst & $N_{coll}$ & syst 1 & syst 2 & syst 3 & syst T\\
\hline
45-50\% & 65 & 4 & 99 & 9 & 3 & 5 & 11\\
35-45\% & 93 & 5 & 164 & 12 & 5 & 8 & 15\\
25-35\% & 138 & 6 & 286 & 18 & 9 & 14 & 25\\
15-25\% & 200 & 8 & 483 & 28 & 15 & 24 & 40\\
6-15\% & 276 & 9 & 762 & 35 & 23 & 38 & 57\\
0-6\% & 344 & 11 & 1040 & 47 & 31 & 52 & 77\\
\hline
\end{tabular}
\caption{
\label{TabNpartNcoll200}
List of centrality parameters extracted for each of the fractional cross
section bins used in the analysis of Au+Au at $\sqrt{s_{_{NN}}}$=200~GeV.
Bins are labeled by the percentage of the total inelastic cross section with
smaller numbers being more central.  The systematic error in $N_{part}$ is found
as described in the text.  There are three components of the systematic error
in $N_{coll}$: 1) the propagation of the uncertainty in $N_{part}$ through the
power law function, 2) the value from the systematic uncertainty in the fit,
3) an estimate of the systematic uncertainty in the Glauber model itself,
and T) total found by summing contributions in quadrature.}
\end{center}
\end{table}

\subsection{Centrality determination in d+Au collisions} 
\label{SecAppC-2}

The centrality determination for d+Au collisions at $\sqrt{s_{_{NN}}}$=200~GeV
incorporates the same considerations necessary for Au+Au collisions, however
the details of the analysis techniques differ.

The initial definition of a `valid collision' took on a different form, as the
most basic event selection could not be defined cleanly from the timing signals of
the Paddles.  The asymmetric nature of the collision system resulted in very
different particle multiplicities impinging on the two symmetrically located
Paddle trigger counters, which caused an overall reduction in the timing
resolution compared to that of the Au+Au collision system.  Previous
requirements of a ZDC detector timing coincidence were also no longer possible
due to the unacceptable bias that would be imposed on the dataset.  To ensure
that the events analyzed were real collisions occurring in a usable proximity to
the detector, a reconstructed collision vertex was required.

Lower total multiplicities precluded the high resolution track-based vertex
reconstruction algorithms
developed for Au+Au collisions ($\sigma_z\leq$0.04~cm for central collisions) as it became increasingly inefficient.  A more
efficient, but less accurate ($\sigma_z\sim$0.8~cm for central collisions)
vertex reconstruction method based on global averaging techniques across the
entire Octagon was created.  This selection coupled with the intrinsic
triggering of the system was estimated to have an overall efficiency of 83\% for
d+Au collisions at 200~GeV.  This high efficiency data set was used for more
global physics analyses, such as the centrality dependence of the
$dN_{ch}/d\eta$ distribution.

In addition to the Octagon-based vertex determination, a new on-line fast vertex position derived
from the T0 \v{C}erenkov counters (see Appendix~\ref{SecAppA}) was developed for
the d+Au collision data.  This T0 time-difference-based method was utilized
as a primary trigger for some of the data sets.  In addition, the fully
calibrated T0 signals were used in
an off-line vertex-finding algorithm which, combined with the new Octagon-based algorithm discussed above,
provided a very clean event selection.  An additional benefit of on-line
vertex triggering from the T0 detectors was the enhancement of the fraction of data occurring
near the center of the detector ($| z |<20$~cm).  A Paddle-triggered
dataset in d+Au allowed data to be written for collisions occurring within
approximately 2~m of the center of the interaction region.  The additional (T0) requirement forced a larger bias on the
data than the Paddles and Octagon vertices alone and further reduced the overall
trigger+vertex efficiency to ~49\%, but resulted in a much higher fraction of
usable data that provided significantly improved statistics necessary for many
Spectrometer-based analyses.

\begin{figure}[htb] 
\begin{center}

\includegraphics[width=0.86\textwidth]{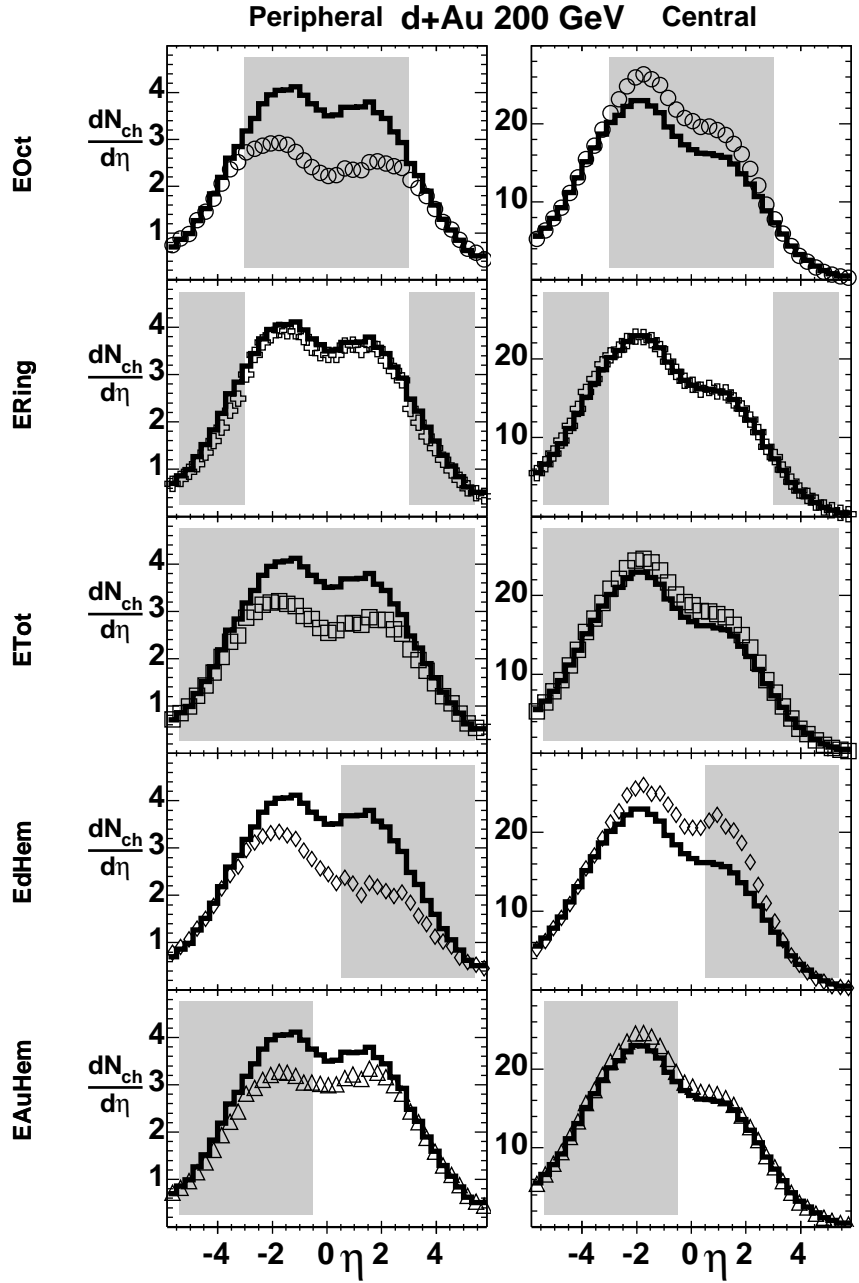}

\caption{ \label{WP42_dNdeta_eta_dAu_Data_MC} 
Reconstructed MC simulated pseudorapidity distributions (open symbols) for d+Au collisions at
$\sqrt{s_{_{NN}}}$=200~GeV for peripheral (left panels) and central (right panels) collisions where the
centrality definition is taken from different regions of pseudorapidity (see text
for discussion).  MC simulations shown utilized the HIJING event generator
coupled to a complete GEANT simulation of the PHOBOS detector.
The unbiased HIJING output (truth values) is shown as histograms.  The shaded areas
indicate the pseudorapidity region covered by each centrality measure.}

\end{center}
\end{figure}

The more significant challenge in the d+Au data analysis was to extract the
centrality dependence of various physics analyses without the centrality
measure itself directly influencing the outcome as a result of strong
auto-correlations.  This issue is not a major factor when measuring quantities
for a minimum-bias configuration \cite{Bac04j}, but it becomes a significant
consideration for any detailed studies requiring a centrality definition.  In
these cases, unlike Au+Au collisions, the centrality determination for d+Au
collisions was found to be strongly dependent on the choice of pseudorapidity
region utilized in the analysis.  This fact is illustrated with MC simulated
data in Fig.~\ref{WP42_dNdeta_eta_dAu_Data_MC}, where strong auto-correlation biases are seen in
the reconstructed pseudorapidity distributions for four of the five different
centrality methods explored.  Specifically a suppression of midrapidity yields
($| \eta | <3$) in the reconstructed spectrum for peripheral collisions is
observed (left column of Fig.~\ref{WP42_dNdeta_eta_dAu_Data_MC}).  The opposite effect is observed for
central collisions, i.e.\ enhancement at midrapidity (right column of Fig.~\ref{WP42_dNdeta_eta_dAu_Data_MC}).
This study utilized five different centrality definitions that each covered
different regions of pseudorapidity: the Octagon detector ($E_{Oct}, | \eta|
\le 3.0$), the Ring detectors ($E_{Ring}, 3.0\le | \eta| \le 5.4$), the
combined coverage of Octagon and Rings ($E_{Tot}, | \eta | \le 5.4$), the
deuteron direction ($E_{dHem}, 0.5\le\eta\le5.4$) and the gold direction ($E_{AuHem},
-5.4\le\eta\le-0.5$).  Both HIJING and AMPT based MC simulations indicated that
a centrality measure based on the signals in the Ring counters provided the
least bias on the measurement.

\begin{figure}[ht]
\begin{center}

\includegraphics[width=0.80\textwidth]{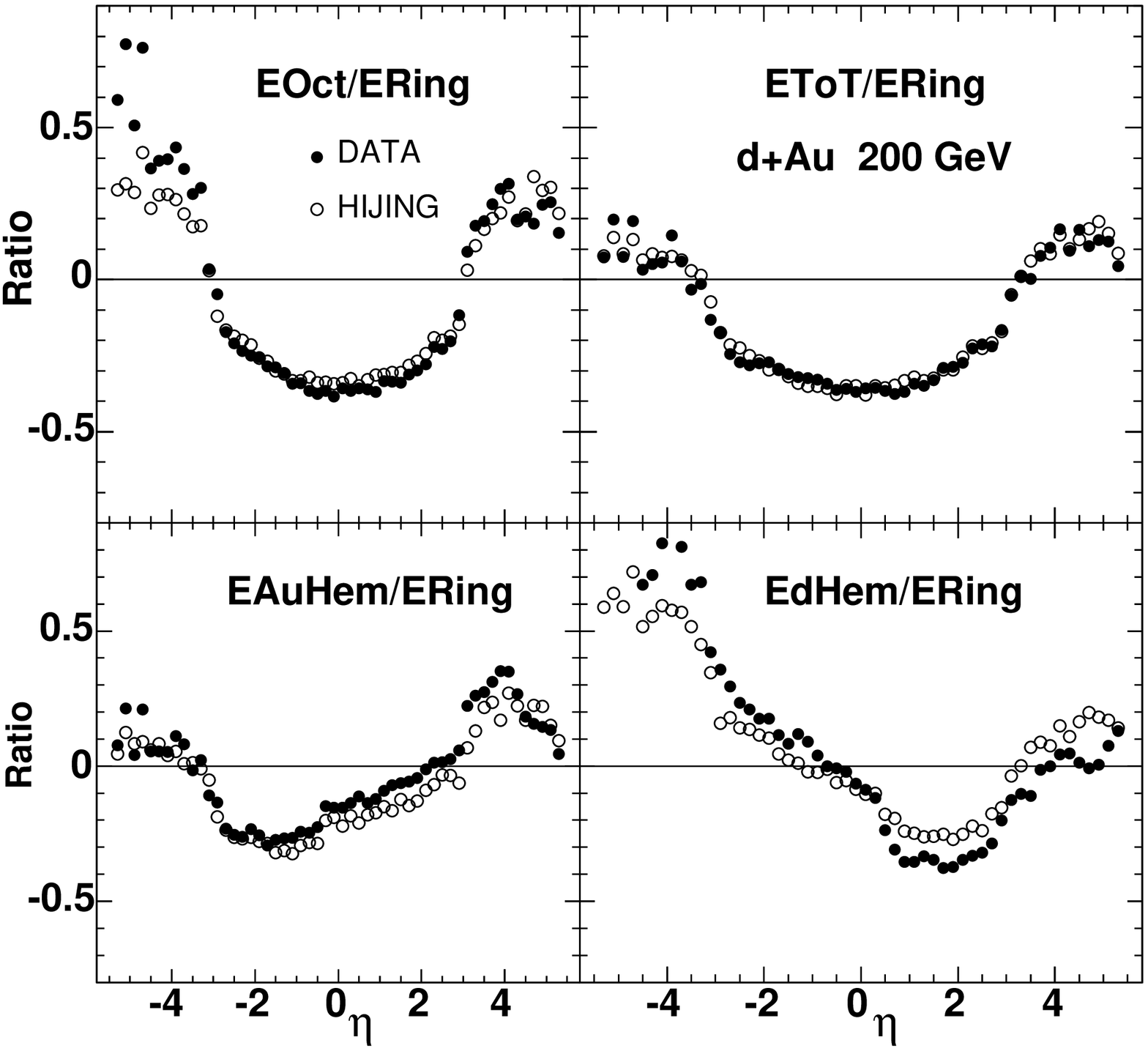}

\caption{ \label{WP43_TrigRatio_dAu_4panel}
Ratios of reconstructed
dN$_{ch}/d\eta$ distributions in d+Au collisions at $\sqrt{s_{_{NN}}}$=200~GeV for both data and MC simulations using different
centrality measures, each of which is selecting on the same percentile of
central collisions.  The good agreement in these ratios gives confidence that
the MC simulations are providing a good basis on which to study the effects of
biases created in the data that result from using different regions of
pseudorapidity for the centrality determination. }

\end{center}
\end{figure}

Additional support for using the MC based simulations to select
the best centrality measure is given in Fig.~\ref{WP43_TrigRatio_dAu_4panel}.  Ratios of the
reconstructed $dN_{ch}/d\eta$ distributions obtained from four centrality
measures relative to that obtained using the $E_{Ring }$ variable are
shown for both MC simulations and data.
These ratios are found to be in very good agreement.  This information, which
is based on data and MC simulation independently, provides the necessary
confidence that using the Ring detectors for the centrality measure will
provide the most accurate experimental result.  It is important to point out that this
study only provided guidance as to the choice of the Rings for the experimental
centrality measure, and the final experimentally measured $dN_{ch}/d\eta$
distributions do not rely on the details of the MC simulation.

\begin{figure}[ht]
\begin{center}

\includegraphics[width=0.65\textwidth]{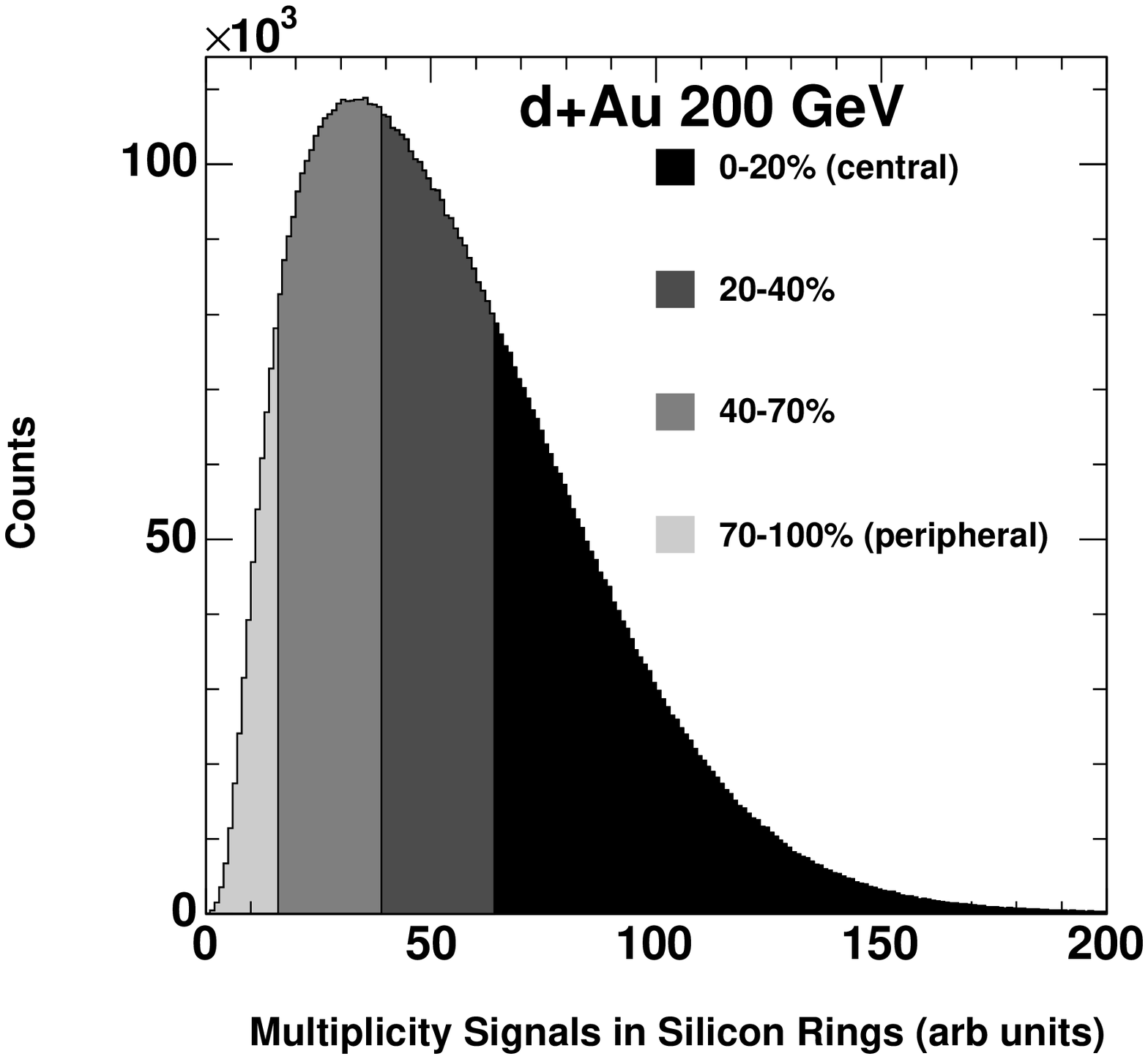}

\caption{ \label{WP44_TrigSlices_dAu}
Illustration of how the centrality is defined for a d+Au collision. The entire
cross section range is used in the analysis.  The shaded bands represent bins
in percentile of cross section based on the multiplicity signals in the Ring
detectors.  The data are shown for an online vertex (T0) restricted data set
(see text).}

\end{center}
\end{figure}

The choice of the Ring detectors for use in the centrality determination for
d+Au collisions, along with the extracted efficiency, allows for the creation of
centrality bins based on percentage of cross section.  For d+Au, a centrality
determination is desired over the entire range of peripheral to central
collisions.  Thus, corrections must be made to both the location of the $E_{Ring}$ bins
and the extracted $N_{part}$ values to properly account for the inefficiencies in
detecting peripheral collisions.  These corrections were made based on
extensive MC simulations using both the HIJING and AMPT event generators.  For
the case of the T0 triggered dataset, the centrality determination was
reanalyzed in terms of the efficiency associated with each cross section bin
and the associated $N_{part}$.  The additional requirement of hits (particles) in
the T0 detectors serves to push the average $N_{part}$ higher, with
the largest shifts for lower centrality classes.  An example of the resulting
centrality bins on the Ring signal distributions for four-bins of
cross section are shown in Fig.~\ref{WP44_TrigSlices_dAu}.  The decreasing efficiency
for more peripheral collisions is immediately evident.

Systematic uncertainties on the deduced average $N_{part}$ values for each
percentile bin of cross section were determined with additional
simulations.  In these studies, the $N_{part}$ distribution taken directly from
Glauber model calculations was matched to the measured centrality related variable, i.e. $E_{Ring}$, distribution from data, and
the average corresponding $N_{part}$ was extracted for each centrality bin.
Many different effects, including various types of detector resolution 
smearing, possible non-linear dependencies of the measured centrality variable
on $N_{part}$, and different deuteron wave functions, were included and the 
analysis repeated.  These studies showed that the mean value of $N_{part}$
estimated from the full HIJING (or AMPT) + GEANT detector simulation was
reasonable and the systematic error on $N_{part}$ reaches $\sim$30\% for the
most peripheral centrality bin, where the overall bias is greatest.


\begin{thebibliography}{99}

\bibitem{Kas93} H. Kastrup, P. Zerwas, eds., QCD 20 Years Later (World Scientific, Singapore, 1993).

\bibitem{PDG04} S. Eidelman, et al., 
(Particle Data Group),
Phys. Lett. B 592 (2004) 1.

\bibitem{Gro73} D. J. Gross, F. Wilczek,
Phys. Rev. Lett. 30 (1973) 1343.

\bibitem{Pol73} H. Politzer, 
Phys. Rev. Lett. 30 (1973) 1346.

\bibitem{Nob04}The work in references \cite{Gro73} and \cite{Pol73} was recognized 
by the 2004 Nobel Prize in Physics; http://nobelprize.org/physics/laureates/2004/.

\bibitem{Col75} J. C. Collins, M. J. Perry, 
Phys. Rev. Lett. 34 (1975) 1353.

\bibitem{Shu80} E. V. Shuryak, 
Phys. Reports  61 (1980) 71.

\bibitem{Sat81} H. Satz, ed., Statistical Mechanics of Quarks and Hadrons, (North-Holland, 1981).

\bibitem{Gro81} D. J. Gross, R. D. Pisarski, L. G. Yaffe,
Rev. Mod. Phys. 53 (1981) 43.

\bibitem{Pag75} H. Pagels,
Phys. Reports  16 (1975) 219.

\bibitem{Mar78} W. Marciano, H. Pagels, 
Phys. Reports  36 (1978) 137.

\bibitem{Pis84} R.D. Pisarski, F. Wilczek, 
Phys. Rev. D 29 (1984) 338.

\bibitem{Raj00b} K. Rajagopal,
Acta Phys. Pol. B 31 (2000) 3021;
\mbox{arXiv:hep-ph/0009058} (2000).

\bibitem{Kar02a} F. Karsch, 
Nucl. Phys. A 698 (2002) 199.

\bibitem{Sch03} T. Sch\"afer, 
\mbox{arXiv:hep-ph/0304281} (2003).

\bibitem{Alf01} M. G. Alford, 
Ann. Rev. Nucl. Part. Sci. 51 (2001) 131.

\bibitem{Raj00a} K. Rajagopal, F. Wilczek, 
\mbox{arXiv:hep-ph/0011333} (2000).

\bibitem{Bay76} G. Baym, S. A. Chin,
Phys. Lett. B 62 (1976) 241.

\bibitem{Cha77}G. Chapline, M. Nauenberg,
Phys. Rev. D 16 (1977) 450.

\bibitem{Cha73}G. F. Chapline, M. H. Johnson, E. Teller, M. S. Weiss,
Phys. Rev. D 8 (1973) 4302.

\bibitem{QM04} See Proceedings of the 
International Conference on Ultra-Relativistic Nucleus-Nucleus Collisions:Quark Matter; 
J. Phys. G 30 (2004) S633; 
Nucl. Phys. A 715 (2003) 1c; 
{\em ibid}. A 698 (2002) 1c; 
{\em ibid}. A 661 (1999) 1c; 
{\em ibid}. A 638 (1998) 1c; 
{\em ibid}. A 610 (1996) 1c; 
{\em ibid}. A 590 (1995) 1c; 
{\em ibid}. A 566 (1994) 1c; 
{\em ibid}. A 544 (1992) 1c; 
{\em ibid}. A 525 (1991) 1c; 
{\em ibid}. A 498 (1989) 1c; 
Z. Phys. C 38 (1988) 1; 
Nucl. Phys. A 461 (1987) 1c; 
``Quark Matter '84'', K. Kajantie, ed. (Springer Verlag, 1985); 
Nucl. Phys. A 418 (1984) 1c; 
Quark Matter Formation and Heavy Ion Collisions, Proceedings of the Bielefeld Workshop 1982, M. Jacob, H. Satz, eds. (WSPC, Singapore, 1983); 
Phys. Reports  88 (1982) 321; 
Workshop on future relativistic heavy ion experiments, R. Bock, R. Stock, eds., (Darmstadt, 1980).

\bibitem{Jac04}P. M. Jacobs, X. -N. Wang,
Prog. Part. Nucl. Phys. 54 (2005) 443;
\mbox{arXiv:hep-ph/0405125} (2004).

\bibitem{Hwa03} R. C. Hwa, X. -N. Wang, eds., Quark Gluon Plasma 3 (World Scientific, Singapore, 2003).

\bibitem{Kol03} P. F. Kolb, U. Heinz, 
in Quark-Gluon Plasma 3 (World Scientific, Singapore, 2003); 
\mbox{arXiv:nucl-th/0305084} (2003).

\bibitem{Ris04}D. H. Rischke,
Prog. Part. Nucl. Phys. 52 (2004) 197.

\bibitem{Wpa04} White Papers from BRAHMS, PHENIX, and STAR, submitted to
Nucl. Phys. A.

\bibitem{Kan03} K. Kanaya, 
Nucl. Phys. A 715 (2003) 233c.

\bibitem{Plu84} M. Pl\"{u}mer, S. Raha, R. M. Weiner,
Nucl. Phys. A 418 (1984) 549c.

\bibitem{Shu04a} E. V. Shuryak, 
Nucl. Phys. A 750 (2005) 64.

\bibitem{Shu04c}E. V. Shuryak,
Prog. Part. Nucl. Phys. 53 (2004) 273.

\bibitem{Hor91} G. T. Horowitz, A. Strominger, 
Nucl. Phys. B 360 (1991) 197.

\bibitem{Gub98}S. S. Gubser, I. R. Klebanov, A. A. Tseytlin, 
Nucl. Phys. B 534 (1998) 202.

\bibitem{Kar03} F. Karsch, S. Datta, E. Laermann, P. Petreczky, S. Stickan, I. Wetzorke, 
Nucl. Phys. A 715 (2003) 701.

\bibitem{Shu04b} E. V. Shuryak, I. Zahed,
Phys. Rev. C 70 (2004) 021901R.

\bibitem{Kaj03}K. Kajantie, M. Laine, K. Rummukainen, Y. Schr\"{o}der,
 Phys. Rev. D 67 (2003) 105008.

\bibitem{Kar04} F. Karsch,
J. Phys. G  30 (2004) S887.

\bibitem{Eli04}M. D'Elia, M. P. Lombardo, 
\mbox{arXiv:hep-lat/0409010} (2004).

\bibitem{Eji04} S. Ejiri, C. R. Allton, M. Doering, S. J. Hands, O. Kaczmarek, F. Karsch, E. Laermann, K. Redlich,
\mbox{arXiv:hep-lat/0408046} (2004).

\bibitem{Fod04}Z. Fodor, S. D. Katz,
Jour. High Ener. Phys. 0404 (2004) 50;
\mbox{arXiv:hep-lat/0402006} (2004).

\bibitem{For04} P. de Forcrand, O. Philipsen,
 Prog. Theor. Phys. Suppl. 153 (2004) 127.

\bibitem{Dav04} C. T. H. Davies, et al., 
(HPQCD and UKQCD),
Phys. Rev. Lett. 92 (2004) 022001.

\bibitem{Lae03}E. Laermann, O. Philipsen,
Ann. Rev. Nucl. Part. Sci. 53 (2003) 163. 

\bibitem{Kar02b}F. Karsch,
Lect. Notes Phys. 583 (2002) 209.

\bibitem{Bac03c} B. B. Back, et al., 
(PHOBOS), 
Phys. Rev. Lett. 91 (2003) 052303.

\bibitem{Bac00a} B. B. Back, et al., 
(PHOBOS),
Phys. Rev. Lett. 85 (2000) 3100.

\bibitem{Bac02a} B. B. Back, et al., 
(PHOBOS),
Phys. Rev. C 65 (2002) 031901R.

\bibitem{Bac02b} B. B. Back, et al., 
(PHOBOS),
Phys. Rev. Lett. 88 (2002) 022302.

\bibitem{Bac02c} B. B. Back, et al., 
(PHOBOS),
Phys. Rev. C 65 (2002) 061901R.

\bibitem{Bac04e} B. B. Back, et al., 
(PHOBOS),
Phys. Rev. C 70 (2004) 021902R.

\bibitem{Afa02} S. V. Afanasiev, et al.,
(NA49), 
Phys. Rev. C 66 (2002) 054902.

\bibitem{Ant04} T. Anticic, et al., 
(NA49),
Phys. Rev. C 69 (2004) 024902.

\bibitem{Ahl98a} L. Ahle, et al.,
(E866),
Phys. Rev. C 57 (1998) 466R.

\bibitem{Bac02e} B. B. Back, et al.,
(E917),
Phys. Rev. C 66 (2002) 054901.

\bibitem{Ahl00} L. Ahle, et al.,
(E866),
Phys. Lett. B 490 (2000) 53.

\bibitem{Kla03}
J.~L.~Klay, et al.,
(E895),
Phys.\ Rev.\ C 68 (2003) 054905.

\bibitem{Dun99}J. Dunlop,
Mass. Inst. of Tech. PhD. Thesis (1999).

\bibitem{Bac04j} B. B. Back, et al., 
(PHOBOS),
Phys. Rev. Lett. 93 (2004) 082301.

\bibitem{Esk02} K. J. Eskola,
Nucl. Phys. A 698 (2002) 78c.

\bibitem{Bas99a} S. A. Bass, et al., 
Nucl. Phys. A 661 (1999) 205c.

\bibitem{Wan99}X.-N. Wang, in \cite{Bas99a}, 210c (1999).

\bibitem{Zha99} B. Zhang, in \cite{Bas99a}, 220c (1999).

\bibitem{Sor99} H. Sorge, in \cite{Bas99a}, 217c (1999); \mbox{arXiv:nucl-th/9905008} (1999).

\bibitem{Ble99} M. Bleicher, in \cite{Bas99a}, 218c (1999).

\bibitem{Cas99} W. Cassing, in \cite{Bas99a}, 222c (1999).

\bibitem{Dre99} H. J. Drescher, in \cite{Bas99a}, 216c (1999).

\bibitem{Cap99} A. Capella, A. Kaidalov, J. Tran Thanh Van,
Heavy Ion Phys. 9 (1999) 169; 
\mbox{arXiv:hep-ph/9903244} (1999).

\bibitem{Ran99} J. Ranft,
\mbox{arXiv:hep-ph/9911213} (1999).

\bibitem{Arm00} N. Armesto, C. Pajares, 
Int. J. Mod. Phys. A 15 (2000) 2019;
\mbox{arXiv:hep-ph/0002163 (2000}).

\bibitem{Jeo00} S. Jeon, J. Kapusta, 
Contest on the RHIC predictions, June 2000.

\bibitem{Esk00} K. J. Eskola, K. Kajantie, P. V. Ruuskanen, K. Tuominen,
Nucl. Phys. B 570 (2000) 379.

\bibitem{Bas99b} S. A. Bass, et al.,
Phys. Rev. C 60 (1999) 021902.

\bibitem{Sta99} J. Stachel, in \cite{Bas99a}, 226c (1999).

\bibitem{Kra01} A. Krasnitz, R. Venugopalan,
Phys. Rev. Lett  86 (2001) 1717.

\bibitem{Kha01a} D. Kharzeev, E. Levin, 
Phys. Lett. B 523 (2001) 79.


\bibitem{Kha03} D. Kharzeev, E. Levin, L. McLerran, 
Phys. Lett. B 561 (2003) 93.

\bibitem{Kha04} D.~Kharzeev, Y.V.~Kovchegov, K.~Tuchin,
Phys. Lett. B 599 (2004) 23.

\bibitem{Bac03d} B. B. Back, et al., 
(PHOBOS),
Phys. Rev. Lett. 91 (2003) 072302.

\bibitem{Bac04f} B. B.~Back, et al.,
(PHOBOS),
Phys. Rev. C 70 (2004) 061901(R).

\bibitem{Ars04} I.~Arsene, et al., 
(BRAHMS),
Phys. Rev. Lett. 93 (2004) 242303.

\bibitem{Liu04} M.X. Liu, et al., 
(PHENIX),
J. Phys. G 30 (2004) S1193.

\bibitem{Acc04} A. Accardi, 
\mbox{arXiv:nucl-th/0405046} (2004).

\bibitem{Adc01} K. Adcox, et al., 
(PHENIX),
Phys. Rev. Lett. 87 (2001) 052301.

\bibitem{Ada04c} J. Adams, et al.,
(STAR), 
Phys. Rev. C 70 (2004) 054907. 

\bibitem{Bac04a} B. B. Back, et al.,
(PHOBOS), 
Phys. Lett. B 578 (2004) 297.

\bibitem{Bac04c} B. B. Back, et al.,
(PHOBOS), 
Phys. Rev. C 70 (2004) 051901(R).

\bibitem{Adl04a} S. S. Adler, et al.,
(PHENIX), 
Phys. Rev. C 69 (2004) 034909.

\bibitem{Bjo83} J. D. Bjorken, 
Phys. Rev. D  27 (1983) 140.

\bibitem{Fer50}E. Fermi,
Prog. Theor. Phys. 5 (1950) 570.

\bibitem{Hag65}R. Hagedorn,
Nuovo Cim. Suppl. 3 (1965) 147.

\bibitem{Hag67}R. Hagedorn, J. Ranft,
Nuovo Cim. Suppl. 6 (1968) 169.

\bibitem{Bac01a} B. B. Back, et al.,
(PHOBOS), 
Phys. Rev. Lett. 87 (2001) 102301.

\bibitem{Bac03a} B. B. Back, et al., 
(PHOBOS),
Phys. Rev. C 67 (2003) 021901R.

\bibitem{Bea96} I. G. Bearden, et al., 
(NA44),
Phys. Lett. B  388 (1996) 431.

\bibitem{Bac99} J. B\"{a}chler, et al., 
(NA49),
Nucl. Phys. A 661 (1999) 45.

\bibitem{Ahl99} L. Ahle, et al., 
(E866),
Phys. Rev. C 60 (1999) 064901.

\bibitem{Ahl98b} L. Ahle, et al.,
(E866),
Phys. Rev. Lett. 81 (1998) 2650.

\bibitem{Bec01} F. Becattini, J. Cleymans, A. Ker\"{a}nen, E. Suhonen, 
K. Redlich,
Phys. Rev. \mbox{C 64} (2001) 024901.

\bibitem{Bec96} F. Becattini,
Z. Phys. C 69 (1996) 485.

\bibitem{Bra99} P. Braun-Munzinger, I. Heppe, J. Stachel, 
Phys. Lett. B 465 (1999) 15.

\bibitem{Bac04b} B. B. Back, et al., 
(PHOBOS),
Phys. Rev. C 70 (2004) 011901R.

\bibitem{Gyu94}M. Gyulassy, X. N. Wang, 
Comput. Phys. Commun. 83 (1994) 307; HIJING v1.383 used for d+Au, and v1.35 used for Au+Au.

\bibitem{Sor95}H. Sorge, 
Phys. Rev. C 52 (1995) 3291; version 2.4 including rope formation was used.

\bibitem{Lin01}Z. W. Lin, S. Pal, C. M. Ko, B. A. Li, B. Zhang, 
Phys. Rev. C 64 (2001) 011902.

\bibitem{Zha00}B. Zhang, C. M. Ko, B. A. Li, Z. Lin, 
Phys. Rev. C 61 (2000) 067901.

\bibitem{Ber04} C. Bernard, et al., 
(MILC),
Phys. Rev. D 71 (2005) 034504. 

\bibitem{Bec04} F. Becattini, M. Gazdzicki, A. Ker\"{a}nen, J. Manninen and
R. Stock, 
Phys. Rev. C 69 (2004) 024905.

\bibitem{Bac02d} B. B. Back, et al., 
(PHOBOS),
Phys. Rev. Lett. 89 (2002) 222301.

\bibitem{Bac04h} B. B.~Back, et al.,
(PHOBOS),
submitted to Phys. Rev. C (RC); \mbox{arXiv:nucl-ex/0407012} (2004).

\bibitem{Ack01} K. H. Ackermann, et al., 
(STAR),
Phys. Rev. Lett. 86 (2001) 402.

\bibitem{Huo04} P. Huovinen, private communication of a calculation using the results of P. F. Kolb, P. Huovinen, U. W. Heinz, H. Heiselberg,
Phys. Lett. B 500 (2001) 232.


\bibitem{Bus93} W. Busza, 
in Particle Production in Highly Excited Matter, H. H. Gutbrod,
J. Rafelski, eds., (Plenum Press, Proceedings of a NATO Advanced Study Institute
on Particle Production in Highly Excited Matter, 1993) p. 149.

\bibitem{Col85}  J. C. Collins, D. E. Soper, G. Sterman,
Nucl. Phys. B 261 (1985) 104.

\bibitem{Col88} J. C. Collins, D. E. Soper, G. Sterman, 
in Perturbative QCD, 
A.H. Mueller, ed., (World Scientific, 1988); Adv.\ Ser.\ Direct.\ High Energy
Phys. 5 (1988) 1.

\bibitem{Adc02} K. Adcox, et al., 
(PHENIX),
Phys. Rev. Lett. 88 (2002) 022301.

\bibitem{Adc03} K. Adcox, et al., 
(PHENIX),
Phys. Lett. B 561 (2003) 82.

\bibitem{Bac04d} B. B. Back, et al.,
(PHOBOS), 
Phys. Rev. Lett. 94 (2005) 082304.

\bibitem{Cro75} J. W. Cronin, H. J. Frisch, M. J. Shochet, J. P. Boymond, P. A. Pirou\'{e}, R. L. Sumner,
Phys. Rev. D 11 (1975) 3105.
 
\bibitem{Ant79} D. Antreasyan, J. W. Cronin, H. J. Frisch, M. J. Shochet, L. Kluberg, P. A. Pirou\'{e}, R. L. Sumner,
Phys. Rev. D 19 (1979) 764. 

\bibitem{Acc02}A. Accardi,
\mbox{arXiv:hep-ph/0212148} (2002).

\bibitem{Vit03}I. Vitev.
Phys. Lett. B 562 (2003) 36.

\bibitem{Ars03} I. Arsene, et al., 
(BRAHMS),
Phys. Rev. Lett. 91 (2003) 072305.

\bibitem{Adl03b} S. S. Adler, et al., 
(PHENIX),
Phys. Rev. Lett. 91 (2003) 072303.

\bibitem{Ada03} J. Adams, et al.,
(STAR), 
Phys. Rev. Lett. 91 (2003) 072304.

\bibitem{Adl03a} C. Adler, et al., 
(STAR),
Phys. Rev. Lett. 90 (2003) 082302.

\bibitem{Ada04b} J. Adams, et al., 
(STAR),
Phys. Rev. Lett. 93 (2004) 252301.

\bibitem{Adl03c} S. S. Adler, et al., 
(PHENIX),
Phys. Rev. Lett. 91 (2003) 182301.

\bibitem{Ada04a} J. Adams, et al., 
(STAR),
Phys. Rev. Lett. 92 (2004) 052302.

\bibitem{Mol03} D. Molnar, S. A. Voloshin, 
Phys. Rev. Lett. 91 (2003) 092301.

\bibitem{Bac03e} B. B. Back, et al., 
(PHOBOS),
submitted to Phys. Rev. C(RC); \mbox{arXiv:nucl-ex/0301017} (2003).

\bibitem{Kla01} J. Klay, 
U.C. Davis PhD. Thesis (2001).

\bibitem{Eid04} S. Eidelman, et al.,
(Particle Data Group),
Phys. Lett. B 592 (2004) 1.

\bibitem{Ver02} G. I. Veres, E\"otv\"os Lor\'and University PhD. Thesis,\\
http://na49info.cern.ch/cgi-bin/wwwd-util/NA49/NOTE?292 (2002).

\bibitem{Bre82} A. E. Brenner, et al., 
Phys Rev D 26 (1982) 1497.

\bibitem{Bas80a} M. Basile, et al., 
Phys. Lett. B 92 (1980) 367.

\bibitem{Bas80b} M. Basile, et al., 
Phys. Lett. B 95 (1980) 311.

\bibitem{Bac01c} B. B. Back, et al., 
(E917),
Phys. Rev. Lett. 86 (2001) 1970.

\bibitem{Mue83} A. H. Mueller,
Nucl. Phys. B 213 (1983) 85.

\bibitem{Bac04i}B. B. Back, et al.,
(PHOBOS),
 submitted to Phys. Rev. Lett.; \mbox{arXiv:nucl-ex/0409021} (2004).

\bibitem{Bea03} I. G. Bearden, et al.,
(BRAHMS),
Phys. Rev. Lett. 93 (2004) 102301.

\bibitem{Bus84} W. Busza, A. S. Goldhaber, 
Phys. Lett. B 139 (1984) 235.

\bibitem{Bus88} W. Busza, R. Ledoux, 
Ann. Rev. Nucl. Part. Sci.  38 (1988) 119.

\bibitem{Aln87} G. J. Alner, et al., 
(UA5),
Phys. Rep. 154 (1987) 247.

\bibitem{Eli80} J. E. Elias, et al.,
(E178), 
Phys. Rev. D 22 (1980) 13.

\bibitem{Bus75} W. Busza, et al., 
(E178), 
Phys. Rev. Lett. 34 (1975) 836.

\bibitem{Bus77} W. Busza, 
Acta Phys. Pol. B 8 (1977) 333.

\bibitem{Bia76} A. Bia\l as, M. Bleszy\'{n}ski, W. Czy\.{z}, 
Nucl. Phys. B 111 (1976) 461.

\bibitem{Whi74} J. Whitmore, 
Phys. Rep. 10 (1974) 273.

\bibitem{Whi76} J. Whitmore, 
Phys. Rep. 27 (1976) 187.

\bibitem{Bog74} H. B{\o}ggild, T. Ferbel,
Ann. Rev. Nucl. Sci. 24 (1974) 247.

\bibitem{Bac01b} B. B. Back, et al.,
(PHOBOS), 
Phys. Rev. Lett. 87 (2001) 102303.

\bibitem{Bri90} D. H. Brick, et al.,
(E565/570), 
Phys. Rev. D 41 (1990) 765.

\bibitem{Dem84} C. DeMarzo, et al.,
(NA5), 
Phys. Rev. D 29 (1984) 2476.

\bibitem{Hal77} C. Halliwell, et al., 
(E178), 
Phys. Rev. Lett. 39 (1977) 1499.

\bibitem{Bar83} D. S. Barton, et al., 
(E451),
Phys. Rev. D 27 (1983) 2580.

\bibitem{Aln86} G. J. Alner, et al., 
(UA5),
Z. Phys. C 33 (1986) 1.

\bibitem{Ste00} H. Stenzel, 
(ALEPH),
Contributed paper to ICHEP2000
(2000).

\bibitem{Sjo94} PYTHIA manual, T. Sj\"ostrand, 
Comp. Phys. Comm. 82 (1994) 74. JETSET 7.4 is currently part of the PYTHIA code.

\bibitem{Bac04l} B. B.~Back, et al.,
(PHOBOS),
Phys. Rev. C 71 (2005) 021901(R).

\bibitem{Ros75} A. M. Rossi, G. Vannini, A. BussiŽère, E. Albini, D. D'Alessandro,
and G. Giacomelli,
Nucl. Phys. B 84 (1975) 269.

\bibitem{Gue76} K. Guettler, et al.,
Nucl. Phys. B 116 (1976) 77.

\bibitem{Ban83} M. Banner, et al.,
(UA2),
Phys. Lett. B 122 (1983) 322.

\bibitem{Agu91} M. Aguilar-Benitez, et al.,
Z. Phys. C 50 (1991) 405.

\bibitem{Arn82} G. Arnison, et al., 
(UA1),
Phys. Lett. B 118 (1982) 167.

\bibitem{Alb90} C. Albajar, et al.,
(UA1), 
Nucl. Phys. B 335 (1990) 261.

\bibitem{Boc96} G. Bocquet, et al.,
(UA1), 
Phys. Lett. B 366 (1996) 434.

\bibitem{Ben69} J. Benecke, T. T. Chou, C. N. Yang, E. Yen, 
Phys. Rev. 188 (1969) 2159.

\bibitem{Cho70} T. T. Chou, C.-N. Yang, 
Phys. Rev. Lett. 25 (1970) 1072.

\bibitem{Fey69} R. P. Feynman,
Phys. Rev. Lett. 23 (1969) 1415.

\bibitem{Fey72} R. P. Feynman, Photon-Hadron Interactions (W.A. Benjamin Inc., 1972)
p. 237-249.

\bibitem{Sjo01} T. Sj\"ostrand, P. Ed\'{e}n, C. Friberg, L. L\"onnblad, G. Miu,
S. Mrenna, E. Norrbin, 
Comp. Phys. Comm. 135 (2001) 238.

\bibitem{Sjo03} T. Sj\"ostrand, L. L\"onnblad, S. Mrenna, P. Skands,
\mbox{arXiv:hep-ph/0308153} (2003).

\bibitem{Tho77} W. Thom\'{e}, et al., 
Nucl. Phys. B 129 (1977) 365.

\bibitem{Abr99}	P. Abreu, et al., 
(DELPHI),
Phys. Lett. B 459 (1999) 397.

\bibitem{And76} B. Andersson, 
in Proceedings  VII International Colloquium on Multiparticle Reactions, Tutzing, 1976, and references therein.

\bibitem{Got74} K. Gottfried, 
in Proceedings  V International Colloquium on Multiparticle Reactions, Uppsala, 1974.

\bibitem{Nik81} N. N. Nikolaev, 
Sov. J. Part. Nucl. 12 (1981) 63, and references therein.

\bibitem{Sto75} L. Stodolsky, 
in Proceedings  VI International Colloquium on Multiparticle Reactions, Oxford, 1975.

\bibitem{Ott78} I. Otterlund, et al., 
Nucl. Phys. B 142 (1978) 445, and references therein.

\bibitem{Fre87} S. Fredriksson, et al.,
Phys. Rep. 144 (1987) 187, and references therein.

\bibitem{Bea04}
I.~G.~Bearden, et al.,  
(BRAHMS),
\mbox{arXiv:nucl-ex/0403050} (2004).

\bibitem{Bea01} I. G. Bearden, et al., 
(BRAHMS),
Phys. Lett. B 523 (2001) 227.

\bibitem{Bea02} I. G. Bearden, et al., 
(BRAHMS),
Phys. Rev. Lett. 88 (2002) 202301.

\bibitem{Kol01b} P. F. Kolb, 
Proceedings of the 17th Winter Workshop on Nuclear
Dynamics, Park City Utah 2001. Acta Phys. Hung. New Ser. Heavy Ion Phys. 15 (2002) 279; \mbox{arXiv:nucl-th/0104089} (2001).

\bibitem{Bac04g} B. B.~Back, et al.,
(PHOBOS),
accepted for publication in Phys. Rev. Lett.; \mbox{arXiv:nucl-ex/0406021} (2004).

\bibitem{Han54} R. Hanbury-Brown, R. Q. Twiss, 
Phil. Mag. Ser. 7, Vol. 45, No. 366 (1954) 663.

\bibitem{Han56} R. Hanbury-Brown, R. Q. Twiss, 
Nature  178 (1956) 1046.

\bibitem{Tom02} B. Tomasik, U. A. Wiedemann,
in Quark-Gluon Plasma 3 (World Scientific, Singapore, 2003); 
\mbox{arXiv:hep-ph/0210250}.

\bibitem{Ris96} D. Rischke, M. Gyulassy, 
Nucl. Phys. A 597 (1996) 701.

\bibitem{Hir02} T. Hirano, K. Tsuda, 
Phys. Rev. C 66 (2002) 054905.

\bibitem{Sof02} S. Soff, 
\mbox{arXiv:hep-ph/0202240} (2002).

\bibitem{Pra86} S. Pratt, 
Phys. Rev. D 33 (1986) 1314.

\bibitem{Bac04k} B. B. Back, et al.,
(PHOBOS),
submitted to Phys. Rev. C (RC); \mbox{arXiv:nucl-ex/0409001} (2004).

\bibitem{Ada04d} J. Adams, et al.,
(STAR),
Phys. Rev. Lett. 93 (2004) 012301.

\bibitem{Adl04b} S. S. Adler, et al.,
(PHENIX),
Phys. Rev. Lett. 93 (2004) 152302.


\bibitem{Bak03} M. D. Baker, 
Proc. of the Eighteenth Lake Louise Winter Institute; \mbox{arXiv:nucl-ex/0309002} (2003).

\bibitem{Ren04} T. Renk,
Phys. Rev. C 70 (2004) 021903R.

\bibitem{Tea03} D. Teaney,
\mbox{arXiv:nucl-th/0301099} (2003).

\bibitem{App98} H. Appelsh\"{a}user, et al.,
(NA49),
Eur. Phys. J. C 2 (1998) 661.

\bibitem{Ans89} R. E. Ansorge, et al.,
Z. Phys. C 43 (1989) 357.

\bibitem{Kha01b} D. Kharzeev, M Nardi,
Phys. Lett. B 507 (2001) 121.

\bibitem{Kha01c} D. Kharzeev, E. Levin, M. Nardi,
\mbox{arXiv:hep-ph/0111315} (2001).

\bibitem{Bre95} A. Breakstone, et al.,
Z. Phys. C 69 (1995) 55.

\bibitem{Dri82} D. Drijard, et al., 
Nucl. Phys. B 208 (1982) 1.

\bibitem{Bac03b} B. B. Back, et al., 
(PHOBOS),
Nucl. Inst. Meth. A 499 (2003) 603.

\bibitem{Bac00b} B. B. Back, et al., 
(PHOBOS),
Nucl. Inst. Meth. A 447 (2000) 257.

\bibitem{Bac99a} B. B. Back, et al., 
(PHOBOS),
Nucl. Phys. B Proc. Sup. 78 (1999) 245.

\bibitem{Bac98} B. B. Back, et al., 
(PHOBOS),
Nucl. Inst. Meth. A 419 (1998) 549.

\bibitem{Ber89} G. F. Bertsch, 
Nucl. Phys. A 498 (1989) 173.

\bibitem{Hei96} U. Heinz, 
Nucl. Phys. A 610 (1996) 264.

\bibitem{Yan78} F. Yano, S. Koonin, 
Phys. Lett. B 78 (1978) 556.

\bibitem{Pod83} M. Podgoretskii, 
Sov. J. Nucl. Phys. 37 (1983) 272.

\end{thebibliography}
\end{document}